\documentclass{aa}
\usepackage[varg]{txfonts}
\usepackage{natbib}
\usepackage{graphicx}
\usepackage{mathabx}
\usepackage{amsmath}

\usepackage{color}
\usepackage{multirow}
\usepackage{longtable}
\usepackage{pdflscape}
\usepackage{rotating}

\bibpunct{(}{)}{;}{a}{}{,}

\begin{document}

\title{X-shooter spectroscopy of young stars with disks}
\subtitle{The TW Hydrae association as a probe of the final stages of disk accretion\thanks{Based on observations collected at the European Southern Observatory under ESO programs 084.C--0269(B), 085.C--0238(A), 086.C--0173(A), 087.C--0244(A), 287.C--5039(A), 089.C--0143(A), 093.C--0097(A), and 095.C--0147(A).}}

\author{L. Venuti\inst{1, 2, 3, }\thanks{NASA Postdoctoral Program (NPP) fellow.} \and B. Stelzer\inst{1, 4} \and J.~M. Alcal\'a\inst{5} \and C.~F. Manara\inst{6} \and A. Frasca\inst{7} \and R. Jayawardhana\inst{2} \and S.~Antoniucci\inst{8} \and C.~Argiroffi\inst{9, 4} \and A.~Natta\inst{10} \and B.~Nisini\inst{8} \and S.~Randich\inst{11} \and A.~Scholz\inst{12}}

\institute{Eberhard Karls Universit\"at, Institut f\"ur Astronomie und Astrophysik, Sand 1, 72076 T\"ubingen, Germany\\ e-mail: laura.venuti@nasa.gov \and Department of Astronomy, Cornell University, Space Sciences Building, Ithaca, NY 14853, USA \and NASA Ames Research Center, Moffett Blvd, Mountain View, CA 94035, USA \and INAF -- Osservatorio Astronomico di Palermo, Piazza del Parlamento 1, 90134 Palermo, Italy \and INAF -- Osservatorio Astronomico di Capodimonte, Salita Moiariello 16, 80131 Napoli, Italy \and European Southern Observatory, Karl-Schwarzschild-Strasse 2, 85748 Garching bei M\"{u}nchen, Germany \and INAF -- Osservatorio Astrofisico di Catania, via S. Sofia, 78, 95123 Catania, Italy \and INAF -- Osservatorio Astronomico di Roma, via di Frascati 33, 00078 Monte Porzio Catone, Italy \and Dipartimento di Fisica e Chimica, Universit\`a degli Studi di Palermo, Piazza del Parlamento 1, 90134 Palermo, Italy \and Dublin Institute for Advanced Studies, 31 Fitzwilliam Place, Dublin 2, Ireland \and INAF -- Osservatorio Astrofisico di Arcetri, Largo E. Fermi 5, 50125 Florence, Italy \and SUPA, School of Physics \& Astronomy, University of St Andrews, North Haugh, St Andrews KY169SS, UK}

\date{Received  / Accepted }

\abstract {Measurements of the fraction of disk-bearing stars in clusters as a function of age indicate protoplanetary disk lifetimes $\lesssim$\,10~Myr. However, our knowledge of the time evolution of mass accretion in young stars over the disk lifespans is subject to many uncertainties, especially at the lowest stellar masses ($M_\star$).}{We investigate ongoing accretion activity in young stars in the TW Hydrae association (TWA). The age of the association ($\sim$8--10~Myr) renders it an ideal target to probe the final stages of disk accretion, and its proximity ($\sim$50~pc) enables a detailed assessment of stellar and accretion properties down to brown dwarf masses.}{Our sample comprises eleven TWA members with infrared excess, amounting to 85\% of the total TWA population with disks. Our targets span spectral types between M0 and M9, and masses between 0.58 and 0.02~$M_\odot$. We employed homogeneous spectroscopic data from 300 to 2500~nm, obtained synoptically with the X-shooter spectrograph, to derive individual extinction, stellar parameters, and accretion parameters for each object simultaneously. We then examined the luminosity of Balmer lines and forbidden emission lines to probe the physics of the star--disk interaction environment.}{Disk-bearing stars represent around 24\% of the total TWA population. We detected {signatures of ongoing} accretion for 70\% of our TWA targets {for which accurate measurements of the stellar parameters could be derived}. This implies a fraction of accretors of 13--17\% across the entire TWA (accounting for the disk-bearing, potentially accreting members not included in our survey). The spectral emission associated with these stars reveals a more evolved stage of these accretors compared to younger PMS populations studied with the same instrument and analysis techniques (e.g., Lupus): {\it (i)} a large fraction ($\sim$50\%) exhibit nearly symmetric, narrow H$\alpha$ line profiles; {\it (ii)} over 80\% of them exhibit Balmer decrements consistent with moderate accretion activity and optically thin emission; {\it (iii)} less than a third exhibit forbidden line emission in [O\,I]\,6300\AA, indicative of winds and outflows activity; {\it (iv)} only one sixth exhibit signatures of collimated jets. However, the distribution in accretion rates ($\dot{M}_{\rm acc}$) derived for the TWA sample follows closely that of younger regions (Lupus, Chamaeleon~I, $\sigma$~Orionis) over the mass range of overlap ($M_\star$\,$\sim$\,0.1--0.3\,$M_\odot$). An overall correlation between $\dot{M}_{\rm acc}$ and $M_\star$ is detected, best reproduced by the function $\dot{M}_{\rm acc} \propto M_\star^{2.1\pm0.5}$. }{At least in the lowest $M_\star$ regimes, stars that still retain a disk at ages $\sim$8--10~Myr are found to exhibit statistically similar, albeit moderate, accretion levels as those measured around younger objects. This ``slow'' $\dot{M}_{\rm acc}$ evolution apparent at the lowest masses may be associated with longer evolutionary timescales of disks around low-mass stars, suggested by the mass-dependent disk fractions reported in the literature within individual clusters.}


\titlerunning{X-shooter spectroscopy of young stellar objects: disk accretion in the TWA}

\maketitle

\section{Introduction} \label{sec:introduction}

Disks are a ubiquitous result of the earliest stages of star formation, and surround the majority of low-mass stars at an age of $\sim$1~Myr. Over the next few million years, the interaction between the inner disk and the central object is regulated by the process of magnetospheric accretion (see \citealp{hartmann2016} for a recent review). This process has a profound impact on the inner disk evolution and on the fundamental properties of the nascent star, such as mass and angular momentum. It also has a direct impact on planet formation, as it alters the respective gas and dust contents of the disk, modifies the local disk structure, and can trigger or halt planet migration within the disk \citep{alexander2014, morbidelli2016}.

Observational studies of disk frequencies in young star clusters as a function of age \citep[e.g.,][]{hernandez2007, fedele2010, ribas2014} indicate that disks are typically cleared within a timescale of $\sim$10~Myr. However, the evolution of disk accretion during this time frame is less well constrained. Indeed, while it has been shown that the characteristic timescale of disk accretion (as inferred from counting the fraction of accreting stars in populations of different ages; \citealp{jayawardhana2006}, \citealp{fedele2010}) is slightly shorter than that of dust dissipation in the inner disk, the actual age dependence of disk accretion remains to be elucidated. 

Extensive compilations of mass accretion rates ($\dot{M}_{\rm acc}$) in pre-main sequence (PMS) stellar populations have been published in recent years, making use of several accretion tracers: i) the continuum excess emission which arises from the accretion shock at the stellar surface \citep[e.g.,][]{herczeg2008, ingleby2013, alcala2014, alcala2017, manara2017}; ii) the $U$-band flux excess as a probe of the total accretion luminosity, $L_{\rm acc}$ \citep[e.g.,][]{rigliaco2011, manara2012, venuti2014}; iii) the H$\alpha$ line emission \citep[e.g.,][]{demarchi2010, kalari2015, sousa2016, biazzo2019} or other emission lines \citep[e.g.,][]{mohanty2005, nguyen2009, antoniucci2011, antoniucci2014, caratti2012} arising from the heated gas in the accretion columns. Such surveys have enabled, for instance, detailed assessments of the relationship between $\dot{M}_{\rm acc}$ and the stellar mass ($M_\star$), a key observational constraint to validate theories on the dynamics of accretion and of disk evolution \citep{alexander2006, dullemond2006, hartmann2006, vorobyov2008, ercolano2014}. However, many of the well-studied PMS populations only sample the first few ($\lesssim$\,3--5) Myr of the early stellar evolution (but note the works of \citealp{ingleby2014} and \citealp{rugel2018} on accretion in older PMS stars). Tentative relationships between $\dot{M}_{\rm acc}$ and age have been extrapolated from intra-cluster accretion surveys, by combining $\dot{M}_{\rm acc}$ measurements and age estimates of individual cluster members \citep[e.g.,][]{sicilia_aguilar2010, demarchi2010, manara2012, antoniucci2014, hartmann2016}. Such studies have suggested an age dependence of $\dot{M}_{\rm acc}(t)$\,$\propto$\,$t^{-\left(1.3 \pm 0.3\right)}$; however, this approach is subject to the strong uncertainty on individual age estimates for young stars \citep{soderblom2014}, and to the interplay between age-related effects and mass-related effects in $\dot{M}_{\rm acc}$.

Given its proximity and age, the TW~Hydrae association (TWA) is an ideal target to probe disk accretion during its final stages, when gas-rich disks evolve into debris disks. Located at a distance of about 50~pc \citep{weinberger2013}, it is the closest stellar association containing PMS stars with active accretion disks. The most recent census of the TWA \citep{gagne2017} encompasses 55 {\it bona fide} members or high-likelihood candidate members, with or without disks, distributed in a total of 34 star systems (some comprising multiple stars). The bulk of these sources have spectral type (SpT) between late-K and late-M\,/\,early-L. Recent age estimates for the TWA oscillate between $\sim$8~Myr \citep[e.g.,][]{ducourant2014, herczeg2014, donaldson2016} and $\sim$10~Myr \citep{bell2015}, and the association exhibits a disk fraction of nearly 25\% in the M-type spectral class \citep{schneider2012, fang2013}. Signatures of ongoing disk accretion in TWA members have been reported by, e.g., \citet{muzerolle2001}, \citet{scholz2005}, \citet{jayawardhana2006}, \citet{stelzer2007}, \citet{herczeg2009}, \citet{looper2010}. The proximity of the association renders each of its members, down to masses $\lesssim 0.05\,M_\odot$, accessible for detailed individual characterization. In spite of a limited number statistics, the TWA therefore provides key observational constraints in undersampled areas of the multi-parameter space (e.g., stellar mass, disk mass, accretion rate, stellar multiplicity) that governs the evolution of young stars and their disks.

We have recently conducted an extensive spectroscopic survey of TWA members with signatures of disks, making use of the wide-band spectrograph X-shooter \citep{vernet2011} at the ESO Very Large Telescope (VLT). Thanks to its broad synoptic wavelength coverage, which extends from $\sim$300~nm to 2500~nm, X-shooter is especially suited to investigate accretion in young stars: it provides access to a rich variety of accretion diagnostics (excess continuum, Balmer series, multiple optical and near-infrared emission lines), and enables measuring the extinction, photospheric flux, stellar parameters, and accretion parameters simultaneously, with no ambiguity related to the intrinsic variability of young stars \citep[e.g.,][]{costigan2014, venuti2015}. Systematic surveys of accretion in various star-forming regions have been conducted with X-shooter in the past few years (see \citealp{rigliaco2012} in the $\sim$6~Myr-old $\sigma$~Orionis; \citealp{manara2015} in the 1--3~Myr-old $\rho$~Ophiucus; \citealp{alcala2017} in the $<$\,3~Myr-old Lupus; \citealp{manara2017} in the 2--3~Myr-old Chamaeleon~I; \citealp{rugel2018} in the $\sim$11~Myr-old $\eta$~Chamaeleontis). The extent and homogeneity of such surveys has enabled, for instance, assessing the presence of a bimodality in the $\dot{M}_{\rm acc} - M_\star$ distribution at masses $<$\,2~$M_\odot$ and ages $\leq$\,3~Myr, with a steeper $\dot{M}_{\rm acc} - M_\star$ relationship in the lower-mass regime than in the higher-mass regime, and a break in the relationship at $M_\star \sim 0.2-0.3\,M_\odot$ \citep{alcala2017, manara2017}. In addition, a synergy between X-shooter and ALMA \citep{ansdell2016} surveys provided first observational evidence of a correlation between $\dot{M}_{\rm acc}$ and disk dust masses $M_{\rm disk,\,dust}$ in the Lupus cloud \citep{manara2016} and in the Chamaeleon~I region \citep{mulders2017}.

To complement the efforts mentioned above and extend them towards the final stages of disk accretion, we present here the X-shooter view of the accretion process across the TWA. The analysis was conducted with the same approach and methods pursued in previous X-shooter surveys of accretion in younger star-forming regions. This enables a direct and internally consistent comparison of our results to the accretion properties of the $\rho$~Oph, Lupus, Cha~I, $\sigma$~Ori, and $\eta$~Cha populations, aimed to constrain the evolution of disk accretion across the pre-main sequence phase, from subsolar to brown dwarf stellar masses. 

Our paper is organized as follows. Section~\ref{sec:data} provides a description of the observations used in this work, and of their reduction. In Sect.\,\ref{sec:param} we describe the procedure adopted to derive the stellar and accretion parameters for our targets. The resulting picture of accretion across the TWA is presented in Sect.\,\ref{sec:TWA_acc}, and information on the accretion dynamics and accretion/ejection processes on individual TWA members is presented in Sect.\,\ref{sec:acc_dynamics}. Our results are discussed in Sect.\,\ref{sec:discussion}, in the context of the evolution of disk accretion and of the impact of stellar parameters, disk parameters, and environmental conditions. Our conclusions are summarized in Sect.\,\ref{sec:conclusions}.

\section{Sample, observations, and data reduction} \label{sec:data}

\subsection{Target selection} \label{sec:targets}

Our sample comprises 14 stars, selected among the confirmed or putative TWA members upon showing evidence of circumstellar material from mid-infrared excess in {\it Spitzer}/IRAC \citep{fazio2004} or WISE \citep{wright2010} bands. A complete list of the targets is reported in Table~\ref{tab:TWA_targets};
\begin{table*}
\caption{Log of the X-shooter observations.}
\label{tab:TWA_targets}
\centering
\resizebox{\textwidth}{!} {
\begin{tabular}{l c c c c c c c}
\hline\hline
\multirow{2}*{Name} & \multirow{2}*{R.A.\,[hh:mm:ss]} & \multirow{2}*{Dec\,[dd:mm:ss]} & \multirow{2}*{Membership} & \multirow{2}*{Program} & \multirow{2}*{Date} & {Exp.\,time\,[s]} & {Slit\,["]} \\
 & & & & & & {UVB/VIS/NIR} & {UVB/VIS/NIR} \\
\hline
{TWA\,1} & {11:01:51.917} & {-34:42:17.00} & {1} &  {085.C-0238} & {2010-04-07} & \multicolumn{1}{l}{2\,$\times$\,(10/10/10)} &  {0.5/0.4/0.4}\\
 {\mbox{ "}} & " & " & " & " &  " & \multicolumn{1}{l}{2\,$\times$\,(60/60/60)} & " \\
TWA\,3A & 11:10:27.767 & -37:31:51.79 & 2 & 085.C-0238 & 2010-04-07 & \multicolumn{1}{l}{2\,$\times$\,(100/100/100)} & 0.5/0.4/0.4\\
TWA\,3B &  &  & 2 & 086.C-0173 & 2011-01-13 & \multicolumn{1}{l}{2\,$\times$\,(150/150/152)} & 0.5/0.4/0.4\\
TWA\,8A & 11:32:41.246 & -26:51:55.95 & 1 & 086.C-0173 & 2011-01-12 & \multicolumn{1}{l}{2\,$\times$\,(200/200/200)} & 0.5/0.4/0.4\\
TWA\,8B & 11:32:41.165 & -26:52:09.04 & 1 & 086.C-0173 & 2011-01-13 & \multicolumn{1}{l}{2\,$\times$\,(150/150/152)} & 1.0/0.9/0.4\\
TWA\,22 & 10:17:26.892 & -53:54:26.52 & 4 & 086.C-0173 & 2011-01-13 & \multicolumn{1}{l}{2\,$\times$\,(150/150/152)} & 1.0/0.9/0.4\\
{TWA\,27} & {12:07:33.468} & {-39:32:53.00} & {1} & 084.C-0269 & 2010-03-23 & \multicolumn{1}{l}{4\,$\times$\,(900/900/900)} & {1.0/0.9/0.9}\\
{\mbox{ "}} & " &" & " & 089.C-0143 & 2012-04-19 & \multicolumn{1}{l}{4\,$\times$\,(900/850/900)} & " \\
TWA\,28 & 11:02:09.833 & -34:30:35.53 & 1 & 084.C-0269 & 2010-03-23 & \multicolumn{1}{l}{4\,$\times$\,(900/900/900)} & 1.0/0.9/0.9\\
{TWA\,30A} & {11:32:18.317} & {-30:19:51.81} & {1} & {087.C-0244} & {2011-04-23} & \multicolumn{1}{l}{2\,$\times$\,(300/300/300)} & 0.5/0.4/0.4\\
 {\mbox{ "}} & " & " & " & " & " & \multicolumn{1}{l}{2\,$\times$\,(450/450/450)} & 1.0/0.9/0.9\\
 {\mbox{ "}} & " & " & " & 287.C-5039 & 2011-07-15 &  \multicolumn{1}{l}{4\,$\times$\,(230/140/300)} & 0.8/0.9/0.9\\
{TWA\,30B} & {11:32:18.223} & {-30:18:31.65} & {1} & 087.C-0244 & 2011-04-23 & \multicolumn{1}{l}{4\,$\times$\,(900/900/900)} & {1.0/0.9/0.9}\\
 {\mbox{ "}} & " & " & " & 287.C-5039 & 2011-07-15 & \multicolumn{1}{l}{4\,$\times$\,(490/400/580)} & " \\
TWA\,31 & 12:07:10.894 & -32:30:53.72 & 4 & 093.C-0097 & 2014-04-18 & \multicolumn{1}{l}{4\,$\times$\,(500/560/560)} & 1.0/0.9/0.9\\
TWA\,32 & 12:26:51.355 & -33:16:12.47 & 1 & 093.C-0097 & 2014-04-18 & \multicolumn{1}{l}{4\,$\times$\,(470/535/560)} & 1.0/0.9/0.9\\
TWA\,40 & 12:07:48.362 & -39:00:04.40 & 2 & 095.C-0147 & 2015-06-01 & \multicolumn{1}{l}{8\,$\times$\,(900/850/900)} & 1.6/1.5/1.2\\
J1247-3816 & 12:47:44.290 & -38:16:46.40 & 3 & 095.C-0147 & 2015-05-28 & \multicolumn{1}{l}{4\,$\times$\,(900/850/900)} & 1.6/1.5/1.2\\
\hline
\end{tabular}
}
\tablefoot{
``{Name}'' = target identifier following the convention of \citet{gagne2017}. ``{R.A., Dec}'' = J2000 target coordinates, as listed in \citet{gagne2017}. A single set of coordinates is reported for the TWA\,3 system, composed of the spectroscopic binary TWA\,3A and of the tertiary component TWA\,3B, separated by about $1\farcs55$ \citep[][and references therein]{kellogg2017}. ``{Membership}'' = membership likelihood to the TWA, compiled in \citet{gagne2017}, coded as: 1 $\rightarrow$ {\it bona fide} member; 2 $\rightarrow$ high-likelihood candidate member; 3 $\rightarrow$ candidate member; 4 $\rightarrow$ likely contaminant from nearby moving group or association. ``{Program}'' = ID of the ESO X-shooter program during which the observations were obtained. ``{Date}'' = date of the observations. ``{Exp.\,time}'' = number of exposures (performed in nodding mode) times the total integration time per exposure for each arm. ``{Slit}'' = slit width used for the observations in each of the three arms of the spectrograph.
}
\end{table*}  
several sources are part of multiple systems. Our targets include TWA\,3B, companion of the spectroscopic binary TWA\,3A. This star is reported to be non-accreting from earlier Balmer continuum or H$\alpha$ emission measurements \citep[e.g.,][]{herczeg2009}. Furthermore, no evidence of infrared excess up to wavelengths of $\sim$10\,$\mu$m has been detected in earlier surveys \citep{jayawardhana1999b, jayawardhana1999a}, which indicates that the disk around this star (if present) is devoid of material in the inner regions. For these reasons, we excluded TWA\,3B from any statistical analysis regarding the accretion activity in the TWA, but we report and discuss all parameters derived from our data on this target in Appendix~\ref{app:TWA3B}.

To our knowledge, and based on the latest census of the TWA \citep{gagne2017}, our survey of accreting sources in the association is $\sim$85\% complete in regard to its disk-hosting population. Two of the targets in our list (TWA\,22 and TWA\,31) were recently disputed as TWA members, and proposed to be members of the $\beta$~Pictoris moving group and of the Lower Centaurus Crux association, respectively \citep{gagne2017}. This would place them at slightly older ages than the TWA, 24~Myr \citep{bell2015} and 17--23~Myr \citep{mamajek2002}, respectively. We used the newest astrometric data from Gaia DR2, together with radial velocity measurements for our targets, to derive their space velocities with respect to nearby young stellar associations. This analysis indicates that, while TWA\,22 appears a more likely member of the $\beta$~Pictoris association than of the TWA, TWA\,31 falls among the confirmed TWA members on $u,v,w$ diagrams. For consistency with the lower membership probability calculated by \citet{gagne2017} for these two objects, in the following we keep them separate from the other targets when discussing the statistical inferences from our survey. Therefore, after excluding TWA\,3B, the final sample of TWA members that are the focus of our study amounts to eleven stars with infrared excess. We note that our results would not change appreciably if all objects were included at every step of the analysis. 

\subsection{Observations and data processing}

The observations were conducted in the course of multiple programs between 2010 and 2015, as reported in Table~\ref{tab:TWA_targets}. Programs 084.C-0269, 085.C-0238, 086.C-0173, 087.C-0244, and 089.C-0143 (PI Alcal\'a) were performed as part of the X-shooter guaranteed-time observations (GTO; \citealp{alcala2011}) granted to the Italian Istituto Nazionale di AstroFisica (INAF) as member of the X-shooter consortium. Programs 093.C-0097 and 095.C-0147 (PI Stelzer) were performed later to expand the GTO survey of the TWA. More than one spectrum were acquired for some objects in the target list (TWA\,1, TWA\,27, TWA\,30), either on the same date but using different exposure time (TWA\,1) or slit width  (TWA\,30), or on different epochs (TWA\,27) to constrain their time variability. TWA\,30A and TWA\,30B were additionally observed with X-shooter in the Director's Discretionary Time (DDT) program 287.C-5039 (PI Sacco).

X-shooter comprises three spectroscopic arms, operating simultaneously in three different spectral windows: UVB (300--555~nm), VIS (545--1035~nm), and NIR (995--2480~nm). For the brightest sources in our sample, observations were acquired using slit widths of 0.5"/0.4"/0.4" in the UVB/VIS/NIR arms, respectively, yielding a resolution R\,$\sim$\,9700/18400/11600. Fainter sources were instead observed using slit widths of 1.0"/0.9"/0.9" (corresponding to R\,$\sim$\,5400/8900/5600) or 1.6"/1.5"/1.2" (corresponding to R\,$\sim$\,3200/5000/4300), respectively, in the UVB/VIS/NIR arms. Observations were conducted in nodding mode, either using a one-cycle A--B pattern (two exposures), or a two-cycle A--B--B--A pattern (four exposures). For the faintest source in the target list (TWA\,40), two cycles of observations with the A--B--B--A nodding mode were performed, and the resulting frames were combined to increase the signal-to-noise ratio (S/N).

The spectra were reduced using the latest version of the X-shooter pipeline \citep{modigliani2010} available at the end of each observing program: versions 1.0.0, 1.3.0 and 1.3.7 were used for the data from periods P084 through P089, and version 2.8.4 was employed for periods P093 and P095. The data reduction steps include bias (UVB, VIS) or dark current (NIR) subtraction, flat-field correction, wavelength calibration, measurement of the instrument resolving power, determination of the instrument response and global efficiency, sky subtraction, and flux calibration. The final one-dimensional spectra were then extracted from the pipeline output two-dimensional spectra by using the {\em specred} package in the Image Reduction and Analysis Facility (IRAF\footnote{IRAF is distributed by the National Optical Astronomy Observatories, which are operated by the Association of Universities for Research in Astronomy, Inc., under cooperative agreement with the National Science Foundation.}) software \citep{tody1986}. This step was performed with the IRAF task {\em apall}, which involves visual optimization of the aperture for flux extraction around the bright signature of the source in the two-dimensional spectrum, and correction for deviations of the source signature from a straight baseline across the spectrum.

\subsection{Telluric correction and flux calibration}

The extracted one-dimensional spectra were subsequently corrected for the presence of telluric features in their VIS and NIR segments. For the observing programs in cycles P089 and earlier, the telluric correction was carried out as described in Appendix~A of \citet{alcala2014}. Namely, the spectra of the target stars were divided by the spectra of telluric standard stars, observed very close in time to the science target at a similar airmass, after normalizing the telluric spectra to their continuum and removing any stellar lines from the normalized spectra. A scaling factor was introduced in the procedure to account for small differences in airmass between the observing epochs of target star and telluric standard. For spectra acquired during the more recent observing programs, the telluric correction was handled with the {\em molecfit} tool \citep{smette2015, kausch2015}. A transmission spectrum of the Earth's atmosphere, accounting for the altitude of the observing site and for the local conditions of temperature, pressure and humidity at the time of the observation, is computed by fitting an atmospheric radiative transfer model to selected telluric regions of the science spectrum, and then using the best-fit parameters (spectral line profile, molecular column densities) to calculate the transmission spectrum over the entire wavelength range. The telluric-corrected spectrum is then obtained by dividing the science spectrum by the generated transmission curve. To verify the consistency between the two methods of telluric correction, we ran {\em molecfit} on several spectra of young stars acquired by our group in different regions, and originally corrected following the method in \citet{alcala2014}. We then compared the telluric-corrected spectra obtained for a given source with the two methods, and could ascertain that the two corrections overlap correctly within the intrinsic spectrum noise.

To account for potential flux losses within the slit, we recalibrated the spectra to the tabulated $J$-band photometry for the objects from the 2MASS survey \citep{skrutskie2006}. We converted the observed magnitudes to fluxes, using the zero-point flux provided by the SVO Filter Profile Service \citep{rodrigo2012}, and multiplied the NIR part of the spectrum by the ratio between the $J$-band flux and the average spectrum flux over the $J$-bandpass. We then used the overlapping wavelength regions between the NIR and VIS and between the VIS and UVB parts of the spectra to align them in flux. The match was performed by computing the point-by-point flux ratio between two neighboring spectrum segments across the overlapping region, and then taking as corrective factor the median of the measured point-by-point flux ratios, after ascertaining that the VIS/NIR and UVB/VIS flux ratios calculated point by point are not dependent on wavelength.

\begin{figure*}
\centering
\includegraphics[width=\textwidth]{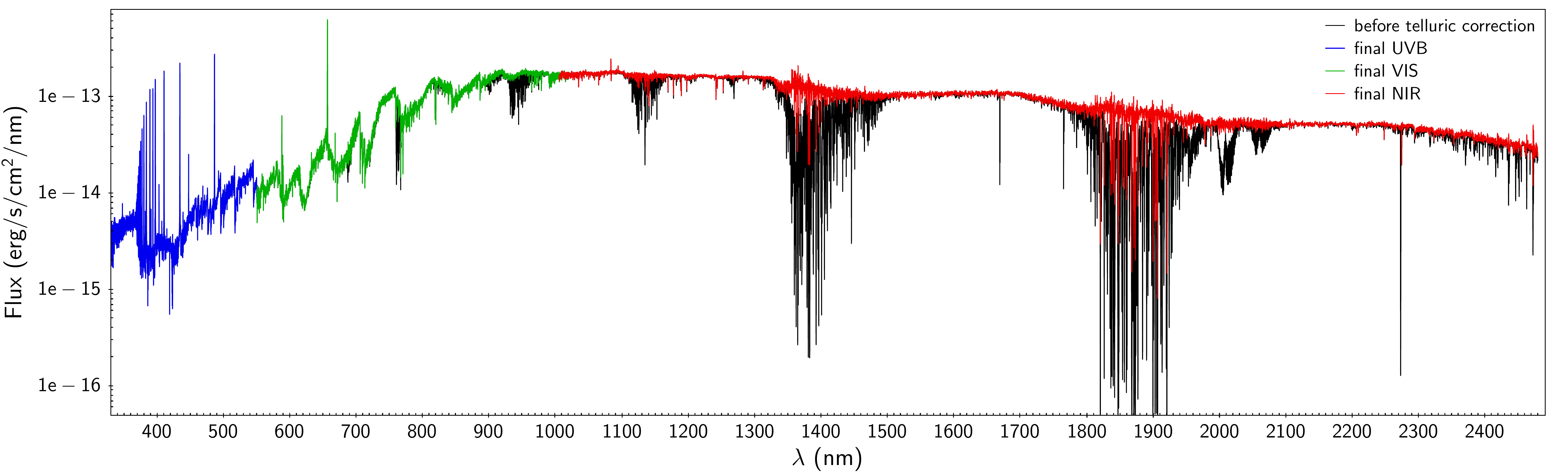}
\caption{UVB (blue), VIS (green), and NIR (red) segments of the final processed X-shooter spectrum of TWA\,32, after performing the telluric correction with {\em molecfit} and removing the most prominent spikes. The source spectrum after the pipeline reduction but before the telluric correction is shown in black for comparison.}
\label{fig:TWA32_reduced_telluric_correction}
\end{figure*}

In Fig.\,\ref{fig:TWA32_reduced_telluric_correction}, one example of the reduced and telluric-corrected spectra from our sample is shown for illustration purposes.

\section{Results}

\subsection{Stellar and accretion parameters} \label{sec:param}

\subsubsection{SpT, $A{\rm _V}$, $L_\star$, and $L_\mathrm{acc}$} \label{sec:multicomponent_fit}

To estimate the SpT, extinction ($A_{\rm V}$), stellar luminosity ($L_\star$), and $L_{\rm acc}$ of all but the faintest sources in our sample (J1247-3816 and TWA\,40; see below), we adopted the same method introduced in \citet{manara2013}. In brief, a multi-component fit to the observed spectrum was carried out. The fit includes: i) a set of photospheric templates, consisting of X-shooter spectra of well-characterized, non-accreting young stars, to reproduce the photospheric and chromospheric emission of PMS stars with SpT between late-K and late-M \citep{manara2013a, manara2017b}; ii) a standard interstellar reddening law {with R$_V$ = 3.1} \citep{cardelli1989}, and a grid of $A_{\rm V}$ values from 0 to several magnitudes of extinction; iii) an isothermal hydrogen slab\footnote{Model in which the accretion luminosity is assumed to arise from a small area of constant temperature and density (a ``slab'').} model, with varying temperature, electron density, and optical depth, to reproduce the emission from the accretion shock. {For each point of the model grid explored (i.e., a different combination of photospheric template spectrum, A$_V$ value, slab parameters described above, and two normalization factors for the photospheric and slab emission), the model spectrum is compared to the observed spectrum in several spectral features: flux ratio at the Balmer jump (i.e., between $\sim$360 and 400~nm), slope of the Balmer continuum (332.5--360~nm), slope of Paschen continuum (398--477~nm), and the continuum flux level in selected, narrow ($\sim$2~nm--wide) wavelength intervals across the spectrum (at $\sim$360~nm, $\sim$460~nm, $\sim$703~nm, $\sim$707~nm, $\sim$710~nm, and $\sim$715~nm). A $\chi^2$-like function\footnote{Function that measures, for each point of the model grid, the residuals between the observed spectral features and their representation in the model spectrum, relative to the local observational uncertainty (see Eq.\,1 of \citealp{manara2013}).} is computed to measure the degree of agreement between each model fit and the observed spectrum.}
The best-fit parameters are determined by minimizing the $\chi^2$-like {distribution}. Parameters derived simultaneously via this procedure will not be impacted by the degeneracy between circumstellar extinction and accretion, which affect the continuum spectrum from the UV to the NIR in a very different way: extinction reduces the observed stellar luminosity and is more efficient at shorter wavelengths, therefore making the source appear redder, hence colder, than it actually is; accretion produces a luminosity excess that peaks in the UV, which makes the source appear bluer, hence hotter and brighter, than it actually is. 

$L_\star$ and $L_{\rm acc}$ values were scaled using the individual distances to TWA members computed from Gaia DR2 \citep{gaia_dr2} parallaxes. These distances are in most cases consistent with the distances listed for each object in \citet{gagne2017}, within the combined uncertainty from the new Gaia parallax errors and from the errors on the previous distance estimates. For two sources, the absolute difference between new and old distance values is more than three times larger than the total uncertainty: these are TWA\,27 (previous distance $52.8 \pm 1.0$~pc, new distance $64.4 \pm 0.7$~pc) and TWA\,22 (previous distance $17.5 \pm 0.2$~pc, new distance $19.61 \pm 0.12$~pc). Another two sources (J1247-3816 and TWA\,28) exhibit a change in estimated distance about twice as large as the combined uncertainty. The change in distance is more marked for J1247-3816 (previous distance 68~pc, new distance 85~pc); however, this is also the source with the largest uncertainty on the distance estimate among our sample. The values of individual distances adopted here and the derived parameters for each star are reported in Table~\ref{tab:TWA_parameters}.

Results of the procedure described above are illustrated in Fig.\,\ref{fig:TWA_fit_atlas}. TWA\,30B appears as a severely underluminous object due to its edge-on configuration \citep{looper2010}; we therefore exclude this star from Fig.\,\ref{fig:TWA_fit_atlas} and from subsequent analyses involving the estimated $L_\star$. {For the remaining targets (except J1247-3816 and TWA\,40, as discussed in the next paragraph), we evaluated the significance of the $L_{\rm acc}$ detection by comparing the luminosity of the observed spectrum to the normalized spectrum of the photospheric template in the Balmer continuum region ($\lambda$$\sim$332.5--360~nm). Namely, we sampled this spectral window in 2.5~nm--wide intervals, and for each interval we compared the median flux of the science target to the median flux of the photospheric template. The large noise that affects this region of the spectra, particularly for the photospheric templates, hampers a strict definition of a quantitative threshold to separate ``accretion--dominated'' from ``chromospheric--dominated'' objects. We therefore decided to sort our objects into three groups: i) {\it bona fide} accretors (5 objects), when the median flux measured for the science target across the Balmer continuum is systematically higher than that measured for the photospheric template, and the difference between the two is larger than the total rms spectral noise; ii) possible accretors (4), when the science targets exhibit systematically higher median flux than the photospheric templates across the Balmer continuum, but with a flux difference comparable to the total rms noise; iii) upper limit (1), when the median flux measured for the science object is higher than that measured for the photospheric template only in some portions of the Balmer continuum. Separate lists for the objects that fall into each group are reported in Table~\ref{tab:TWA_parameters}.}

For the two faintest sources of our sample (J1247-3816 and TWA\,40), no solution could be determined with the multi-component fit, due to poor S/N in the UVB part of the spectrum. We therefore followed the approach of \citet[][also used for consistency checks in \citealp{stelzer2013} and \citealp{alcala2014, alcala2017}]{manara2015} to estimate the SpT and $A_{\rm V}$ of these two objects. We used photospheric templates of varying SpT from the grid of X-shooter spectra of young non-accreting stars from \citet{manara2013a, manara2017b}; these were artificially reddened following the reddening law of \citet{cardelli1989} for R$_V$ = 3.1, with $A_{\rm V}$ allowed to vary between 0 and 3~mag. At each step in $A_{\rm V}$, we then normalized the spectrum of the reddened template to the observed spectrum around $\lambda \sim 1025$~nm, and computed a $\chi^2$-like function which samples the discrepancy between the observed spectrum and the reddened template in 24 spectral windows, each a few nm wide, distributed between $\lambda\,=\,700$~nm and $\lambda\,=\,1700$~nm. The best (SpT, $A_{\rm V}$) solution was then determined via $\chi^2$-like minimization. 

Independent estimates of SpT were derived for the whole sample by computing a variety of spectral indices \citep{allers2007, jeffries2007, riddick2007, scholz2012, rojas-ayala2012, herczeg2014}, which probe the depth of $T_{\rm eff}$--dependent molecular bands in the VIS and NIR spectra. The best SpT was computed as the average of the individual SpT estimates from the various spectral indices, after excluding those estimates that would fall outside the range of validity specified in the literature for the corresponding spectral index. An {\it a posteriori} consistency check between the SpT estimates from our multi-component fit and those from the spectral indices analysis allowed us to identify which sets of spectral indices work best in a given SpT range. In turn, this analysis allowed us to better constrain the (SpT, $A_{\rm V}$) solution for J1247-3816 and TWA\,40 from the spectral fit. The $L_\star$ of these two stars was then estimated from the luminosity of the template corresponding to the (SpT, $A_{\rm V}$) solution, with a scaling factor that accounts for the flux and squared distance ratio between the target and the template.

\begin{figure}
\resizebox{\hsize}{!}{\includegraphics{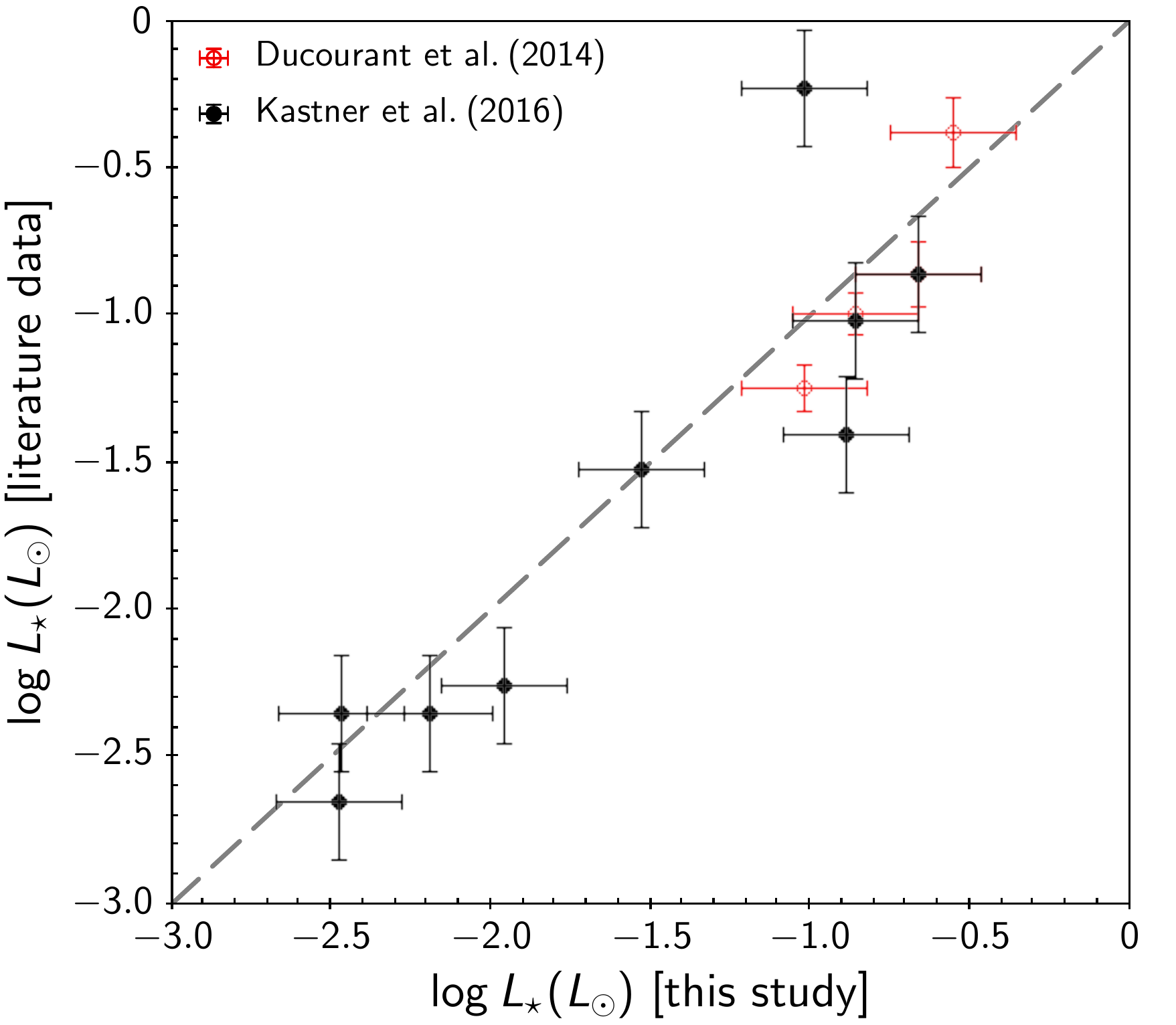}}
\caption{Comparison between our $L_\star$ estimates (see Sect.\,\ref{sec:multicomponent_fit} and Table~\ref{tab:TWA_parameters}) and the values reported by \citet[][red symbols]{ducourant2014} and \citet[][black symbols]{kastner2016}, corrected for difference in distance between the value adopted here and the value listed in those references for individual objects. The error bars trace the individual uncertainties associated with the $L_\star$ values in the compilation of \citet{ducourant2014}, and a typical uncertainty of $\pm$0.2~dex in the other cases, for illustration purposes. The identity line is dashed in gray to guide the eye.}
\label{fig:TWA_Lbol_comp}
\end{figure}

In Fig.\,\ref{fig:TWA_Lbol_comp} we compare our measurements of $L_\star$ with results from previous surveys and compilations for objects in common \citep{ducourant2014, kastner2016}, rescaled to the values of distances adopted in this study. The diagram shows an overall agreement between different sets of measurements for the bulk of sources, with most $L_\star$ estimates lying close to the identity line on Fig.\,\ref{fig:TWA_Lbol_comp}. 

\begin{sidewaystable*}
\caption{$J$-band photometry, individual distances, and stellar and accretion parameters derived for sources in Table~\ref{tab:TWA_targets}, {except TWA\,3B (see Appendix~\ref{app:TWA3B})}.}
\label{tab:TWA_parameters}
\centering
\resizebox{\textwidth}{!} {
\begin{tabular}{l c c c c c c c c c c c c c c c c c c c c c c c}
\hline\hline
{Name} & $J$ & err & $d$ & err & {SpT} & $A_{\rm V}$ & $\log{L_\star}$ & $\log{L_{\rm acc}^{slab}}$ & $\log{\widetilde{L}_{\rm acc}^{lines}}$ & $T_{\rm eff}$ & err & $\log{g}$ & err & RV & err & $v \sin{i}$ & err & $r$ & $M_\star$ & err & $R_\star$ & $\log{\dot{M}_{\rm acc}^{slab}}$ & $\log{<\dot{M}_{\rm acc}^{lines}>}$\\
 & \multicolumn{2}{c}{$[mag]$} & \multicolumn{2}{c}{$[pc]$} & & $[mag]$ &  $[L_\odot]$ &  \multicolumn{2}{c}{$[L_\odot]$} & \multicolumn{2}{c}{$[K]$} & & & \multicolumn{2}{c}{$[km/s]$} & \multicolumn{2}{c}{$[km/s]$} & (6200\AA) & \multicolumn{2}{c}{$[M_\odot]$} & $[R_\odot]$ &  \multicolumn{2}{c}{$[M_\odot/yr]$}\\
\hline
\multicolumn{24}{l}{}\\
\multicolumn{24}{l}{{\textit{Bona fide accretors:}}}\\
\multicolumn{24}{l}{}\\
{\small TWA\,1\,{\it [10~s]}} & 8.217 & 0.017 & 60.10 & 0.14 & M0.5 & 0.0  & -0.55 & -1.52 & -1.54 & {3880} & 30 & {4.4} & {0.2} & 13.5 & 1.4 & 7 & 3 & 0.8 & 0.58 & 0.09 & 1.22 & -8.59 & -8.62\\
{\small TWA\,1\,{\it [60~s]}} & '' & '' & '' & '' & M0.5 & 0.0  & -0.55 & -1.57 & -1.60 & {3870} & {40} & {4.46} & {0.13} & 13.1 & 1.4 & 7 & 2 & 0.8 & " & " & '' & -8.64 & -8.68\\
{\small TWA\,27\,{\it [P084]}} & 13.00 & 0.02 & 64.4 & 0.7 & M9 & 0.0  & -2.19 & -5.10 & -4.55 & 2640 & 20 & 3.75 & 0.14 & 10 & 4 & 30 & 10 & 0.0 & 0.019 & 0.006 & 0.35 & -11.23 & -10.68\\
{\small TWA\,27\,{\it [P089]}} &'' & '' & '' & '' &  M9 & 0.0  & -2.47 & -5.22 & -4.92 & 2730 & 100 & {3.8} & {0.2} & 27 & 4 & 25  & 10 & 0.0 & " & "  & '' & -11.35 & -11.05\\
{\small TWA\,28} & 13.03 & 0.02 & 59.9 & 0.7 & M9 & 0.0  & -2.47 & -6.00 & -5.15 & 2660 & 70 & 4.1 & 0.3 & 18 & 3 & 25 & 10 & 0.0 & 0.020 & 0.005 & 0.29 & -12.24 & -11.38\\
{\small TWA\,31} & 13.048 & 0.018 & 81.1 & 0.6 & M5 & 0.7  & -1.96 & -3.28 & -3.13 & 2990 & 50 & 4.49 & 0.11 & 7 & 2 & $<8$ & & 1.2 & 0.078 & 0.018 & 0.39 & -9.97 & -9.82\\
{\small TWA\,32} & 10.69 & 0.02 & 63.7 & 1.4 & M5.5 & 0.0  & -1.68 & -3.85 & -3.43 & 2800 & 70 & 3.98 & 0.11 & 7 & 2 &  $<8$ & & 0.3 & 0.070 & 0.015 & 0.57 & -10.35 & -9.93\\
\multicolumn{24}{l}{}\\
\multicolumn{24}{l}{{\textit{Possible accretors:}}}\\
\multicolumn{24}{l}{}\\
{\small TWA\,30A\,{\it [P087,\,ns]}} & 9.64 & 0.02 & 48.0 & 0.3 & M4.5 & 2.2  & -0.89 & -3.21 & -3.75 & 3110 & 80 & 4.36 & 0.15 & 12.8 & 1.5 & 28.2 & 1.7 & 0.3 & 0.15 & 0.02 & 1.26 & -9.70 & -10.23\\
{\small TWA\,30A\,{\it [P087,\,ls]}} & '' & '' & '' & '' & M4.5 & 2.4  & -0.89 & -3.39 & -3.92 & 3050 & 60 & 4.42 & 0.13 & 14 & 3 & 19 & 11 & 0.3 & "  & '' & " & -9.88 & -10.40\\
{\small TWA\,30A\,{\it [P287]}} & '' & '' & '' & '' & M4.5 &  & &  & -3.49 & 3070 & 80 & 4.34 & 0.13 & 11 & 3 & 25 & 9 & 0.3 & " & '' & " & & -9.97\\
{\small TWA\,22} & 8.554 & 0.013 & 19.61 & 0.12 & M5 & 0.1  & -1.74 & -4.86 & -4.31 & 2830 & 100 & 4.09 & 0.11 & 16 & 2 & $<8$ & & 0.1 & 0.082 & 0.018 & 0.51 & -11.46 & -10.92\\
\small{TWA\,8A} & 8.337 & 0.017 & 46.27 & 0.19 & M3 & 0.2  & -0.66 & -3.06 & -2.96 & {3370} & {90} & {4.6} & {0.2} & 7.1 & 1.2 & 9 & 6 & 0.3 & 0.29 & 0.04 & 1.35 & -9.79 & -9.72\\
{\small TWA\,8B} & 9.84 & 0.02 & 46.5 & 0.2 & M5.5 & 0.0  & -1.52 & -4.36 & -4.15 & 2830 & 80 & 4.04 & 0.10 & 10 & 2 & $<8$ & & 0.1 & 0.074 & 0.018 & 0.68 & -10.80 & -10.58\\
\multicolumn{24}{l}{}\\
\multicolumn{24}{l}{{\textit{Upper limit:}}}\\
\multicolumn{24}{l}{}\\
{\small TWA\,3A} & 7.651 & 0.009 & 36.62 & 0.16 & M4.5 & 0.0  & -0.86 & -3.74 & -3.42 & 3180 & 80 & 4.23 & 0.13 & -1/+29 & 3/3 & $<8$/$<8$  &  & 0.2 & 0.15 & 0.02 & 1.30 & -10.21 & -9.89\\
\multicolumn{24}{l}{}\\
\multicolumn{24}{l}{{\textit{Underluminous:}}}\\
\multicolumn{24}{l}{}\\
{\small TWA\,30B\,{\it [P087]}} & 15.35 & 0.05 & 46.3 & 0.6 & M4 &  & -4.46 & & -7.43 & 3180 & 50 & 4.35 & 0.14 & 18 & 3 & 30 & 8 & 0.4 & & & & &\\
{\small TWA\,30B\,{\it [P287]}} & '' & '' & '' & '' & &  & & & -5.39 & 3410 & 100 & 4.4 & 0.3 & 19 & 4 & 30 & 15 & 0.2 & & & & &\\
\multicolumn{24}{l}{}\\
\multicolumn{24}{l}{{\textit{No slab model solution:}}}\\
\multicolumn{24}{l}{}\\
{\small TWA\,40} & 15.49 & 0.06 & 67 & 4 & M9.5/L0 & 1.0  & -2.16 & & & 2500 & 70 & {4.4} & {0.2} &  20 & 7 & & & 0.0 & 0.015 & 0.002 & 0.44 & & -13.34$^{\dag}$\\
{\small J1247-3816} & 14.79 & 0.03 & 85 & 4 & M8 & 0.35  & -3.00 & & -5.84 & 2660 & 30 & 4.50 & 0.15 & 14 & 6 & $<8$ & & 0.0 & 0.060 & 0.010 & 0.15 & & -12.85\\
\hline
\end{tabular}
}
\tablefoot{
{The groupings into ``{\it bona fide} accretors'', ``possible accretors'' and `` upper limit'' are referred to the measured Balmer continuum emission from the multi-component fit (see Sect.\,\ref{sec:multicomponent_fit}). {Objects in the possible accretors group are listed in decreasing order of strength of the measured Balmer continuum flux excess above the photospheric level.} Objects for which no solution could be derived from the multi-component fitting procedure are listed under ``no slab model solution''.} A blank space is reported for a given parameter when no accurate estimate could be derived from the corresponding spectrum. ``{Name}'' = Target identifier, as reported in Table~\ref{tab:TWA_targets}. The {\it Pxxx} notation following some of the target names identifies the program during which the corresponding observation was conducted for sources with multiple spectra in our sample. The {\it ns} and {\it ls} notations associated with TWA\,30A indicate the observations obtained with narrower slit and larger slit, respectively, during program 087.C-0244. ``{\it J}, err'' = $J$-band apparent magnitudes from 2MASS, and associated uncertainties. ``{\it d}'' = Individual distances derived from Gaia DR2 parallaxes. ``{SpT, $A_{\rm V}$, $\log{L_\star}$, $\log{L_{\rm acc}^{slab}}$}'' = parameters corresponding to the best solution of the spectrum fitting procedure (see Sect.\,\ref{sec:multicomponent_fit}). ``{$\log{\widetilde{L}_{\rm acc}^{lines}}$}'' = median estimate determined from the luminosity of emission lines (see Sect.\,\ref{sec:Lacc_Lline}). Typical uncertainties amount to 0.2~dex on $\log{L_\star}$, 0.25~dex on $\log{L_{\rm acc}^{slab}}$, and 0.17~dex on $\log{\widetilde{L}_{\rm acc}^{lines}}$. ``{$T_{\rm eff}$, err, $\log{g}$, err, RV, err, $v \sin{i}$, err, $r$}'' = parameters corresponding to the best fit to the spectra obtained with ROTFIT (see Sect.\,\ref{sec:ROTFIT}), and associated uncertainties. The two values of RV and $v \sin{i}$ reported for TWA\,3A correspond to the measurements extracted separately for the two components of the spectroscopic binary by deblending the two peaks observed in the absorption lines. ``$M_\star$, err, $R_\star$, $\log{\dot{M}_{\rm acc}^{slab}}$'' = parameters computed as in Sect.\,\ref{sec:mass_rad_macc}. Typical uncertainties amount to 20\% on $R_\star$, and 0.3~dex on $\log{\dot{M}_{\rm acc}^{slab}}$. ``$\log{<\dot{M}_{\rm acc}^{lines}>}$'' = accretion rate estimate derived from $\log{\widetilde{L}_{\rm acc}^{lines}}$. Typical uncertainties amount to 0.2~dex.
\\{\it Notes on individual sources}: TWA\,30B is an edge-on disk source \citep{looper2010}, therefore the derived $L_\star$ is a severe underestimate, and no stellar parameters could be estimated from it. \\$^\dag$ = upper limit.
}
\end{sidewaystable*}  

\subsubsection{$T_{\rm eff}$, $\log{g}$, radial velocity, $v \sin{i}$, veiling} \label{sec:ROTFIT}

To determine the atmospheric parameters ($T_{\rm eff}$, $\log{g}$) of the stars and their radial velocities (RV), projected rotational velocities ($v \sin{i}$), and veiling ($r$), we made use of the ROTFIT code \citep{frasca2017}. In brief, BT-Settl synthetic spectra \citep{allard2012}, degraded to match the instrument resolution, were used as templates to fit the observed spectra around spectral features particularly sensitive to the parameters of interest. Both observed spectra and templates were normalized to the local continuum before the fitting procedure, which was implemented via a $\chi^2$ routine. The fitting regions were chosen primarily in the VIS window, and selected so as to avoid any contamination from accretion or chromospheric signatures. The rotational broadening effects were simulated in the synthetic templates by convolving them with a rotational profile across a grid of $v \sin{i}$ values. The effects of veiling were described as $\left(F_\lambda/F_C\right)_r = \left(F_\lambda/F_C + r\right)/\left(1+r\right)$, where $F_\lambda$ represents the line flux, $F_C$ the continuum flux, and $r$ was set as a free parameter over a grid of test values\footnote{We also estimated the veiling $r$ from the multi-component fit results, following the definition of \citet{herczeg2008}. We could ascertain that the two approaches yield consistent results on a relative scale, albeit with some discrepancies on the individual $r$ values that can be of the order of 50\%. These discrepancies reflect the larger uncertainty associated with the veiling estimates from the multi-component fit parameters, due to the fact that the multi-component fit procedure, by construction, does not take into account the line emission but only the continuum emission.}.

\begin{figure}
\resizebox{\hsize}{!}{\includegraphics{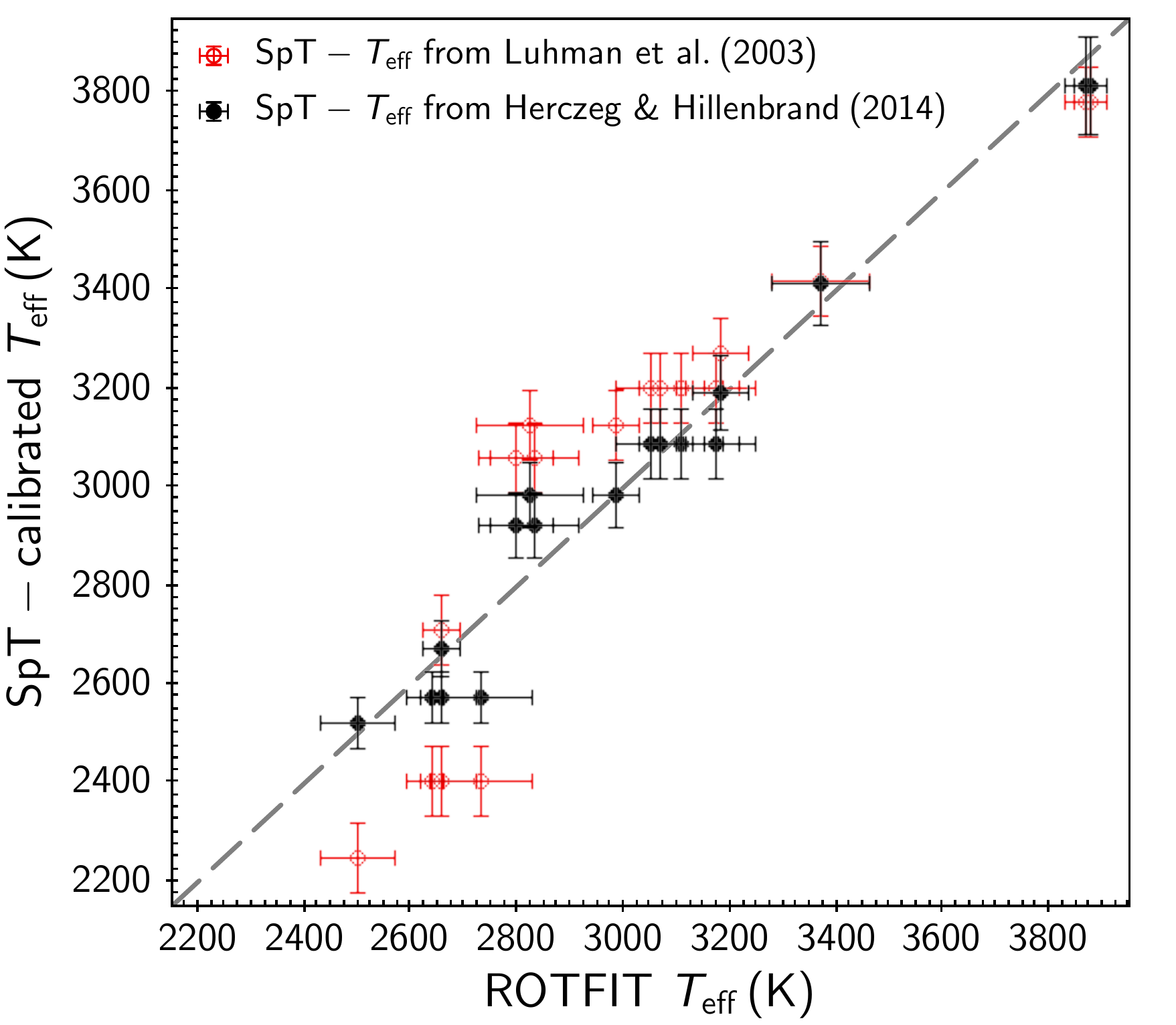}}
\caption{Comparison between the $T_{\rm eff}$ estimates derived with ROTFIT for all targets and spectra, and the $T_{\rm eff}$ corresponding to the SpT estimated for each source (Table~\ref{tab:TWA_parameters}) using the calibrations of \citet[][red symbols]{luhman2003} and of \citet[][black symbols]{herczeg2014}. The $x$-axis uncertainties are those derived from the fit; the $y$-axis uncertainties correspond to half the difference in $T_{\rm eff}$ between consecutive spectral subclasses in the corresponding SpT--$T_{\rm eff}$ scale. The identity line is dashed in gray to guide the eye.}
\label{fig:TWA_Teff_scales}
\end{figure}

The derived parameters are reported in Table~\ref{tab:TWA_parameters}. Fig.\,\ref{fig:TWA_Teff_scales} illustrates the comparison between the $T_{\rm eff}$ estimates derived with ROTFIT, and the values of $T_{\rm eff}$ that would be assigned to our sources based on the SpT derived in Sect.\,\ref{sec:multicomponent_fit} and the SpT--$T_{\rm eff}$ calibration scales proposed for PMS stars by \citet{luhman2003} and \citet{herczeg2014}. As can be observed on the diagram, the latter calibration provides a better match to the $T_{\rm eff}$ measurements obtained via spectral fitting. This is in agreement with the result obtained by \citet{frasca2017} on a statistically larger sample of objects; we therefore adopt the \citeauthor{herczeg2014}'s (\citeyear{herczeg2014}) scale throughout this work. Our estimates of SpT are consistent with those reported in the compilation of \citet{gagne2017} within half or one spectral subclass for all of our targets.
Comparison plots between our RV and $v\sin{i}$ measurements and previous estimates reported in the literature are shown in Appendix~\ref{sec:rv_vsini_plots}. Despite a few discrepant cases, the majority of objects fall along the identity line on both diagrams, indicating that the various sets of parameters are overall consistent with each other.

\subsubsection{Stellar mass, radius, and $\dot{M}_{\rm acc}$} \label{sec:mass_rad_macc}

\begin{figure*}
\centering
\includegraphics[width=\textwidth]{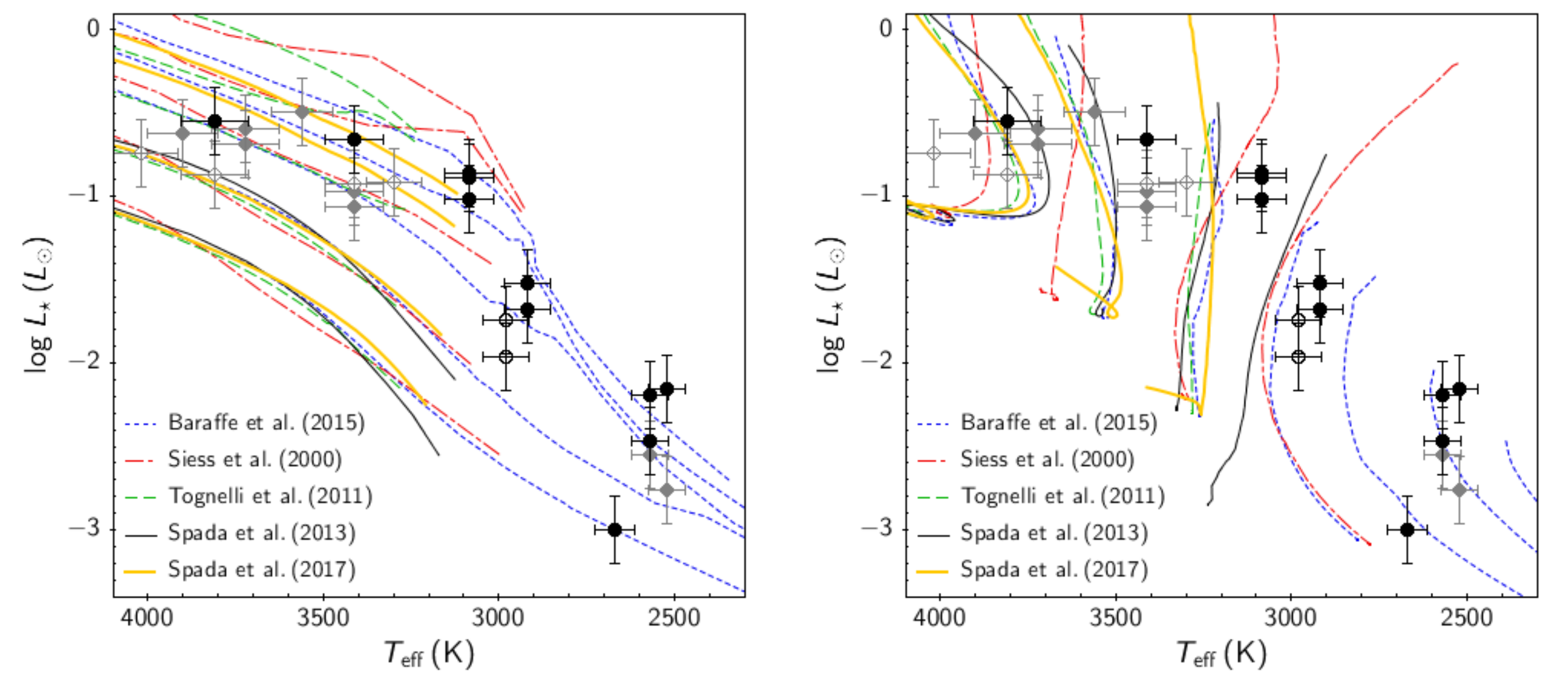}
\caption{HR diagram for stars in Table~\ref{tab:TWA_parameters} (black dots) compared with disk--free TWA objects investigated in \citet{manara2013a} and \citet{stelzer2013} (gray diamonds), rescaled to Gaia DR2 distances and \citeauthor{herczeg2014}'s (\citeyear{herczeg2014}) SpT--based $T_{\rm eff}$. Filled symbols identify probable TWA members, open symbols targets with disputed membership. Errors on $T_{\rm eff}$ and $\log{L_\star}$ are as in Fig.~\ref{fig:TWA_Teff_scales} ($y$-axis) and Fig.\,\ref{fig:TWA_Lbol_comp} ($x$-axis), respectively. {\it Left}: Theoretical isochrones from different model grids (\citealp{baraffe2015}, dotted blue line; \citealp{siess2000}, dash-dot red line; \citealp{tognelli2011}, dashed green line; \citealp{spada2013}, solid black line; \citealp{spada2017}, solid yellow line) are shown for comparison on the diagram. The illustrated isochrone ages are, from top to bottom, {1~Myr, 2~Myr, 5~Myr, 20~Myr, and 100~Myr (the isochrones at 1, 2, and 5~Myr are not available from the models of \citealp{spada2013}, and the 5~Myr isochrone is not available from the models of \citealp{spada2017})}. {\it Right}: Mass tracks from the same model grids are illustrated for comparison. Displayed masses are, from right to left: 0.01~$M_\odot$, 0.02~$M_\odot$, 0.05~$M_\odot$, 0.1~$M_\odot$, 0.2~$M_\odot$, 0.4~$M_\odot$, and 0.6~$M_\odot$ \citep{baraffe2015}; 0.1~$M_\odot$, 0.2~$M_\odot$, 0.4~$M_\odot$, 0.6~$M_\odot$ \citep{siess2000, spada2013}; 0.2~$M_\odot$, 0.4~$M_\odot$, 0.6~$M_\odot$ \citep{tognelli2011, spada2017}.}
\label{fig:TWA_HR_models}
\end{figure*}

Fig.\,\ref{fig:TWA_HR_models} shows the HR diagram populated by our targets\footnote{TWA\,30B is omitted from the diagram, as it would artificially be placed well below the 1~Gyr isochrone as a result of the $L_\star$ underestimate in Table~\ref{tab:TWA_parameters}. }, using the $L_\star$ values reported in Table~\ref{tab:TWA_parameters}, and the $T_{\rm eff}$ values associated with our SpT estimates following the scale of \citet{herczeg2014}. A comparison between the location of the sources on the diagram and theoretical model isochrones, illustrated in the left panel of Fig.\,\ref{fig:TWA_HR_models}, would suggest that our sample spans the age range between 1--2~Myr and 10~Myr. This picture would also indicate that TWA\,22 and TWA\,31, whose membership to the TWA has recently been challenged (see Sect.\,\ref{sec:data}), are slightly older than the bulk of TWA members. 

Taken at face value, the data in Fig.\,\ref{fig:TWA_HR_models} (left) appear inconsistent with a quoted age of 10~Myr for the TWA, which seems to be rather an upper limit to the age distribution of stars belonging to the association. This property appears to be shared by the sample of disk-free TWA members investigated with the same method in \citet{manara2013a} and \citet{stelzer2013}, shown here for comparison purposes. Disk--free objects might tend to exhibit older ages than our targets, but a statistically significant comparison is hampered by the small number of objects and by the incompleteness of the sample (based on \citeauthor{gagne2017}'s \citeyear{gagne2017} census, only $\sim$25\% of the disk--free stars in the TWA are shown in Fig.\,\ref{fig:TWA_HR_models}). Several studies \citep[e.g.,][]{hillenbrand2008, pecaut2016} have also shown that lower-mass stars tend to appear younger on PMS isochrone grids than higher-mass stars. This bias may amount to an artificial increase in logarithmic age by $\sim$0.7~dex over 1~dex in $\log{M_\star}$  for solar-type stars at ages of a few Myr \citep{venuti2018}, although those studies do not typically extend below $M_\star \sim 0.2-0.3\,M_\odot$, and the effect is not necessarily seen in our Fig.\,\ref{fig:TWA_HR_models}. On the other hand, a clear $T_{\rm eff}$--dependent age trend, with cooler objects in the very low-mass and brown dwarf regime appearing younger than earlier-type M-stars, is seen on the $\log{g}$ vs. $T_{\rm eff}$ diagram in Fig.\,\ref{fig:TWA_Teff_logg_ROTFIT}, which also suggests a typical age for the TWA older than indicated on the HR diagram.

However, Fig.\,\ref{fig:TWA_HR_models} also shows that the absolute age calibration is somewhat dependent on the model grid. This point is well discussed in \citet{herczeg2015}, who illustrate in particular how different model grids applied to the same sample of TWA members can yield disparate age estimates between 5 and 10--15~Myr. Different ages proposed by different authors in the literature may also be affected by the limited number of objects included in each study. In addition, as shown in Fig.\,\ref{fig:TWA_HR_models}, most model tracks do not cover the region at the youngest ages and at the lowest $M_\star$ ($\leq 0.1\,M_\odot$), where it appears that small uncertainties in $L_\star$ would correspond to significant uncertainties on the individual age estimate (i.e., of the order of, or larger than, the age estimate itself). One of our TWA targets, J1247-3816, would seem significantly older than the others based on its location on the HR diagram ($\log{L_\star}\left[L_\odot\right] \sim -3$); this is one of the two sources for which no solution could be determined via the multi-component fit (see Sect.\,\ref{sec:multicomponent_fit}), hence its $L_\star$ estimate may be subject to a larger uncertainty. The other source without multi-component fit (TWA\,40) appears to be the youngest and lowest-mass object of the sample on Fig.\,\ref{fig:TWA_HR_models}; also in this case the position of the source on the HR diagram may be affected by larger uncertainty than those of other targets. 

For the reasons discussed above, we did not attempt a detailed analysis of individual ages for stars in our sample. Mass tracks from different model grids, instead, appear to be in overall agreement across the region of the HR diagram where they overlap (Fig.\,\ref{fig:TWA_HR_models}, right). We therefore estimated individual stellar masses via interpolation between theoretical mass tracks on the diagram, as detailed in the following. To ensure a uniform analysis across our sample, we adopted the set of models by \citet{baraffe2015}, which cover the entire mass range spanned by our targets, as illustrated in Fig.\,\ref{fig:TWA_HR_models}. We used the $T_{\rm eff}$ and $L_\star$ values of the sources to interpolate their age along the isochrone grid: i) from the $T_{\rm eff}$ coordinate we extracted the corresponding $L_\star^{model}$ along the two closest isochrones to the object's location on the diagram; ii) we interpolated between these two $L_\star^{model}$--age points to determine the age corresponding to the measured $L_\star$.  We then used the derived value of age to extract the corresponding $T_{\rm eff}^{model}$ along the two closest mass tracks to the object's location on the HR diagram, and the actual stellar $T_{\rm eff}$ was used to interpolate between the resulting $T_{\rm eff}^{model}-M_\star^{track}$ points to derive the final estimate of $M_\star$ for our target. Stellar radii ($R_\star$) were calculated using the Stefan-Boltzmann law.

The derived $M_\star$ and $R_\star$ parameters are reported in Table~\ref{tab:TWA_parameters}. These were in turn used, together with the $L_{\rm acc}$ from the multi-component fit, to compute the accretion rate onto the star, $\dot{M}_{\rm acc}$, as in \citet{gullbring1998}:
\begin{equation} \label{eqn:Mdot}
\dot{M}_{\rm acc} \sim 1.25\, \frac{L_{\rm acc} \, R_\star}{G M_\star}\,,
\end{equation} 

\noindent where $G$ is the gravitational constant. This procedure could not be adopted for J1247-3816 and TWA\,40, which lack an $L_{\rm acc}$ measurement from the multi-component fit. For these two objects, an average estimate of $L_{\rm acc}$ was derived from the measured luminosities of several emission lines, as described in detail in Sect.\,\ref{sec:Lacc_Lline}. This procedure, which was applied to the entire sample as a consistency check for accretion parameters determined from different diagnostics, allowed us to obtain an estimate of $\dot{M}_{\rm acc}$ for J1247-3816, and to set an upper limit for the $\dot{M}_{\rm acc}$ on TWA\,40, for which we did not detect any line emission. These two independent measurements of $\dot{M}_{\rm acc}$ derived for all of our targets are also reported in Table~\ref{tab:TWA_parameters}.

\subsection{A picture of accretion in the TWA} \label{sec:TWA_acc}

\subsubsection{Continuum excess luminosity vs. line luminosity} \label{sec:Lacc_Lline}

As mentioned earlier, the intense line emission produced by the heated gas in the accretion columns represents a widely used tracer of accretion, notably because such emission lines (e.g., H$\alpha$, H$\beta$) are more easily detected than the continuum excess luminosity on lower-mass stars and weaker accretors \citep[e.g.,][]{muzerolle2003}. Line emission provides a more indirect diagnostics of accretion onto the star than the excess emission from the accretion shock itself, as the observed line luminosities and line profiles may also bear the imprints of magnetospheric outflows of material, as well as stellar chromospheric activity. Nevertheless, as shown in \citet{herczeg2008} and then in \citet{alcala2014}, the luminosities ($L_{\rm line}$) of several emission lines of H, He, and Ca correlate well with the values of $L_{\rm acc}$, measured simultaneously from the spectral excess continuum emission. The derived empirical calibrations between the two quantities can therefore be used to extract an estimate of $L_{\rm acc}$ and, hence, $\dot{M}_{acc}$ from the measured $L_{\rm line}$ for individual objects (see also \citealp{frasca2017}). In turn, this provides a cross-check for the accretion properties estimated on a given star from the observed excess continuum: disagreements between the two approaches may be symptomatic, for instance, of sources seen close to edge-on, where the visibility of the various accretion diagnostics may differ from case to case \citep[e.g.,][]{alcala2014, sousa2016}.

\begin{figure}
\resizebox{\hsize}{!}{\includegraphics{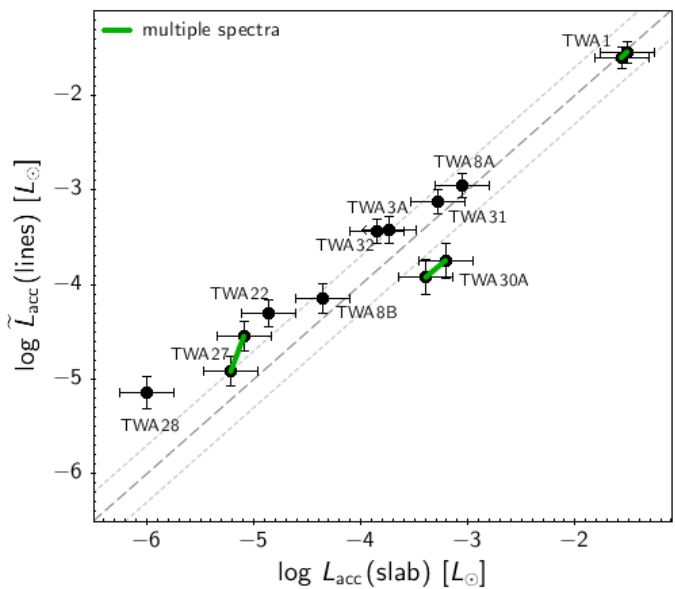}}
\caption{Comparison between the $\log{L_{\rm acc}}$ values measured from fitting the excess continuum with a slab emission model (Sect.\,\ref{sec:multicomponent_fit}), and those derived from the emission line luminosity ($L_{\rm line}$) measurements, using the empirical $L_{\rm line} - L_{\rm acc}$ calibration relationships from \citet{alcala2014}. Error bars on the $x$-axis correspond to a typical uncertainty of 0.25~dex on the $L_{\rm acc}$ measurements from the multicomponent spectral fit; error bars on the $y$-axis follow from error propagation including uncertainties on the measured line fluxes, and uncertainties on the calibration relationships, as listed in \citet{alcala2014}. Green lines connect multiple measurements for the same object. Light dotted lines delimit an area of $\pm$0.3~dex around the equality line (dashed in gray on the diagram).}
\label{fig:TWA_logLacc_slab_logLacc_lines}
\end{figure}

In Fig.\,\ref{fig:TWA_logLacc_slab_logLacc_lines}, $L_{\rm acc}$ measurements inferred from fitting the hydrogen slab to the continuum excess emission are compared to the {typical $L_{\rm acc}$ estimates ($\widetilde{L}_{\rm acc}$) derived from line emission diagnostics. Namely, the dereddened luminosities of several emission lines (H11, H10, H9, H8, Ca\,II\,$\lambda$3934, He\,I\,$\lambda$4026, H$\delta$, H$\gamma$, H$\beta$, He\,I\,$\lambda$5876, H$\alpha$, and Pa$\beta$), when detected, were converted to $L_{\rm acc}$ using the relationships of \citet{alcala2014}}\footnote{Very similar results are obtained when adopting the empirical relationships by \citet{alcala2017}. Here we use the calibrations from \citet{alcala2014} because they yield slightly smaller differences between the estimated $L_{\rm acc}^{slab}$ and $L_{\rm acc}^{line}$ across our sample than \citeauthor{alcala2017}'s (\citeyear{alcala2017}) calibrations.}. {The final $\widetilde{L}_{\rm acc}$\,(lines) estimate for a given object was then calculated as the median of all individual $L_{\rm acc}$ measurements from the detected emission lines, which are reported in Appendix~\ref{sec:Lline_data}.}
A good fraction of the points on Fig.\,\ref{fig:TWA_logLacc_slab_logLacc_lines} are consistent, within the uncertainties, with a range of 0.3~dex around the identity line on the diagram, which corresponds to the rms scatter around the same identity line measured for the Lupus population. This suggests that the empirical $L_{\rm acc} - L_{\rm line}$ calibrations determined by \citet{alcala2014, alcala2017} for the younger Lupus cluster can also be applied to derive a global picture of accretion in older PMS populations like the TWA. A few objects exhibit, however, a more conspicuous discrepancy between the different $L_{\rm acc}$ diagnostics. In the region of the diagram at $\log{L_{\rm acc}} < -4$, we observe two outliers (from left to right, TWA\,28 and TWA\,22) that fall above the diagonal strip (i.e., their $L_{\rm acc}$ estimated from the emission lines is significantly larger than their $L_{\rm acc}$ estimated from the Balmer continuum; a marginal discrepancy is observed for TWA\,27). 

It is interesting to note that all sources that populate the lower $L_{\rm acc}$ end of the distribution (hence presumably more evolved with respect to other accreting stars in our sample) share the property of being located above the identity line on the diagram, although in some cases being marginally consistent with the latter within the uncertainties. We speculate that, at lower accretion regimes, the emission lines may suffer a larger contamination from chromospheric activity, and therefore yield artificially higher values of $L_{\rm acc}$ than the measurement from the Balmer continuum. This effect would be especially pronounced on stars that rotate faster with respect to other accreting members, either at the end of magnetic braking or as a result of initial conditions at different stellar masses \citep{bouvier2014, scholz2018, moore2019}, and therefore exhibit an enhanced chromospheric activity. Unfortunately, we could not find any estimate of rotation period reported in the literature for the two main outliers in this group to test this hypothesis, and the $v\,\sin{i}$ measurement that we have are rather uncertain. However, we note that the rotation period of 0.78~d measured by \citet{lawson2005} for TWA\,8B, which lies at the upper end of the low $\log{L_{\rm acc}}$ tail of objects on Fig.\,\ref{fig:TWA_logLacc_slab_logLacc_lines} ($\log{L_{\rm acc}}\,[L_\odot] \sim -4$), projects this object among the fastest rotators in the TWA (see also \citealp{messina2010}). {Some discrepancy at the lowest $L_{\rm acc}$ may also be symptomatic of less robust $L_{\rm line}$--$L_{\rm acc}$ calibrations at these luminosity regimes, underrepresented in the Lupus sample (see Figs.\,C.5 and following in \citealp{alcala2014}) and where the relative contribution of chromospheric emission is more important.} In addition, small differences between $L_{\rm acc}$\,(slab) and $L_{\rm acc}$\,(lines) might also be accounted for by the uncertainties associated with the best-fit stellar parameters.

In the area of the diagram at higher $L_{\rm acc}$ ($\log{L_{\rm acc}}\,[L_\odot] > -4$) we observe one star, TWA30\,A, located below the diagonal strip (i.e., with $L_{\rm acc}$ estimate from the Balmer continuum larger than that derived from the line emission). This object is reported to be observed in a near edge-on configuration, although not fully edge-on \citep{looper2010}. In this case, depending on the system geometry, it might occur that the regions where the line emission is mostly formed (e.g., the accretion funnels) are more occulted by the disk than the accretion shock at the star surface, thus determining the imbalance between the two approaches.

\begin{figure}
\resizebox{\hsize}{!}{\includegraphics{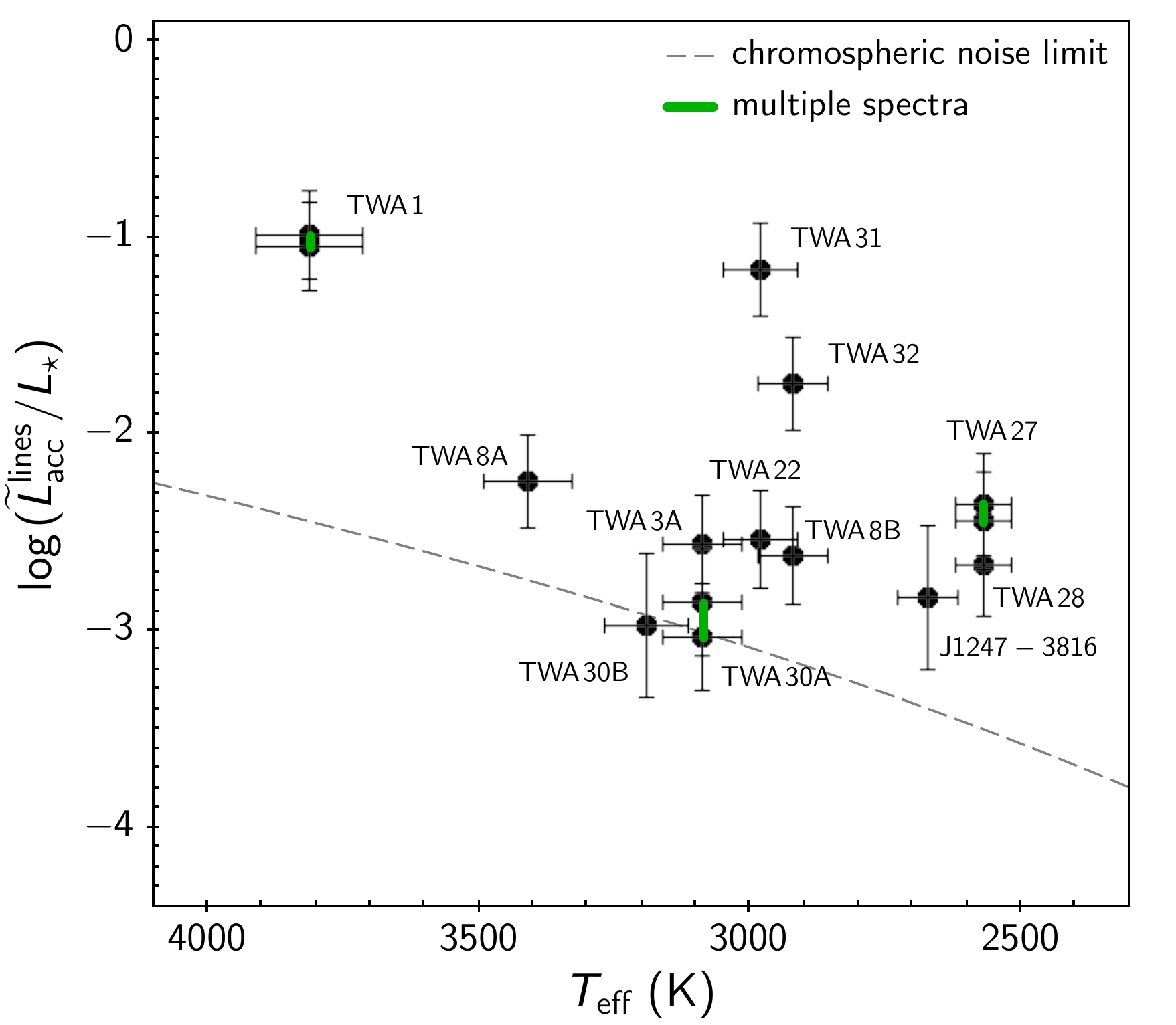}}
\caption{Logarithmic ratio between the median $L_{\rm acc}$ measured from $L_{\rm line}$ (see text) and the stellar $L_\star$, as a function of $T_{\rm eff}$. Green vertical lines connect multiple measurements obtained for objects with more than one spectrum. The gray dashed line indicates the emission level below which the impact of chromospheric emission on the observed accretion signatures becomes dominant.}
\label{fig:TWA_log_Lacc_over_Lbol_vs_Teff}
\end{figure}

Fig.\,\ref{fig:TWA_log_Lacc_over_Lbol_vs_Teff} shows the distribution in $\log{(L_{\rm acc}/L_\star)}$ ratios as a function of stellar $T_{\rm eff}$, where $L_{\rm acc}$ is computed from the measured $L_{\rm line}$. The diagrams also show the $T_{\rm eff}$-dependent ``chromospheric noise limit'' estimated for PMS stars. This corresponds to the amount of accretion luminosity that could be simulated by a non-accreting young star at the corresponding $T_{\rm eff}$/SpT, as a result of its chromospheric activity (see Sect.\,\ref{sec:m_macc_trend}). This limit was calculated following the prescription of \citeauthor{manara2013a} (\citeyear{manara2013a}; see also \citealp{manara2017b}), which was calibrated by measuring the luminosity of several emission lines, typically used as accretion tracers, on a sample of disk-free X-shooter targets in Lupus, $\sigma$~Ori, and TWA. The line traces the approximate emission level below which the contribution from stellar chromospheric activity may become predominant in the observed line profiles. Conversely, stars projected well above this threshold on the diagram can be confidently assumed to be accreting. We stress that this threshold provides only an indication of where the transition occurs: as shown in \citet{manara2013a}, the sample of objects from which the calibration was drawn exhibit a scatter of $\pm$0.2~dex in $\log{(L_{\rm acc}^{\rm noise}/L_\star)}$ around the best-fitting line. {Moreover, while a clear-cut boundary may be traced between strongly accreting objects and disk-free objects, this is likely not the case for weakly accreting objects, which can formally swing across the ``accreting'' vs. ``non-accreting'' threshold \citep[e.g.,][]{cieza2013}.}

Many of the stars in Fig.\,\ref{fig:TWA_log_Lacc_over_Lbol_vs_Teff} fall above the chromospheric noise limit on the diagram, albeit by a smaller margin than observed in younger X-shooter regions (roughly half of them fall within 0.5~dex of the limit in $\log{(L_{\rm acc}/L_\star)}$, whereas more than half of the Lupus targets in \citealp{alcala2014} fall over 1~dex above the threshold). {The same caveat mentioned earlier regarding the applicability of the $L_{\rm line}$--$L_{\rm acc}$ calibrations at the lowest $L_{\rm acc}$ regimes holds here; however, we note that for at least two of the lowest $L_{\rm acc}$ objects (TWA\,27 and TWA\,28), an accreting nature is also suggested by their H$\alpha$ flux following the criteria and empirical boundary for non-accreting objects derived by \citet[][Figs.\,11 and 12]{frasca2015} on the Chamaeleon~I (3~Myr) and $\gamma$~Velorum (5--10~Myr) populations.} Two objects (TWA\,30A and TWA\,30B), instead, exhibit $L_{\rm acc}$ measurements consistent with the chromospheric noise threshold in Fig.\,\ref{fig:TWA_log_Lacc_over_Lbol_vs_Teff}. These two stars are known to be seen in a close to edge-on configuration (near edge-on but not fully edge-on for TWA\,30A; fully edge-on for TWA\,30B; \citealp{looper2010}); it is therefore plausible that their accretion features are partly concealed from view, resulting in a low apparent accretion activity. In these cases, the measured $L_{\rm acc}$ should therefore be interpreted as a lower limit to the true level of accretion. The presence of forbidden emission lines in the spectra of both stars (see Sect\,\ref{sec:forbidden_lines}), on the other hand, indicates an active circumstellar environment with outflows of material.

\subsubsection{The $\dot{M}_{\rm acc}$--$M_\star$ distribution} \label{sec:m_macc_trend}

\begin{figure}
\resizebox{\hsize}{!}{\includegraphics{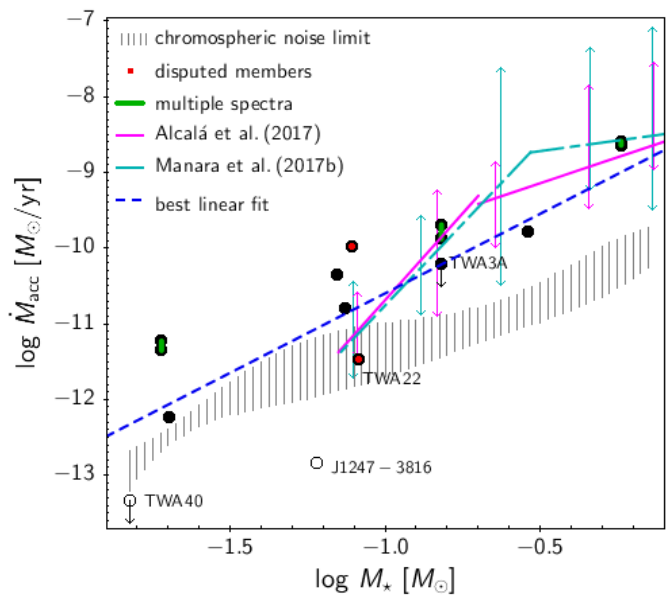}}
\caption{$\dot{M}_{\rm acc}$ vs. $M_\star$ distribution for stars in our sample, where $\dot{M}_{\rm acc}$ is calculated from the measured continuum excess luminosity for all objects (filled dots) except TWA\,40 and J1247-3816 (empty dots). Smaller red dots mark the locations of TWA\,31 (upper point) and TWA\,22 (lower point). Green vertical lines connect multiple $\dot{M}_{\rm acc}$ measurements for the same objects. The shaded gray area traces the estimated chromospheric noise limit for $\dot{M}_{\rm acc}$ detection, as a function of stellar mass, for stars aged between 3 and 10~Myr on \citeauthor{baraffe2015}'s (\citeyear{baraffe2015}) isochrone grid. The dash-dot cyan line traces the bimodal $\dot{M}_{\rm acc}-M_\star$ relationship derived by \citet{manara2017} in the Chamaeleon~I star-forming region, with a break at $M_\star \sim 0.29\, M_\odot$. The solid magenta line traces the bimodal relationship derived in \citet{alcala2017} in Lupus, assuming a break at $M_\star = 0.2 \, M_\odot$. Double arrows mark the 10th--90th percentile range in $\dot{M}_{\rm acc}$ covered by the Lupus (magenta) and Chamaeleon~I (cyan) populations as a function of mass. The dotted blue line traces a linear fit to the datapoints, which accounts for objects dominated by chromospheric emission (labeled on the diagram) as upper limits.}
\label{fig:TWA_M_Mdot}
\end{figure}

Fig.\,\ref{fig:TWA_M_Mdot} illustrates the distribution of $\dot{M}_{\rm acc}$ as a function of $M_\star$. The accretion rates shown are the estimates derived from the continuum excess emission (Sect.\,\ref{sec:multicomponent_fit}), when available; these are complemented by the values derived from the line emission diagnostics (Sect.\,\ref{sec:Lacc_Lline}) for J1247-3816 and TWA\,40 (upper limit). To estimate the mass-dependent, minimum $\dot{M}_{\rm acc}$ we expect to be able to measure above the chromospheric noise on our targets, we followed the approach of \citet{manara2013a, manara2017b} applied to the model tracks of \citet{baraffe2015}. Namely, we took the $T_{\rm eff}$-dependent chromospheric noise limit illustrated in Fig.\,\ref{fig:TWA_log_Lacc_over_Lbol_vs_Teff}, and converted it to a range of $\dot{M}_{\rm acc}$ by using the mass-dependent stellar parameters ($L_\star$, $R_\star$, $T_{\rm eff}$) tabulated in the theoretical models from \citet{baraffe2015} that span the region of the HR diagram covered by our targets in Fig.\,\ref{fig:TWA_logLacc_slab_logLacc_lines}. The result of this computation is shown as a shaded gray strip, which covers the approximate range in chromospheric emission expected, at a given mass, from stars aged between the average projected age of our targets on the HR diagram in Fig.\,\ref{fig:TWA_logLacc_slab_logLacc_lines} ($\sim$3--5~Myr), and the upper estimate of 10~Myr for the nominal age of the TWA. The mass--dependent values of the $\dot{M}_{\rm acc}$ noise threshold at an age of 10~Myr, {calculated using \citeauthor{baraffe2015}'s (\citeyear{baraffe2015}) models,} are also reported in Table~\ref{tab:M_Macc_threshold}.
\begin{table}
\caption{Mass--dependent values of $\dot{M}_{\rm acc}$ noise threshold, computed for stars at an age of 10~Myr using the models of \citet{baraffe2015}.}
\label{tab:M_Macc_threshold}
\centering
\begin{tabular}{c c}
\hline\hline
$M_\star$ & $\log{\dot{M}_{\rm acc, noise}}$ \\
$[M_\odot]$ & \\
\hline
0.02	& -12.538\\
0.03	& -12.242\\
0.04	& -12.189\\
0.05	& -12.061\\
0.06	& -11.980\\
0.07	& -11.886\\
0.08	& -11.798\\
0.09	& -11.725\\
0.1	& -11.693\\
0.11	& -11.643\\
0.13	& -11.540\\
0.15	& -11.457\\
0.17	& -11.371\\
0.2	& -11.283\\
0.3	& -11.034\\
0.4	& -10.814\\
0.5	& -10.625\\
0.6	& -10.461\\
0.7	& -10.291\\
\hline
\end{tabular}
\end{table}
{The location of the $\dot{M}_{\rm acc}$ noise threshold on the diagram in Fig.\,\ref{fig:TWA_M_Mdot} is dependent, to some extent, on the specific model grid adopted for the conversion. However, a quantitative evaluation of the impact that this may have on the classification of our potential accretors is hampered by the fact that most sets of models do not cover $M_\star$ below $\sim$0.1~$M_\odot$, as illustrated in Fig.\,\ref{fig:TWA_HR_models}. A comparison between the minimum $\dot{M}_{\rm acc}$ obtained, as a function of $M_\star$, using alternatively the isochrones by \citet{baraffe2015} and \citet{siess2000} suggests differences of the order of 0.15~dex for $M_\star \sim 0.1-0.7\,M_\odot$. Such margin of uncertainty would not alter our selection of potential accretors above the noise threshold in Fig.\,\ref{fig:TWA_M_Mdot}.}

Three of our sources (from left to right, TWA\,40, J1247-3816, and TWA\,22) fall onto or below the estimated chromospheric noise limit. We therefore conclude that no significant accretion activity was detected for these sources. {TWA\,3A would fall above the semi-empirical $\dot{M}_{\rm acc}$ noise threshold in Fig.\,\ref{fig:TWA_M_Mdot}, but we consider the corresponding $\dot{M}_{\rm acc}$ as an upper limit rather than a true detection for the reason explained in Sect.\,\ref{sec:multicomponent_fit}.} The remaining sources (8/12, excluding TWA\,30B), however, exhibit some accretion activity. The measured $\dot{M}_{\rm acc}$ show a dependence on the stellar mass. Following \citet{venuti2014}, we performed a Kendall's $\tau$ test for correlation \citep{feigelson2012, helsel2012} to assess the presence of a correlation trend and determine the quantitative relationship taking into account the upper limits (values consistent with the chromospheric noise threshold, labeled in Fig.\,\ref{fig:TWA_M_Mdot}) as censored data. Assuming a single power-law description, we obtained a best-fitting trend $\dot{M}_{\rm acc}\, \propto \, M_\star^{2.1 \pm 0.5}$ with a significance of {2.5}\,$\sigma$. This result is consistent with the relationships reported in the literature for younger clusters, albeit with a larger uncertainty that reflects the low number statistics of our sample and the sparse distribution on the diagram. The paucity of objects with $M_\star \gtrsim 0.3\,M_\odot$ in our sample prevents any statistical comparison with the $\dot{M}_{\rm acc}$ measured at those masses in younger regions. However, the levels of accretion observed on very low-mass stars and brown dwarfs in the TWA are similar in value and range to those derived with X-shooter among Lupus and Chamaeleon~I stars. A more thorough discussion of these results in the context of $\dot{M}_{\rm acc}$ evolution is reported in Sect.\,\ref{sec:Macc_evol_mass}.

Recently, \citet{manara2017} and \citet{alcala2017} explored the possibility of a double power-law to reproduce the observed $\dot{M}_{\rm acc}$--$M_\star$ distribution in the Chamaeleon~I and Lupus regions, respectively, using the same type of data and analysis adopted here. They showed that, in both cases, a double power-law, with a break around $M_\star \sim 0.2-0.3\, M_\odot$ and a steeper relationship in the lower-mass regime than in the higher-mass regime, provides a better statistical description to the observed data than a single power-law, although the latter cannot be excluded. This bimodal behavior, tentatively observed also in accretion surveys of other star-forming regions using different methods \citep[e.g.,][]{fang2013b, venuti2014}, has been predicted theoretically as an outcome of the early evolution of circumstellar disks governed by self-gravity and turbulent viscosity \citep[][but see the discussion in \citealp{manara2017} for other possible explanations of this bimodal trend]{vorobyov2009}. The bimodal power law determined in \citet{manara2017} (break at $M_\star \sim 0.29\, M_\odot$) and that computed in \citet{alcala2017} (break at $M_\star = 0.2\, M_\odot$) are both traced for comparison in Fig.\,\ref{fig:TWA_M_Mdot}. Objects in our sample do not provide sufficient parameter coverage to assess whether a bimodal or a unimodal distribution describe the observations better. We simply note that the extrapolation of \citeauthor{manara2017} and \citeauthor{alcala2017}'s relationships (calibrated down to $\log{M_\star}\,[M_\odot] \sim -1.1$) does not appear to match the observed accretion properties of the lowest mass objects in the TWA, although both descriptions could fit the data above $\log{M}_\star\,[M_\odot] \sim -1.3$. This is consistent with indications from \citeauthor{manara2015}'s (\citeyear{manara2015}) study of accretion in very low-mass stars and brown dwarfs in the 1-2~Myr-old $\rho$~Ophiucus, with measured $\dot{M}_{\rm acc}$ that would fall well above the predicted $\dot{M}_{\rm acc}$ at the lowest mass regimes in the broken power-law description of \citet{manara2017} and \citet{alcala2017}.

We note that cases of objects near the brown dwarf regime with $\dot{M}_{\rm acc}$ higher than expected based on the $M_\star - \dot{M}_{\rm acc}$ trend for $M_\star \simeq 0.1-0.3\,M_\odot$ stars were also observed in Lupus. Overall, these data hint at a different behavior of accretion at the lowest stellar masses. \citet{alcala2017} proposed that these objects could be understood in the theoretical framework of \citet{stamatellos2015}, according to which the lowest-mass stars do not form via prestellar core collapse, but as a product of the fragmentation of disks around solar-mass stars. In this scenario, the disk that takes shape around the newly formed substellar object continues to accrete material from the parent disk before starting to evolve independently. This process leads to comparatively more massive disks, relative to the mass of the central object, than those formed via prestellar core collapse, which translates to comparatively higher accretion rates onto the star. On the other hand, \citet{testi2016} used ALMA data to report tentative indications of a lower $M_{\rm disk}/M_\star$ ratio for brown dwarfs than for solar-mass stars in $\rho$~Oph, albeit potentially affected by sample incompleteness and large uncertainties on the stellar parameters adopted.

\subsection{Emission lines as tracers of the star-disk interaction environment} \label{sec:acc_dynamics}

\subsubsection{H$\alpha$ line profiles and Balmer decrements} \label{sec:decrements}

The profiles of several emission lines, in particular H$\alpha$ and others from the Balmer series, have long been investigated in the literature in connection with the physical conditions of the circumstellar environment in PMS stars. In their seminal work, \citet{reipurth1996} established a first thorough classification of H$\alpha$ line profiles in young stars. The authors identified four main types (Type~I--IV) of H$\alpha$ morphologies, related to the dynamics of material infall or ejection that takes place in the inner disk regions. We here applied \citeauthor{reipurth1996}'s (\citeyear{reipurth1996}) classification to sort our H$\alpha$ line profiles empirically based on their shapes. In addition, we implemented a simple Gaussian description to evaluate the deviation of the individual lines from a symmetric profile. Details of this analysis are reported in Appendix~\ref{sec:Halpha_analysis}, and the main inferences are summarized in the following paragraph.

While none of the observed line profiles appears strictly Gaussian, a Gaussian description appears adequate to describe the global H$\alpha$ profile observed on around 50\% of the targets. Overall, 50\% of the H$\alpha$ profiles in our sample fall into the Type~I class definition (symmetric, with no absorption features); Type~II profiles (double-peaked) occur in 37.5\% of cases, while smaller percentages are associated with the Type~III (8.3\%) and Type~IV (4.2\%) classes. This distribution is similar to those found by \citet{antoniucci2017} in Lupus and \citet{sousa2016} in NGC~2264, both aged around 3~Myr. \citet{antoniucci2017} also classified the observed H$\alpha$ profiles in the Lupus region based on their FWHM, and identified two similarly populated, distinct groups: narrow-line objects, with FWHM~$\lesssim$~100~km/s, and wide line profiles, with FWHM~$\sim$~200--300~km/s. Our sample is distributed in a different way than the Lupus sample in this respect: the majority of objects (83\%, including those where chromospheric activity is possibly predominant) are found in the narrow-line group, while only TWA\,1 and the disputed member TWA\,31 exhibit wide line profiles with FWHM~$\sim$~200--250~km/s.

\begin{figure}
\resizebox{\hsize}{!}{\includegraphics{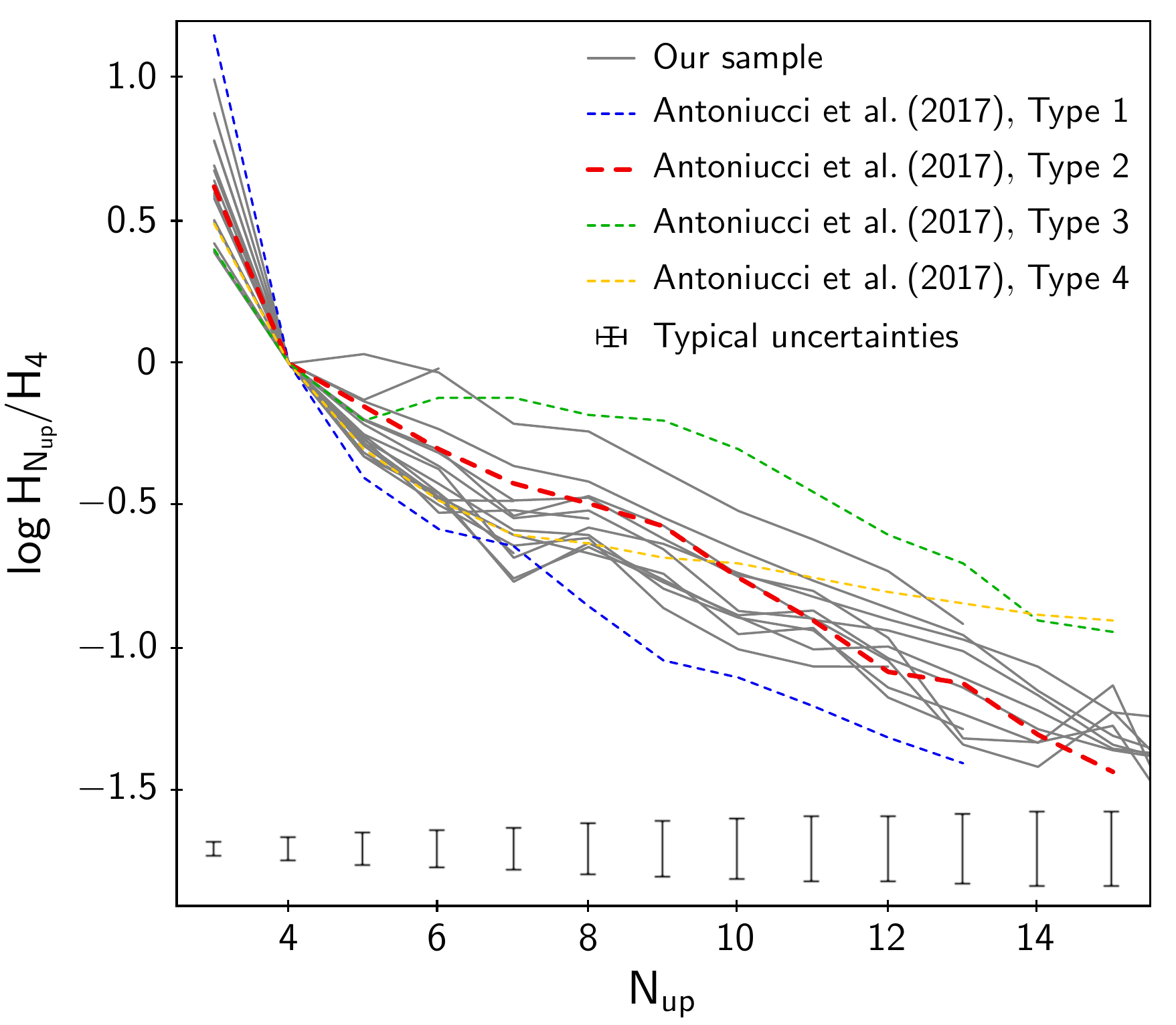}}
\caption{Shapes of Balmer decrements computed for each of our targets (gray lines), defined as $\log{\mbox{Hn/H4}}$ ($\equiv \log{\mbox{Hn/H$\beta$}}$). The prototypes of Balmer decrement types identified in Lupus by \citet{antoniucci2017} are shown with dotted lines for comparison; a thicker stroke is used to emphasize Type 2 (the most common type in their sample and in ours). Typical uncertainties on the $\log{H_{N_{up}}/H_4}$, displayed below the individual Balmer decrement trends, were calculated from the uncertainties associated with the corresponding line flux measurements.}
\label{fig:TWA_Balmer_decrements}
\end{figure}

In the context of the X-shooter survey of the Lupus star-forming region, \citet{antoniucci2017} also undertook a detailed analysis of the Balmer decrements, i.e., of the ratio of fluxes between different HI emission lines in the Balmer series, exhibited by accreting stars in the region. {As shown in their article (see in particular their Table~3), the decrement shapes can be related to the physical properties (e.g., density, temperature) of the emitting gas.} \citet{antoniucci2017} identified four main types of Balmer decrements, defined by the value of $\log{{F_{\rm H9}/F_{\rm H4}}}$ (henceforward $\log{{\rm H9}/{\rm H4}}$). {The different decrement shapes observed were found to correspond to different predictions of the local line excitation calculations by \citet{kwan2011}, which describe line emission under varying physical conditions of winds and accretion flows in T~Tauri stars}. In Fig.\,\ref{fig:TWA_Balmer_decrements}, we illustrate the Balmer decrement shapes computed for our targets, compared with the prototypes of the four Balmer decrement types identified in Lupus by \citet{antoniucci2017}. The luminosities of the Balmer lines from H$\alpha$ to H15 that we measured for our targets are reported in Table~\ref{tab:Balmer_series}. The Balmer line ratios were computed taking as reference the luminosity of the H4 (H$\beta$) line, in analogy with the definition of \citet{antoniucci2017}. Most of the nine objects with measured H9-to-H4 ratio in our sample (83\%) fall into the type~2 category of Balmer decrement shapes, as defined by \citet{antoniucci2017}, {which can be associated with total gas densities $\log{n_H}\sim9.5$ and temperatures in the range 5000--15\,000~K}. This result is similar to the one reported in Lupus (type 2 being the most common category of Balmer decrement shapes), but with a larger rate of occurrence than the 56\% found in that case. One object (TWA\,28) exhibits a type 1 Balmer decrement profile (11\%). This rate, albeit subject to our very limited statistics, is similar to the one found by \citet{antoniucci2017} in Lupus. We could not establish from the literature whether TWA\,28 is also an edge-on system like the Lupus objects that exhibit type~1 Balmer decrement profiles. We note, however, that TWA\,28 is the source with the largest discrepancy between $\log{L_{\rm acc}^{slab}}$ (lower) and $\log{\widetilde{L}_{\rm acc}}^{lines}$ (higher) in Fig.\,\ref{fig:TWA_logLacc_slab_logLacc_lines}, which might be consistent with a geometry where the central region of the star is more obscured than the line-emitting region. Another object (TWA\,27) exhibits a type~2 Balmer decrement at one epoch (P089), and a type~3 Balmer decrement in another epoch (P084). We did not detect any strong variations in the H$\alpha$ FWHM, and measured a variation of only $\sim$0.15--0.25~dex in $\dot{M}_{\rm acc}$ between the two epochs, which is consistent with the typical week-long (that is, on rotational timescales) accretion variability monitored on young stars at ages of a few Myr \citep[e.g.,][]{costigan2014, venuti2014, frasca2018}. However, TWA\,27 exhibits stronger H$\alpha$ emission in P089 than in P084, but stronger H$\beta$ emission in P084 than in P089. We note that this object had been the target of detailed spectroscopic monitoring by \citet{stelzer2007}, who reported generally multi-structured Balmer line profiles, with variations in the shape and relative intensity of the peaks on timescales ranging from hours to months. Evident variations in the H$\alpha$ line morphology for TWA\,27 are also observed between our two epochs, as illustrated in Fig.\,\ref{fig:TWA_Ha_atlas}.

We did not find any straightforward correlation between the Balmer decrement type and the H$\alpha$ line shape or the $\dot{M}_{\rm acc}$ level; the typical $\log{\mbox{H9/H4}}$ in each H$\alpha$ class and $\dot{M}_{\rm acc}$ bin falls always in the type~2 Balmer decrement category, but there might be a tendency for Type~II H$\alpha$ profiles to correspond to slightly larger $\log{\mbox{H9/H4}}$, and for the strongest accretors in our sample to exhibit slightly smaller $\log{\mbox{H9/H4}}$. Following \citeauthor{antoniucci2017}'s (\citeyear{antoniucci2017}) interpretation, the typical Balmer decrement properties observed may indicate that stars in the TWA are accreting at a moderate regime with optically thin emission. None of our targets with Balmer emission lines detected until the order 14 have Balmer decrements of type 4, which were associated by \citeauthor{antoniucci2017} with wide H$\alpha$ line profiles, intense accretion activity, and optically thick emission. However, we did detect two objects (TWA\,1 and TWA\,31) with large H$\alpha$ FWHM, as discussed earlier; these two objects exhibit the largest $\dot{M}_{\rm acc}$ measured at the corresponding $M_\star$ in our sample.

For stars in Lupus, \citet{frasca2017} also investigated the line flux ratios in the Ca\,II infrared triplet (IRT), which are used as tracers of both chromospheric activity and accretion, and likely probe the physical conditions of the emitting gas in a region closer to the photosphere than what traced by the Balmer lines. In our sample, Ca\,II\,8498\AA\mbox{} and   Ca\,II\,8542\AA\mbox{} emission was only detected for seven objects (TWA\,1, TWA\,8A, TWA\,28, TWA\,30B, TWA\,31, and TWA\,32, plus the object TWA\,3B discussed in Appendix~TWA3B). In all of the cases where Ca\,II IRT emission was detected, the measured flux ratios are consistent with optically thick emission that may originate in accretion shocks or chromospheric plages, as illustrated in Fig.\,\ref{fig:fratio_CaIRT_xsho_teff}.

\subsubsection{Accretion/ejection: outflow signatures} \label{sec:forbidden_lines}

Mass accretion processes in PMS stars are interconnected with mass ejection phenomena, such as magnetically powered jets and disk winds. These mechanisms play a crucial role in removing angular momentum from the system, which in turn allows disk material to be transported inwards and accrete onto the star. Jets and winds in young stellar objects (YSOs) are revealed spectroscopically by the presence of forbidden emission lines, which are excited in the heated high-velocity gas. Such lines often exhibit a complex profile with multiple components \citep[e.g.,][]{rigliaco2013, simon2016, mcginnis2018, banzatti2019}: a low velocity component (LVC), centered close to the system velocity and possibly associated with slow disk winds \citep[e.g.,][]{natta2014}, and a high velocity component (HVC), shifted by up to $\sim$200~km/s with respect to the stellar rest frame, and associated with extended collimated jets \citep[e.g.,][]{nisini2018}. 

The [O\,I]\,6300\AA\mbox{} is the most intense among the forbidden emission lines observed in YSOs. As part of the X-shooter Italian GTO survey of star-forming regions, \citet{natta2014} and \citet{nisini2018} recently investigated the connection between the luminosity of the LVC and the HVC of the [O\,I]\,6300\AA\mbox{} line, and the stellar and accretion parameters derived simultaneously and homogeneously for YSOs in Lupus, Chamaeleon, and $\sigma$~Ori. Following their approach, we here analyzed the [O\,I]\,6300\AA\mbox{} emission properties for stars in the older TWA, to test any evolutive trends in the relationships reported by those authors. We identified [O\,I]\,6300\AA\mbox{} emission above the local continuum in four sources: TWA\,1 (both spectra), TWA\,30A (all spectra and epochs), TWA\,30B (both epochs), and disputed member TWA\,31. Our rate of detection is therefore 4/13 objects ($\sim$31\%) across our entire sample, and 3/11 (27\%) including high-likelihood TWA members only. This percentage is lower than the 77\% found in \citet{nisini2018}, and than the 59\% reported by \citet{mcginnis2018} in a similar study conducted with VLT/FLAMES on the NGC~2264 region (3-5~Myr). Even higher rates of detection were reported by \citet{simon2016} in Taurus (91\%, based on 32 disk-bearing objects), and by \citet[][94\%]{banzatti2019} in a composite sample of 65 objects from different star-forming regions and PMS associations.

To recover the LVC and HVC, we performed a Gaussian fit to the observed line profile, after normalizing the flux to the local continuum {and subtracting the photospheric contribution to the line emission}. In the formulation of Eq.\,\ref{eqn:gaussian} (where $a$ is fixed to 1), we explored a range of values for $k$, $\mu$, and $\sigma$, and for each of them we evaluated the goodness-of-fit via the $\chi^2$ statistics. We then repeated the analysis assuming two Gaussians, and selected the best solution (single line component or two line components) as the one with the smallest reduced $\chi^2$. 
\begin{figure}
\resizebox{\hsize}{!}{\includegraphics{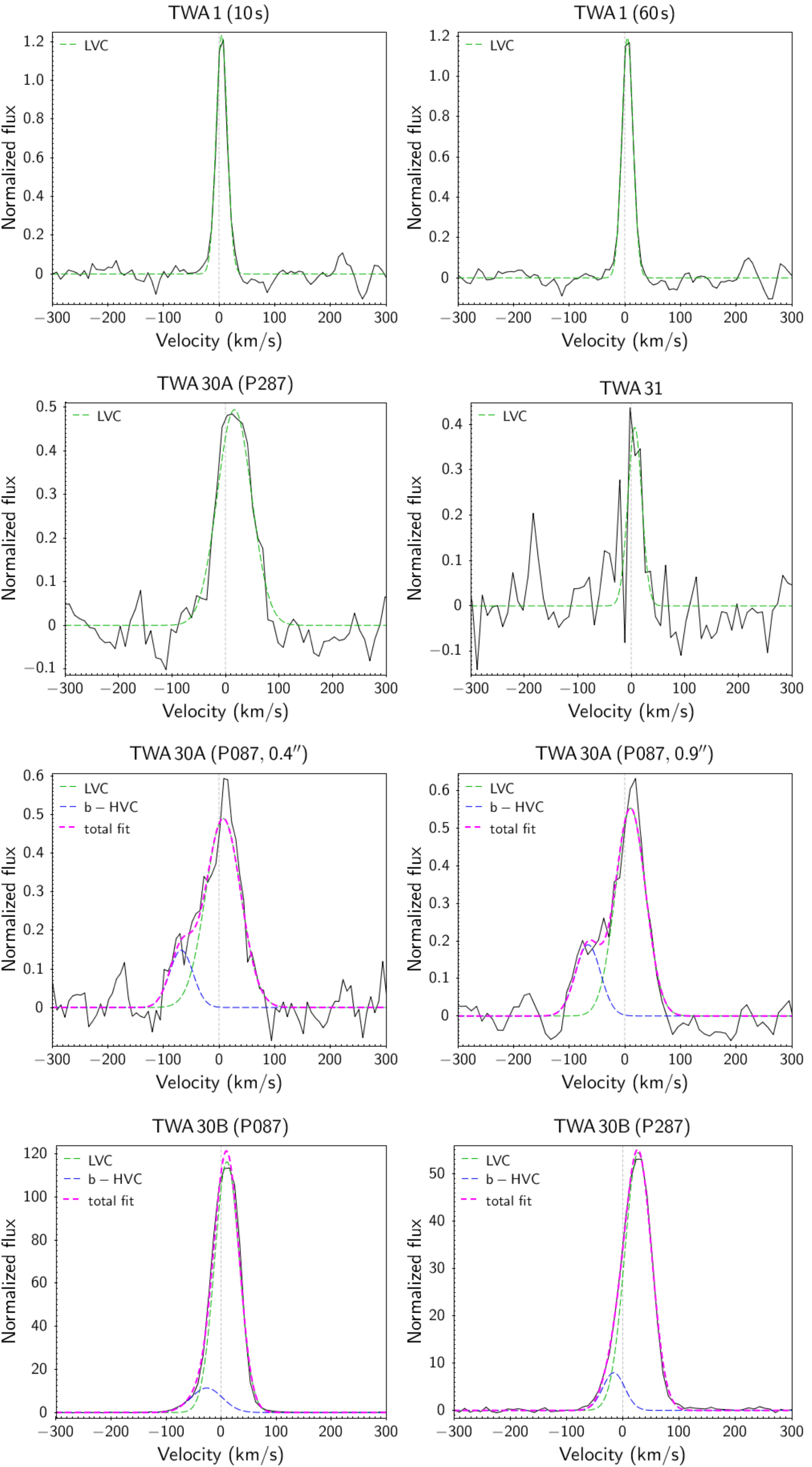}}
\caption{Best single-Gaussian (LVC) or double-Gaussian (LVC+HVC) solution obtained for the [O\,I]\,6300\AA\mbox{} line profile decomposition of TWA\,1, TWA30\,A, TWA\,30B, and TWA\,31. The LVC is traced with a dotted green line, the blueshifted HVC (when present) is traced as a dotted blue line, and the total fit is traced as a dotted magenta line.}
\label{fig:TWA_OI630}
\end{figure}
The results of this procedure are illustrated in Fig.\,\ref{fig:TWA_OI630}. In half of the cases (both spectra of TWA1, TWA30\,A\_P087, and TWA\,31) we could only identify a LVC; we instead extracted two separate components (LVC plus a blueshifted HVC) from the [O\,I]\,6300\AA$\mbox{}$ line profile exhibited by TWA\,30A\_P087 (both spectra) and TWA\,30B (both epochs). The extracted luminosity values for both components are reported in Table~\ref{tab:forbidden_line_emission}. Our rate of detection of an HVC is therefore 2/13 (15\%) across the entire sample, and 2/11 (18\%) for TWA only. For comparison, \citet{nisini2018} reported an HVC for 30\% of their targets in Lupus, Chamaeleon~I and $\sigma$~Ori, while \citet{mcginnis2018} detected HVCs in 17\% of their targets in NGC~2264. 

Figure~\ref{fig:TWA_OI6300_param} shows the comparison between the luminosity measured for the [O\,I]\,6300\AA\mbox{} LVC in the four objects listed above, and the corresponding stellar ($L_\star$, $M_\star$) and accretion ($L_{\rm acc}$, $\dot{M}_{\rm acc}$) parameters. A statistically meaningful comparison with earlier results is prevented by the very small number of objects with [O\,I]\,6300\AA\mbox{} emission in our sample. We can simply observe that the distribution of points derived here on each panel in Fig.\,\ref{fig:TWA_OI6300_param} is globally consistent with the prescriptions obtained by \citet{nisini2018} and \citet{natta2014}, within the dispersion about the best fitting relationships measured for the Lupus, Chamaeleon, and $\sigma$~Ori populations (see Figs.\,4 and 6 of \citealp{nisini2018}).

\section{Discussion} \label{sec:discussion}

\subsection{The evolution of disk accretion between $\sim$1 and 10~Myr} \label{sec:Macc_evol_mass}

\begin{figure}
\resizebox{\hsize}{!}{\includegraphics{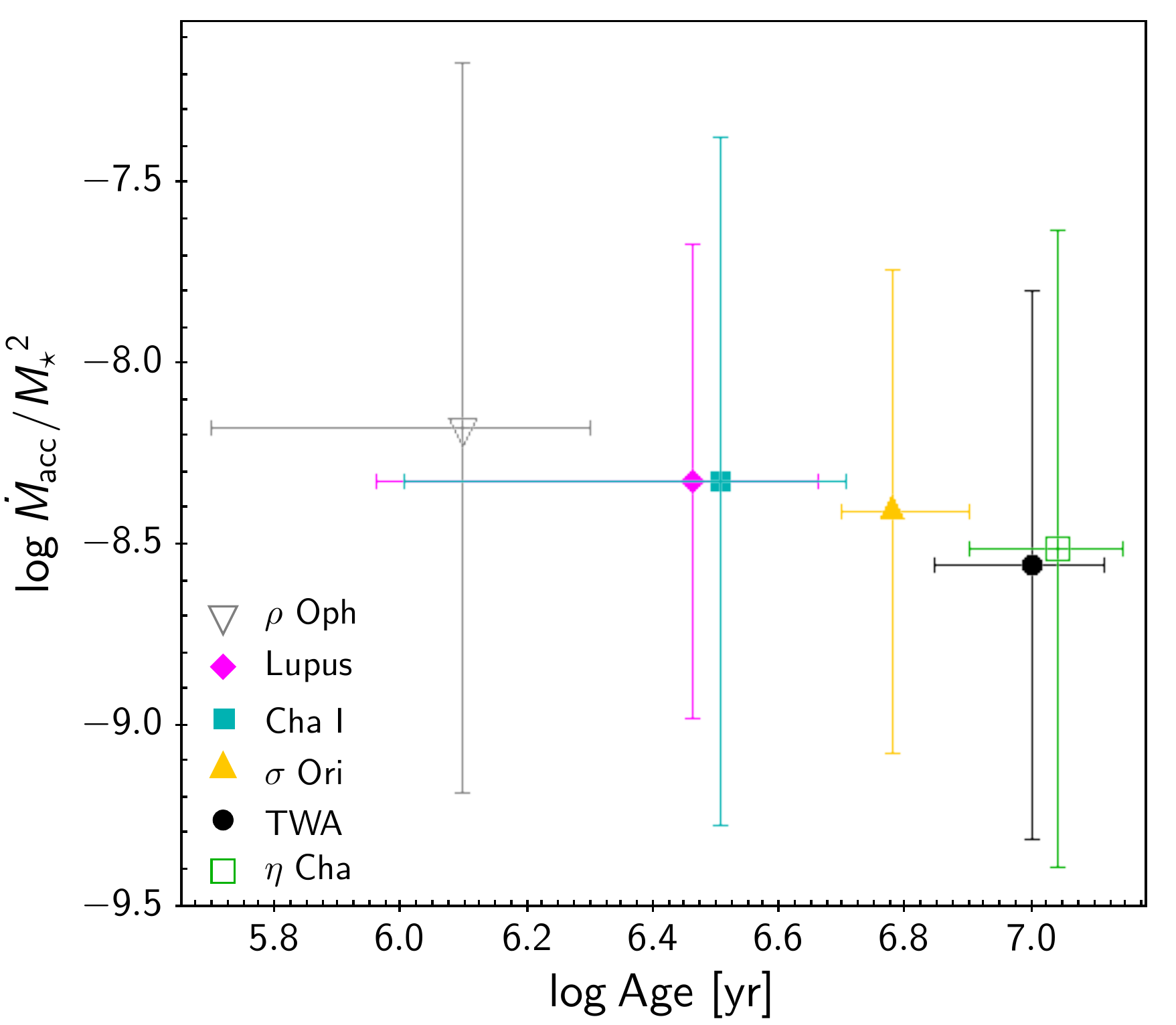}}
\caption{Median age and accretion properties normalized to squared mass for six young regions targeted with X-shooter: $\rho$~Ophiuchi (\citealp{manara2015}; gray empty triangle), Lupus (\citealp{alcala2014, alcala2017}; magenta diamond), Chamaeleon~I (\citealp{manara2017}; cyan square), $\sigma$~Orionis (\citealp{rigliaco2012}; yellow triangle), TWA (this study; black dot), and $\eta$~Chamaeleontis (\citealp{rugel2018}; green empty square). The horizontal error bars mark the measured age spread among cluster members in the case of $\rho$~Oph, Lupus, and Cha~I, and the uncertainty on the nominal age for $\sigma$~Ori \citep{bell2013} and the associations TWA and $\eta$~Cha \citep{bell2015}. The vertical errors mark the logarithmic dispersion in $\dot{M}_{\rm acc}/M_\star^2$ measured across the various populations. For a clearer depiction, the datapoints representing Lupus and Cha~I were shifted by -0.015 and 0.015, respectively, from the nominal median age (3~Myr).}
\label{fig:XS_region_med_age_Macc}
\end{figure}

Fig.\,\ref{fig:XS_region_med_age_Macc} summarizes the global accretion properties measured in the course of X-shooter campaigns on six PMS clusters and stellar associations of different ages: i) $\rho$~Ophiuchi (\citealp{manara2015}; age $\sim$0.5--2~Myr, \citealp{wilking2008}), ii-iii) the previously introduced Lupus and Chamaeleon~I (age $\sim$ 3~Myr, with a spread of 2~Myr), iv) $\sigma$~Orionis (\citealp{rigliaco2012}; age re-estimated as $\sim 6^{+2}_{-1}$~Myr by \citealp{bell2013}), v) the $\eta$~Chamaeleontis association (\citealp{rugel2018}; age $\sim 11\pm3$~Myr, \citealp{bell2015}), and vi) the TWA investigated here (age $\sim 10\pm3$~Myr, \citealp{bell2015}). In spite of the age controversy mentioned in Sect.\,\ref{sec:mass_rad_macc}, we adopt here the age estimates from \citet{bell2013, bell2015} for the older regions because they were obtained with a uniform approach, hence warranting at least a robust relative ordering in age. We note that, while the surveys of Lupus, Cha~I and TWA are estimated to be $\gtrsim$90\% complete in the disk-bearing population of the targeted regions, the X-shooter surveys of $\rho$~Oph and $\sigma$~Ori specifically address the lowest mass component of the cluster populations ($M_\star \lesssim 0.3 M_\odot$), and the survey of $\eta$~Cha encompasses around 68\% of the suspected members of the association. To account for the different distributions in mass of each sample, we followed the approach of \citet[][see also \citealp{antoniucci2014}]{rugel2018}, and show in Fig.\,\ref{fig:XS_region_med_age_Macc} the median $\dot{M}_{\rm acc}$, normalized to $M_\star^2$, measured across each region. {This normalization effectively removes the typical $M_\star$--dependence observed in the $\dot{M}_{\rm acc}$ distributions of PMS populations \citep[$\dot{M}_{\rm acc} \sim M_\star^2$; e.g.,][]{hartmann2016}, and therefore enables a comparison of the average accretion level as a function of cluster age as unaffected by the specific mass coverage as possible.}

Fig.\,\ref{fig:XS_region_med_age_Macc} seems to suggest a marginal decrease in $\dot{M}_{\rm acc}$ with increasing age, the extent of which is, however, appreciably smaller than the 1~$\sigma$ scatter in value measured in any region. To assess the statistical significance of the apparent trend, we created 10\,000 sets of $(\log{\rm Age}, \log{\dot{M}_{\rm acc}/M_\star^2})$ pairs, randomly extracted within the uncertainties illustrated in Fig.\,\ref{fig:XS_region_med_age_Macc} assuming a uniform distribution in $\log{\rm Age}$ and a Gaussian distribution in $\log{\dot{M}_{\rm acc}/M_\star^2}$ around the best values. For each set of simulated measurements, we computed the Pearson's correlation coefficient $r$ \citep[e.g.,][]{press1992}; we then calculated the average and the standard deviation of the resulting $r$ values as an indicator of the strength of the seeming anticorrelation trend, given the uncertainties. The distribution in $r$ that we derived peaks around $-0.2$, with a standard deviation of $0.4$; this result is not dependent on the number of simulated sets adopted, as the same parameters for the distribution in $r$ are obtained when using 100 or 1\,000 iterations instead of 10\,000. The test therefore indicates that, if we account for the uncertainties and for the intrinsic scatter in values within each population, we cannot conclude on whether a decrease in $\dot{M}_{\rm acc}$ is actually observed on Fig.\,\ref{fig:XS_region_med_age_Macc}. Furthermore, the error bars displayed on the diagram do not include additional sources of uncertainty, such as the exact power of $M_\star$ that ought to be adopted to flatten the distribution in $\dot{M}_{\rm acc}$ vs. $M_\star$ within a given population, or the different selection biases (for instance, the inclusion or exclusion of upper limits in $\dot{M}_{\rm acc}$) when computing the averaged statistics for each region.

\begin{figure}
\resizebox{\hsize}{!}{\includegraphics{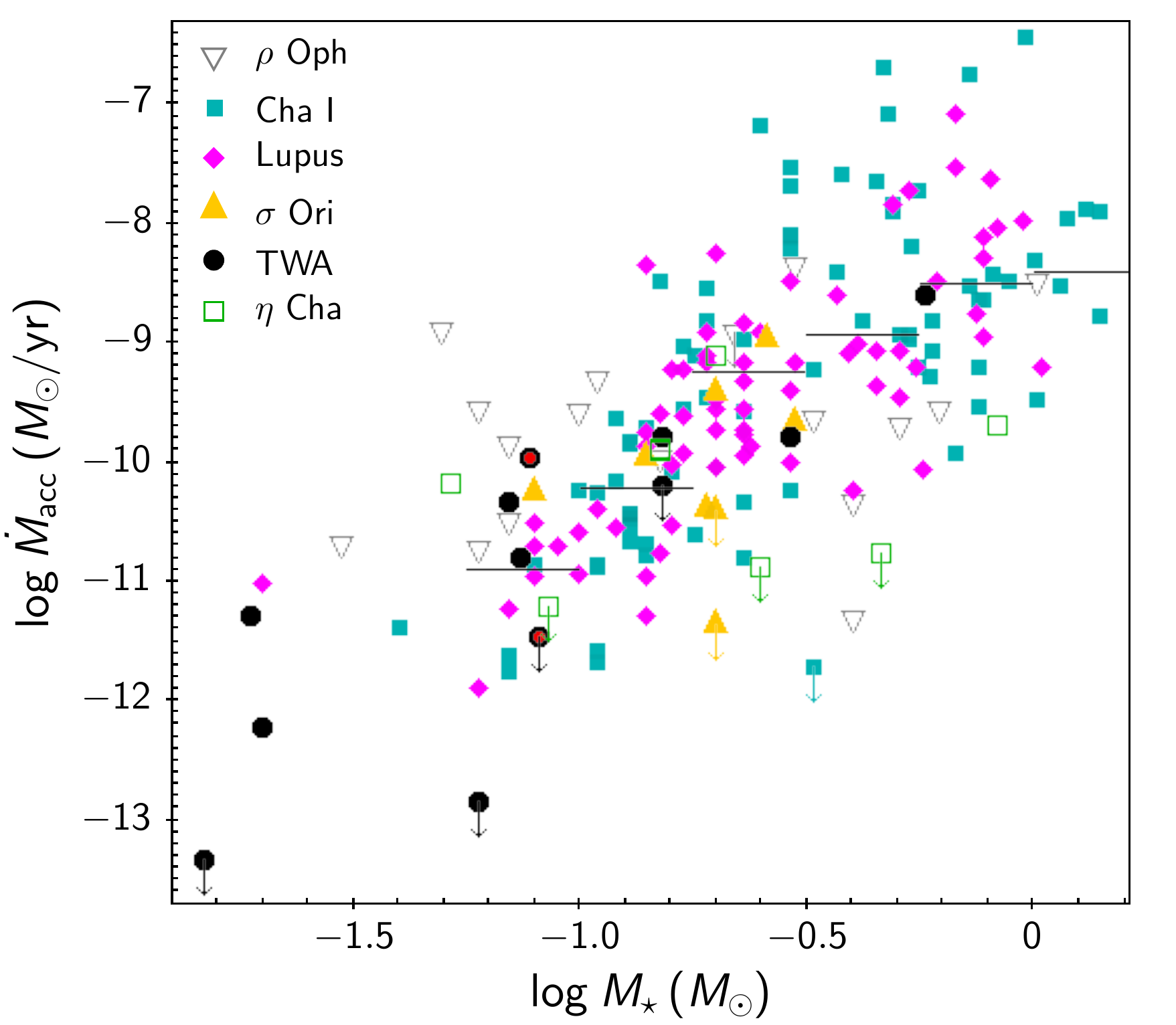}}
\caption{Comparison between the $\dot{M}_{\rm acc}-M_\star$ distributions obtained from X-shooter data for $\rho$~Ophiuchi, Lupus, Chamaeleon~I , $\sigma$~Orionis, TWA, and $\eta$~Chamaeleontis (symbols and sources of data as detailed in the caption of Fig.\,\ref{fig:XS_region_med_age_Macc}). Upper limits in $\dot{M}_{\rm acc}$ are marked by downward arrows. The horizontal lines mark the median $\dot{M}_{\rm acc}$ measured in bins of 0.25 dex in $\log{M_\star}$ for the combined Cha~I and Lupus datasets.}
\label{fig:XS_regions_Macc_comparison}
\end{figure}

Fig.\,\ref{fig:XS_regions_Macc_comparison} illustrates the distribution of individual targets for the six regions presented in Fig.\,\ref{fig:XS_region_med_age_Macc} on the $\dot{M}_{\rm acc}-M_\star$ diagram. A detailed quantitative analysis is hampered by the fact that the distributions in mass pertaining to the six samples are different from each other, and little can be said for the mass regime $M_\star \geq 0.3\,\,M_\odot$, where we have only one star from the TWA (TWA\,1) and a few more from $\rho$~Oph and $\eta$~Cha, with measured $\dot{M}_{\rm acc}$ below the median accretion level detected in the corresponding mass bin for the composite Lupus and Cha~I sample (horizontal line in Fig.\,\ref{fig:XS_regions_Macc_comparison}). At lower masses ($M_\star \sim 0.08-0.25\,M_\odot$), where the samples overlap, stars in $\rho$~Oph appear located systematically above the $\dot{M}_{\rm acc}$ levels measured in the other populations. However, this effect should be interpreted with caution since the accretion rates for $\rho$~Oph targets were measured from the emission lines, contrary to the quasi-totality of other measurements shown on the diagram (which were derived from the Balmer continuum emission), and are subject to larger uncertainties \citep{manara2015}. If we exclude $\rho$~Oph, on the other hand, it appears that the remaining populations cover similar ranges in $\dot{M}_{\rm acc}$ at the lowest stellar masses, with striking evidence for some 5-10 Myr-old accreting stars with measured $\dot{M}_{\rm acc}$ higher than the typical level of accretion detected in the combined Lupus\,+\,Cha~I sample. This comparison might be affected by different completeness limits of the various surveys over the mass range of intersection among the various samples (for instance, the Lupus surveys did not include many young star candidates with $M_\star \lesssim 0.1\,\,M_\odot$, and \citealp{manara2017} quote a completeness limit of SpT\,$\sim$\,M6 for the Cha~I sample). Nevertheless, the selection bias would impact predominantly the low $\dot{M}_{\rm acc}$ regimes at the corresponding $M_\star$, and is unlikely to cause a dearth of stronger accretors in the mass range of interest for the two younger populations. The picture in Fig.\,\ref{fig:XS_regions_Macc_comparison} therefore suggests that very low-mass stars that still retain a disk at the age of the TWA can sustain a prolonged accretion activity with an $\dot{M}_{\rm acc}$ possibly more ``stable'' (i.e., slowly evolving) in the long-term than on more massive stars. This may perhaps be related to a more efficient photoevaporation effect in dissipating material in the inner disks around higher mass stars.

To ascertain whether this conclusion might be affected by the incompleteness of the X-shooter survey of disk--free PMS stars in the TWA (see Sect.\,\ref{sec:mass_rad_macc}), we examined the SpT distribution of all 55 high-likelihood members listed in \citet{gagne2017} with reference to the presence or absence of accretion disk signatures. This analysis indicates that, while less than 10\% of TWA stars with SpT between M0 and M3 exhibit such signatures, the percentage of accreting objects rises to $\sim$40\% among M4--M9 stars. If the evolutionary timescales for disk accretion were the same for very low-mass stars and brown dwarfs ($M_\star \sim 0.02-0.2\,M_\odot$) as for higher-mass stars ($M_\star \sim 0.2-0.6\,M_\odot$), this result would imply that the lowest--mass objects in the TWA are statistically younger (i.e., they were formed at later epochs) than higher--mass objects in the same association. However, this scenario is not supported by the similarity in spatial and kinematic properties between low--mass stars and brown dwarfs in loose star--forming regions such as Taurus \citep{luhman2007}.

We note that the mass regime ($M_\star \leq 0.2\,M_\odot$) where longer timescales for $\dot{M}_{\rm acc}$ evolution appear to hold from our analysis lies below the typical mass range ($M_\star \sim 0.3-1\,M_\odot$) for which tentative $\dot{M}_{\rm acc}(t)$ have been proposed in the literature following a multi-cluster approach \citep[e.g.,][]{hartmann1998, calvet2005, sicilia_aguilar2010, hartmann2016}\footnote{\citet{antoniucci2014} adopted a similar approach on a broader stellar mass range, but they focused on the earlier stages of $\dot{M}_{\rm acc}$ evolution, for populations of the same age as Lupus or Cha~I and younger.}. \citet{hartmann2006} speculated that low-mass disks around the lowest mass stars might be viscously drained on shorter timescales, therefore yielding a stronger $\dot{M}_{\rm acc}$ dependence on stellar age than observed on young stars of mass closer to $1\,M_\odot$. In their analysis of accretion in the Orion Nebula cluster, \citet{manara2012} combined individual $\dot{M}_{\rm acc}$ and stellar age measurements to report a faster temporal decay of $\dot{M}_{\rm acc}$ in very low-mass stars ($M_\star < 0.2\,M_\odot$) than in more massive stars ($M_\star \sim 0.4-1\,M_\odot$). On the other hand, in a similar study of the NGC~2264 region \citet{venuti2014} reported a looser correlation between $\dot{M}_{\rm acc}$ and age for lower-mass stars ($M_\star < 0.4\,M_\odot$) than for higher-mass stars ($M_\star \sim 0.4-1\,M_\odot$), although pointing out the uncertainty associated with model--based individual age determinations for young stars. 

\subsection{Connection between disk mass and mass accretion}

The relationship between the mass accretion rate $\dot{M}_{\rm  acc}$ onto young stars and the mass of their circumstellar disks ($M_{\rm disk}$) is a key observational proxy to understand the leading physical processes in disk evolution. From a theoretical standpoint \citep[see, e.g.,][]{rosotti2017, lodato2017}, it is expected that, in a scenario where the disk has attained a quasi-steady state of viscous evolution, $\dot{M}_{\rm acc}$ should be of the order of ${M}_{\rm  disk}$ over the age $t$ of the star-disk system: $\dot{M}_{\rm  acc} \sim M_{\rm disk}/t$. Other processes that trigger an inside-out disk clearing (such as internal photoevaporation driven by the radiation from the central object) would determine a faster depletion of material from the inner disk region (where accretion takes place) than from the outer disk region; in this case, the theoretical expectation is $\dot{M}_{\rm  acc} < M_{\rm disk}/t$. In the opposite scenario of outside-in disk clearing (which can be triggered by external photoevaporation by nearby massive stars), the disk mass decreases at a faster pace than can be accounted for by viscous accretion onto the star; in this case, $\dot{M}_{\rm  acc} > M_{\rm disk}/t$ \citep{rosotti2017}. Recent studies \citep{manara2016, mulders2017} have combined X-shooter measurements of $\dot{M}_{\rm  acc}$ with ALMA measurements of $M_{\rm disk}$ to infer a correlation between the two parameters for young stars in Lupus and Chamaeleon~I. The observed trend resulted compatible with the linear relationship predicted in the case of viscous disk evolution, albeit with a significant scatter that may reflect contributions from additional physical processes in the transport of material and angular momentum across the disk (grain growth and migration, disk winds), or viscous timescales longer than the average age of the regions \citep{lodato2017}. 

\begin{figure}
\resizebox{\hsize}{!}{\includegraphics{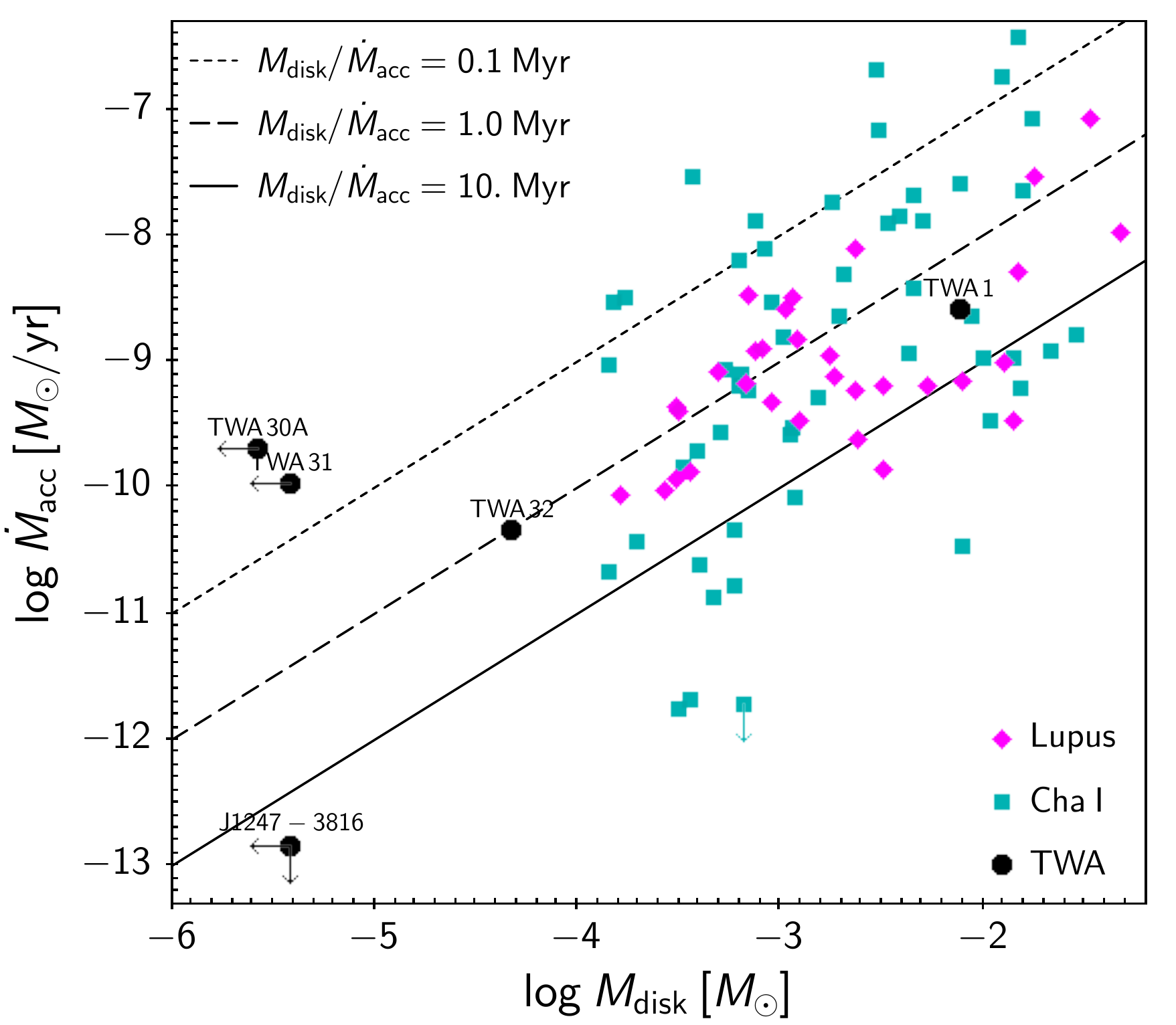}}
\caption{Logarithmic $\dot{M}_{\rm acc}$ vs. disk mass distribution for stars in the TWA, compared to the populations of the Lupus and Chamaeleon~I star-forming regions. The straight lines mark the $\dot{M}_{\rm acc}-M_{\rm disk}$ relationship predicted for different stellar ages in a scenario of viscous quasi-steady disk evolution \citep{rosotti2017}. Upper limits are indicated by correspondingly oriented arrows.}
\label{fig:logMdisk_logMacc_TWA}
\end{figure}

In Fig.\,\ref{fig:logMdisk_logMacc_TWA} we compare the $M_{\rm disk}-\dot{M}_{\rm  acc}$ properties for TWA sources in our sample with the distributions reported for Lupus and Chamaeleon~I. 
For consistency purposes, the values adopted for the latter two are those converted to individual Gaia DR2 distances and to \citeauthor{baraffe2015}'s (\citeyear{baraffe2015}) models. For TWA stars, we adopted the estimates (or upper limits) of $M_{\rm disk,\,dust}$ computed by \citet{rodriguez2015} from ALMA observations, where available; the $M_{\rm disk,\,dust}$ were converted to $M_{\rm disk}$ assuming $M_{\rm disk}\sim100\,M_{\rm disk,\,dust}$, as in \citet{manara2016}. The $M_{\rm disk}$ shown for TWA\,1 is taken from \citet{trapman2017}. With the exception of TWA\,1, the TWA sources exhibit some evolution with respect to the populations of Lupus and Cha~I, being located in the quarter of the diagram at low $\dot{M}_{\rm acc}$ and low $M_{\rm disk}$. However, the vertical scatter among TWA members on the diagram is at least similar to that measured across the statistically larger Lupus and Cha~I samples, rendering the comparison between the observed TWA properties and the theoretical viscous predictions very uncertain. J1247-3816 is the only object of the TWA subsample that exhibits a position on the diagram compatible with the expected locus for disks aged $\sim$10~Myr that have attained a viscous quasi-steady state. TWA\,1 and TWA\,32 lie roughly in the same strip traced by Lupus and Cha~I members (i.e., above the viscous steady-state prediction at the nominal age of the association). TWA\,30A and TWA\,31 appear even above the assumed steady-state locus with a viscous timescale of 0.1~Myr, which crosses the upper envelope of the datapoints distributions pertaining to Lupus and Cha~I. TWA\,30A is, however, a nearly edge-on star-disk system, which may have introduced additional uncertainties on the measured parameters (see Sect.\,\ref{sec:Lacc_Lline}) and on the $M_{\rm disk}$ derivation. As to TWA\,31, its membership to the TWA has recently been disputed, as mentioned in Sect.\,\ref{sec:targets}, although the new classification would place this object at an older age than the TWA and, therefore, would render its position on Fig.\,\ref{fig:logMdisk_logMacc_TWA} even more inconsistent with the predicted locus for viscously evolved disks. 

The scarceness of objects in our sample with definite measurements for Fig.\,\ref{fig:logMdisk_logMacc_TWA} prevents us from drawing any robust conclusion from the comparison with Lupus and Cha~I. What the overall picture appears to hint at is that a purely viscous disk evolution scenario is insufficient to explain all of the observations. This tentative inference is aligned with the conclusions of \citet{ingleby2014}, who compared observations of older (5--10~Myr) stars with significant accretion to the expected decline in $\dot{M}_{\rm acc}$ with age from viscous evolution models, and showed that the latter cannot explain a prolonged accretion activity at the observed rates without assuming very high $M_{\rm disk}$ that would render the disk unstable. 

\subsection{Impact of stellar mass and environmental conditions on $\dot{M}_{\rm acc}$ evolution}

Statistical studies of disk fractions in young star clusters as a function of stellar mass and age have suggested longer evolutionary timescales for disks surrounding lower-mass stars than those around higher-mass stars. \citet{ribas2015} analyzed a large sample of stars from eleven young clusters and stellar associations of age between 1 and 10~Myr, to infer that lower-mass stars are statistically more likely to host a ``protoplanetary'' disk (i.e., a disk which produces an infrared excess at $\lambda < 10\,\,\mu$m) than more massive stars at any given age between 1--3 and 10~Myr. However, the broad mass bins used in that study to define ``low-mass stars'' ($M_\star < 2\,M_\odot$) and ``high-mass stars'' ($M_\star \geq 2\,M_\odot$) do not allow us to connect this picture with the slow $\dot{M}_{\rm acc}$ evolution we infer for $M_\star < 0.3\,M_\odot$ stars from the discussion in Sect.\,\ref{sec:Macc_evol_mass}. 

In their review on protoplanetary disk evolution, \citet{williams2011} pointed out that the disk dissipation timescales for substellar objects appear to be at least comparable to, if not longer than, those pertaining to solar-like objects (SpT of K to early-M), with the caveat that the observational statistics available in the lowest-mass regime is significantly poorer than at higher masses. Nevertheless, some evidence of a mass-dependent disk fraction has been reported also among stars less massive than $\sim$1~$M_\odot$ within a given cluster. \citet{carpenter2006} employed {\it Spitzer} observations from 4.5 to 16~$\mu$m to assess the disk fraction of young stars in the Upper Scorpius region ($\sim$8 Myr-old). They estimated a total disk fraction of 19\% among K0--M5 stars, in contrast to $\lesssim$1\% among massive stars in their sample. Their data also suggest an increasing disk fraction with later SpT among M-type stars themselves, as about 10\% of early M-stars ($<$ M3) were found to exhibit an infrared excess at 8 and 16 $\mu$m, while the percentage amounted to 20--30\% among M4--M5 stars. \citet{scholz2007} complemented that survey by investigating the disk fraction among young brown dwarfs in Upper Sco (SpT = M5--M9), using {\it Spitzer} data between 8 and 24~$\mu$m, and observed signatures of disks around $37\%\pm9\%$ of their targets. This value may be subject to a larger uncertainty than the estimates around higher-mass stars, perhaps reflecting completeness limits at the low-mass end of the cluster population, as suggested by the subsequent work of \citet{riaz2009}, who reported a disk fraction of $11\%^{+9\%}_{-3\%}$ from their own sample of brown dwarfs in Upper~Sco, and a combined disk fraction of $27\%^{+6\%}_{-5\%}$ from their and \citeauthor{scholz2007}'s (\citeyear{scholz2007}) studies. 

More recently, \citet{dawson2013} considerably expanded the census of very low-mass stars and brown dwarfs in Upper~Sco, and derived a disk fraction ($23\%\pm5\%$) consistent with that estimated by \citet{carpenter2006} for K0--M5 stars. On the other hand, \citet{luhman2012} examined multiple disk signatures at {\it Spitzer} and WISE wavelengths and reported a consistent increase in disk fraction with later spectral types, going from $\sim$10\% for K-type stars to $\sim$20\% in M0--M4 stars and $\sim$25\% in M4--M8 stars. A definite increase in disk fraction with decreasing stellar mass in Upper~Sco was also found by \citet{cook2017}, who focused in particular on the substellar mass regime and reported a disk frequency among $M_\star < 0.05\,M_\odot$ objects about 2--3 times higher than among $0.05-0.15\, M_\odot$ objects. {The criteria adopted in each of these recent studies to build a census of substellar mass and brown dwarf members of the cluster (photometric association with the cluster sequence on optical/near-IR color-magnitude diagrams, kinematic association with the cluster, spectral indicators of youth) are independent of the presence or absence of a disk, which was assessed later using separate mid-IR data. Therefore, no significant biases against selection of low-mass stars without disks are expected to affect their inferences.} Overall, these results provide a good statistical indication of a longer disk lifetime around very low-mass stars than around solar-mass objects. The slow evolution in $\dot{M}_{\rm acc}$ that we observe in Figs.~\ref{fig:XS_region_med_age_Macc} and \ref{fig:XS_regions_Macc_comparison} for $M_\star \lesssim 0.3\,M_\odot$ stars between 3 and 10~Myr may be connected to the longer disk lifespans, although \citet{scholz2007} found definite signatures of ongoing accretion only for around 30\% of their targets with clear indications of inner dust disks (note, however, that \citeauthor{scholz2007}'s sample includes objects with $M_\star < 0.1\,M_\odot$, whereas the mass range where our comparison is best performed from the dataset in Fig.\,\ref{fig:XS_regions_Macc_comparison} lies between $M_\star = 0.1$ and 0.3~$M_\odot$).  

Previous studies have also reported evidence for longer timescales of disk evolution in sparse stellar associations as opposed to compact or crowded clusters, where dynamical interactions with nearby objects may play a significant role. \citet{fang2013}, in particular, examined the disk fraction vs. age relation for 19 star-forming regions and young stellar associations (including Cha~I, $\sigma$~Ori, Upper~Sco, and the TWA) to derive the characteristic timescale $\tau_0$ for disk evolution. While the exact numbers are dependent on the specific age estimates for a given young stellar population (see discussion in \citealp{bell2013}), the $\tau_0$ that they estimated from their sample of sparse associations, $4.3\pm0.3$, resulted to be significantly longer than the $2.8\pm0.1$~Myr measured for young star clusters (see also \citealp{fedele2010}). The authors also inferred a tentative connection between the characteristic timescale of disk accretion and the sparseness of the stellar population, showing that at least the primary target of their investigation, the $\epsilon$\,Cha association, hosts a significantly larger fraction of accreting stars than expected based on the accretion timescales estimated by \citet{fedele2010} for their selection of clusters. More studies, combining the information on occurrence of disk-bearing and accreting objects as a function of age with accurate measurements of their accretion rates over a wide range of stellar masses and external conditions, are needed to fully elucidate the dynamics and drivers of accretion evolution in protoplanetary disks.

\section{Conclusions} \label{sec:conclusions}

We have conducted a thorough spectroscopic study of disk accretion in the $\sim$5--10~Myr-old TWA, employing homogeneous data from the X-shooter spectrograph, and making use of well-established methods of reduction and analysis developed earlier during the Italian X-shooter GTO program on star-forming regions \citep{alcala2011}. Our survey encompasses eleven stars with IR signatures of circumstellar material in the TWA, plus two disk-bearing stars recently disputed as TWA members and suggested to belong to nearby PMS associations, and one TWA member with no evidence of dusty inner disk reported in the literature. Our targets account for $\sim$85\% of the estimated population of young stars with disks in the TWA, they range in SpT between M0 and M9, and have masses between 0.58 and 0.02~$M_\odot$. We took advantage of the broad spectral window covered by X-shooter (from $\sim$300 to 2500~nm) to determine individual extinction, stellar parameters, and accretion parameters for each target simultaneously via multi-component fits to several features of the observed spectra. We then examined the shapes and relative strengths of the detected Balmer emission lines, and the detection rate and structure of the forbidden [O\,I]\,6300\AA\mbox{} emission line, to probe the physical conditions of the accretion/ejection mechanisms at play. 

The TWA offers a unique perspective into the final stages of the disk lifetimes around stars from subsolar masses to the brown dwarf regime, often inaccessible in more distant regions. We detected {signatures of ongoing} accretion activity above the chromospheric noise level for around 70\% of our TWA targets (disk-bearing stars), which translates to a fraction of accretors of 13\% to 17\% across the whole TWA population (including both disk-bearing and disk-free objects). This fraction is significantly smaller than the frequencies of accreting stars typically detected in very young (<\,3~Myr-old) PMS populations ($\sim$40--50\%), but larger than those measured for several $>$\,5~Myr-old clusters in earlier studies ($<$5--10\%; see, e.g., \citealp{fedele2010} and references therein, and \citealp{frasca2015}). Our analysis of the line emission detected on our targets indicates that some evolution has taken place with respect to younger star-forming regions investigated previously in a similar manner, most notably the 1--3~Myr-old Lupus:
\begin{enumerate}[(i)]
\item while the morphological classification of H$\alpha$ profiles in the TWA is statistically similar to the distribution in profile types \citep{reipurth1996} found in younger regions (e.g., Lupus, NGC~2264), around 50\% of our targets exhibit H$\alpha$ lines with an overall Gaussian profile, and 83\% exhibit narrow H$\alpha$ lines (FWHM $\lesssim$ 100~km/s) vs. the 44\% narrow-line profiles found by \citet{antoniucci2017} in Lupus;
\item most of our targets (83\%) exhibit Balmer decrements consistent with the most common class identified by \citet{antoniucci2017} in Lupus (type~2, corresponding to moderate accretion regimes with optically thin emission), but this rate of occurrence in the TWA is larger than the one found in Lupus (56\%);
\item forbidden [O\,I]6300\AA\mbox{} emission indicative of disk winds and jets is observed only in 31\% of our targets (significantly lower than the detection rate of 77\% reported by \citealp{nisini2018} in a composite sample from Lupus, Cha~I and $\sigma$~Ori, and the 59\% reported by \citealp{mcginnis2018} in NGC~2264);
\item among our stars with [O\,I]6300\AA\mbox{} emission, only 50\% exhibit the high-velocity component believed to be associated with extended jets (which corresponds to a total detection rate of 15\% in our sample of disk-bearing stars, smaller than the 30\% detected by \citealp{nisini2018}, although comparable to the 17\% detected by \citealp{mcginnis2018}).
\end{enumerate}
Interestingly, though, the distribution in accretion rates that we measured for our targets is similar to that observed in younger regions observed with X-shooter (Lupus, Cha~I, $\sigma$~Ori) at the corresponding stellar mass. The different coverage in $M_\star$ achieved in the various surveys does not allow us to extract any inferences for the $M_\star \gtrsim 0.3\,M_\odot$ regime (where we have only one object, which falls in $\dot{M}_{\rm acc}$ close to the median level measured for Lupus and Cha~I), nor for the $M_\star \lesssim 0.1\,M_\odot$ regime (where there is no substantial representation from the other surveys). However, for the mass range of overlap between the various surveys ($M_\star \simeq 0.1-0.3\,M_\odot$), our results indicate that stars which still retain a disk at the age of the TWA exhibit statistically comparable accretion levels to those measured at ages $\leq 3-5$~Myr. The larger disk fraction observed in the TWA by comparison with other similarly aged star clusters (where the census is often incomplete for M-stars) is consistent with the higher disk fractions typically found among the lowest-mass stars within individual clusters at any age. Together, these results suggest longer evolutionary timescales for accretion disks around very low-mass stars than assumed for their higher-mass counterparts. 

As far as we can tell from our limited statistics, it appears that viscous evolution alone may not be adequate to explain the range in disk and accretion properties observed, and the slow $\dot{M}_{\rm acc}$ evolution on low-mass objects. More observational data of accretion in young, sub-solar mass stars of different ages are needed to properly address the time dependence of $\dot{M}_{\rm acc}$ and establish the intrinsic evolutionary timescales in different mass regimes. In this respect, new insights will soon be provided by the latest X-shooter survey of the $\sim$8~Myr-old Upper Scorpius region (Manara et al., in preparation), whose results appear to be in line with the picture discussed here. 

\begin{acknowledgements}
{We wish to thank the anonymous referee for their careful manuscript reading.} Work of L.V. was supported by the Institutional Strategy of the University of T\"ubingen (Deutsche Forschungsgemeinschaft, ZUK 63). L.V. also acknowledges financial support during her stay at Cornell University Department of Astronomy as a Visiting Scholar in Fall 2018. J.M.A., S.A., and B.N. acknowledge financial support from the project PRIN-INAF 2016 ``The Cradle of Life--GENESIS-SKA'' (General Conditions in Early Planetary Systems for the rise of life with SKA). C.F.M. acknowledges an ESO fellowship. Work of C.F.M. has received funding from the European Union's Horizon 2020 research and innovation program under the Marie Sklodowska-Curie grant agreement No 823823 (DUSTBUSTERS), and was partly supported by the Deutsche Forschungsgemeinschaft (DFG, German Research Foundation) - Ref no. FOR 2634/1 TE 1024/1-1. S.A. acknowledges the support by INAF/Frontiera through the ``Progetti Premiali'' funding scheme of the Italian Ministry of Education, University, and Research.
\end{acknowledgements}

\bibliographystyle{aa}
\bibliography{references}

\begin{appendix}

\section{TWA\,3B} \label{app:TWA3B}

TWA\,3B is a visual companion to the spectroscopic binary TWA\,3A \citep{malo2014}. Previous study have reported weaker H$\alpha$ emission towards TWA\,3B than towards the primary \citep[e.g.,][]{webb1999, jayawardhana2006}, and no evidence of IR excess emission around TWA\,3B down to wavelengths $\sim$10~$\mu$m \citep{jayawardhana1999b}, which suggests the absence of dust within a distance of a few AU from the central star.

We reduced and analyzed the X-shooter spectrum of TWA\,3B in a similar fashion as the other targets in Table~\ref{tab:TWA_targets}, assuming the same distance as for TWA\,3A (as indicated from Gaia DR2), and extracted the following parameters:
\begin{itemize}
\item SpT = M4.5;
\item $A_V = 0.0\,\,mag$;
\item $\log{L_\star}\,[L_\odot] = -1.019$;
\item $T_{\rm eff} = 3170\,K\pm40\,K$;
\item $\log{g} = 4.40\pm0.11$;
\item RV = $8.8\,\,km/s\pm1.0\,\,km/s$;
\item $v\,\sin{i} = 11\,\,km/s\pm10\,\,km/s$;
\item $M_\star = 0.15\,M_\odot\pm0.03\,M_\odot$;
\item $R_\star = 1.08\,R_\odot$.
\end{itemize}
Surprisingly, our best fit model yields $\log{L_{\rm acc}^{slab}}\,[L_\odot] = -3.02$, which corresponds to $\log{\dot{M}_{\rm acc}^{slab}}\,[M_\odot/yr] = -9.57$. {This value is, however, to be considered an upper limit to the true $L_{\rm acc}^{slab}$ according to the definition introduced in Sect.\,\ref{sec:multicomponent_fit}.} This {upper limit estimate} departs significantly from the results of \citet{herczeg2009}, who reported $\log{\dot{M}_{\rm acc}}\,[M_\odot/yr] < -11.1$ for TWA\,3B using Keck~I/LRIS spectra ($\lambda$$\sim$3200--9000\,\AA), although $\sim$0.5~dex of the difference between the two values can be explained by the larger $M_\star$ and smaller $R_\star$ adopted by \citet{herczeg2009} with respect to our estimates. Our accretion parameters determined from the emission lines are instead $\log{\widetilde{L}_{\rm acc}^{lines}}\,[L_\odot] = -3.92$ and $\log{<\dot{M}_{\rm acc}^{lines}>}\,[M_\odot/yr] = -10.47$, closer to the upper limit measured by \citet{herczeg2009}. TWA\,3B exhibits one of the narrowest H$\alpha$ line profiles in our sample, with an overall shape that can be reproduced satisfactorily with a Gaussian curve, consistent with a substantially non-accreting object.

\section{$T_{\rm eff}$, $\log{g}$, RV, and $v\sin{i}$ measurements for TWA stars} \label{sec:rv_vsini_plots}

\begin{figure}
\resizebox{\hsize}{!}{\includegraphics{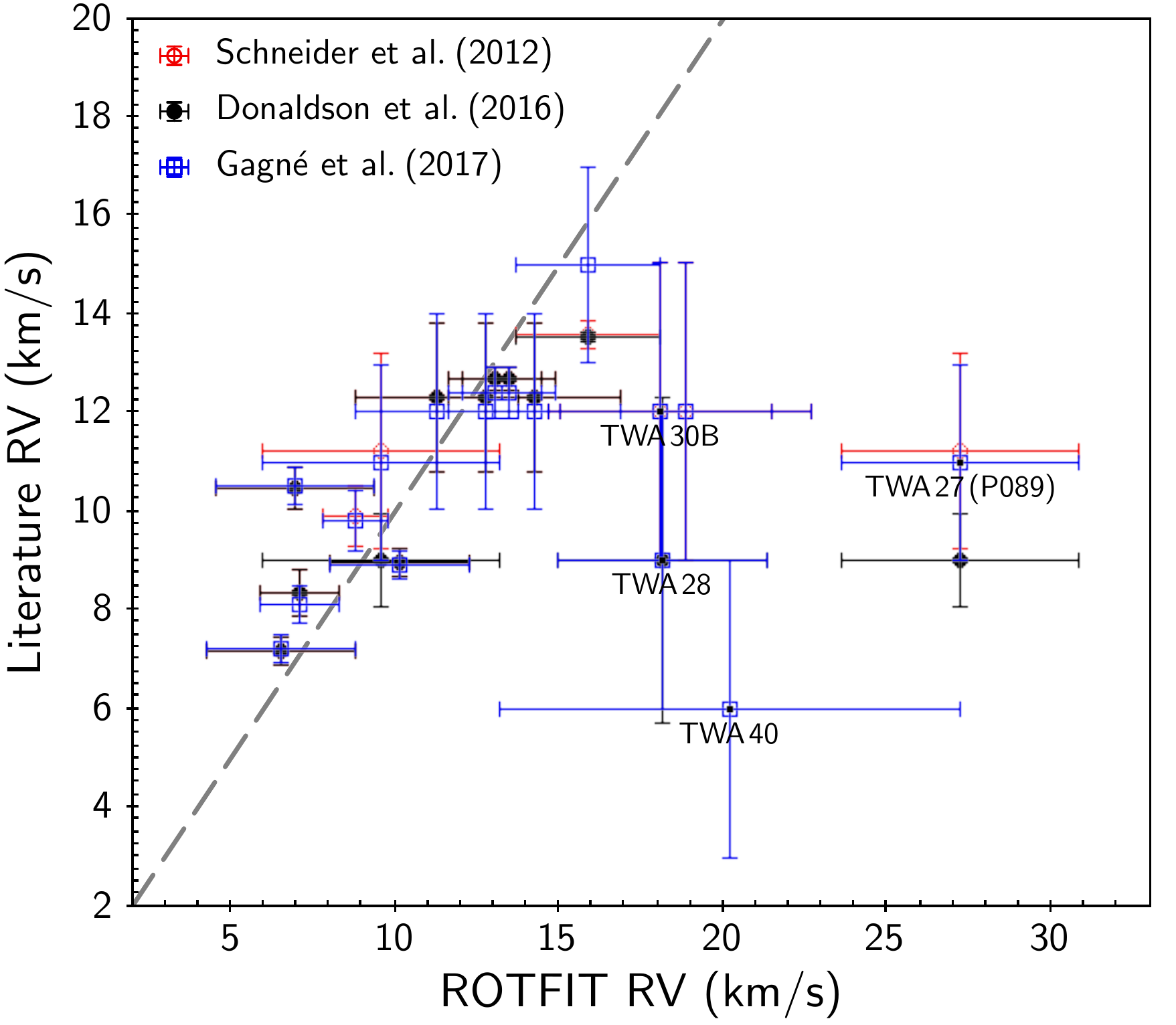}}
\caption{Comparison between our ROTFIT estimates of RV and the values reported by \citet[][red symbols]{schneider2012}, \citet[][black symbols]{donaldson2016}, and \citet[][blue symbols]{gagne2017}, with the respective uncertainties. The equality line is traced in gray to guide the eye.}
\label{fig:TWA_RV_comp}
\end{figure}

In Fig.\,\ref{fig:TWA_RV_comp} we compare our measurements of RV with those reported by \citet{schneider2012}, \citet{donaldson2016}, and \citet{gagne2017} (some objects repeated in their respective literature compilations). We excluded TWA\,3A from the comparison as the values of RV reported in the listed studies are systemic values and not individual estimates for the two components of the spectroscopic binary. The diagram shows overall agreement between our measurements and those reported in the literature for most stars. Significant discrepancies in the estimated RV can be observed for a few objects (see labels in Fig.\,\ref{fig:TWA_RV_comp}), located well below the equality line in Fig.\,\ref{fig:TWA_RV_comp}. These objects tend to exhibit more uncertain RV estimates, both from this study and from the literature, with respect to the other stars of the sample, as indicated by the large error bars on the diagram. We also note that the RV estimate derived from the other spectrum of TWA\,27 (P084) falls well within the point dispersion around the identity line on the diagram. 

\begin{figure}
\resizebox{\hsize}{!}{\includegraphics{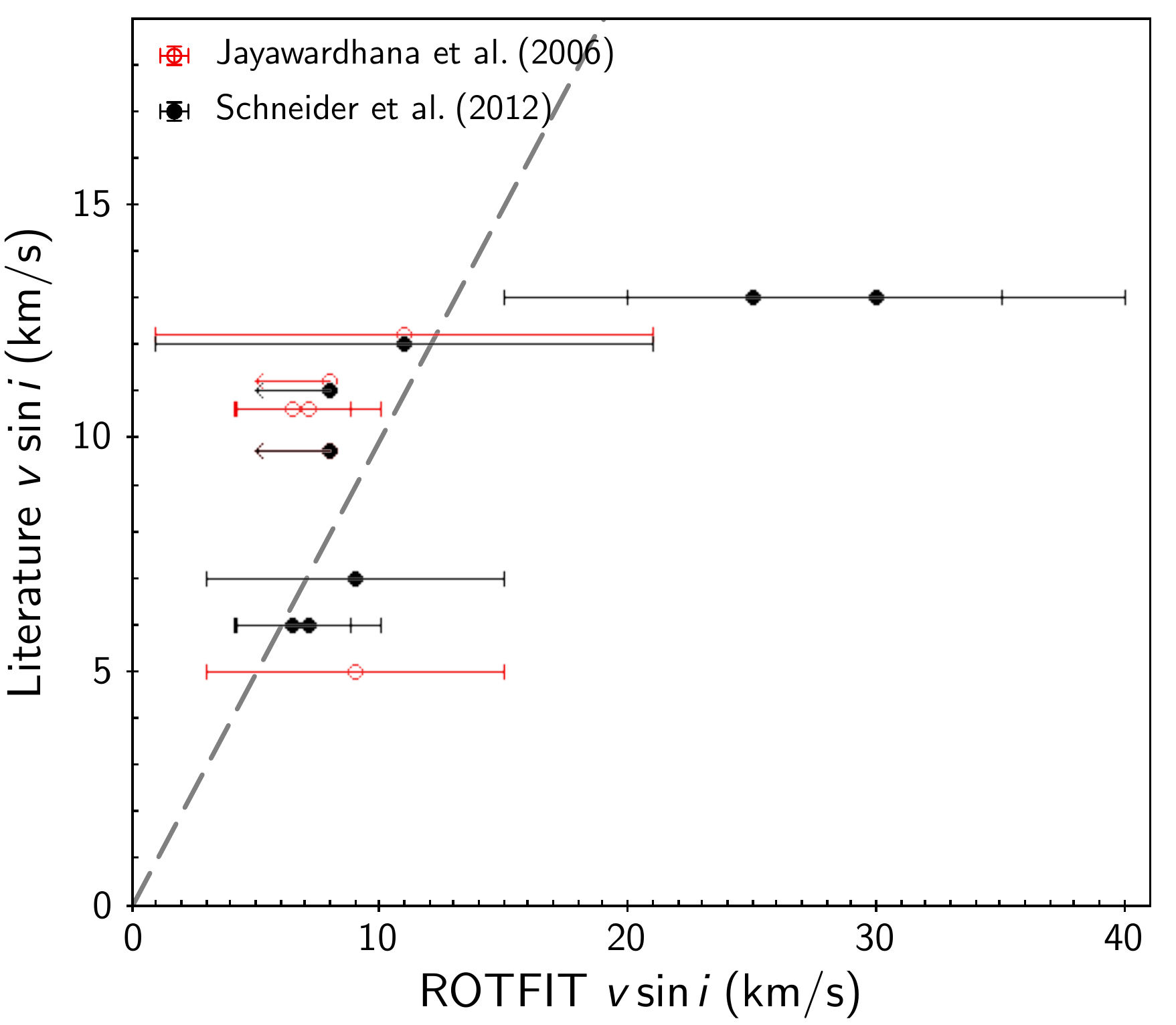}}
\caption{Comparison between our ROTFIT estimates of $v\,\sin{i}$ and the values reported by \citet[][red symbols]{jayawardhana2006} and \citet[][black symbols]{schneider2012} for objects in common. Upper limits are marked by arrows. The equality line is traced in gray to guide the eye.}
\label{fig:TWA_vsini_comp}
\end{figure}

Some objects in our list have $v \sin{i}$ measurements in \citet{jayawardhana2006} and \citet{schneider2012}. As illustrated in Fig.\,\ref{fig:TWA_vsini_comp} (which excludes TWA\,3A for the same reason mentioned above), the three sets of measurements are overall consistent within the estimated uncertainties.

\begin{figure}
\resizebox{\hsize}{!}{\includegraphics{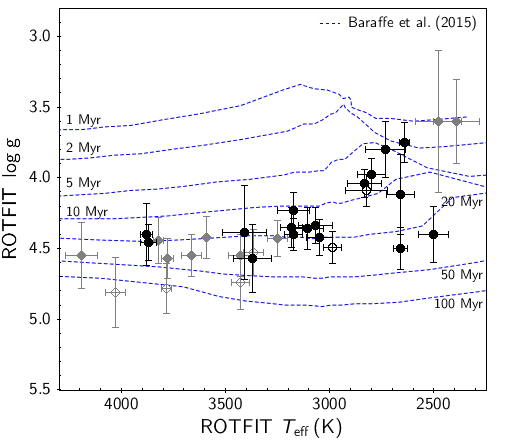}}
\caption{Measurements of $\log{g}$ for TWA stars with disks (this study, black dots) and without disk \citep[][gray diamonds]{stelzer2013}, as a function of their $T_{\rm eff}$, both derived with ROTFIT. Empty symbols identify objects with disputed membership to the TWA in \citet{gagne2017}. PMS isochrones from \citeauthor{baraffe2015}'s (\citeyear{baraffe2015}) model grid are overlaid as blue dotted lines for comparison purposes.}
\label{fig:TWA_Teff_logg_ROTFIT}
\end{figure}

In Fig.\,\ref{fig:TWA_Teff_logg_ROTFIT}, $\log{g}$ measurements for TWA objects are shown as a function of their $T_{\rm eff}$. {Disk-free TWA objects from \citet{stelzer2013} are also shown for comparison purposes. The spectra of the latter have been re-analyzed using the latest version of the ROTFIT code, in order to obtain a homogeneous set of atmospheric parameters for disk-bearing and disk-free stars in the association. The $\log{g}$ values adopted here for stars without disk are, therefore, slightly different than those reported in \citet{stelzer2013} and with smaller associated uncertainties. A list of the newly derived $\log{g}$ parameters for TWA targets in \citet{stelzer2013} is reported in Table~\ref{tab:new_logg_III}}. As discussed in \citet{stelzer2013}, substellar mass objects appear younger on the diagram than earlier-type M-stars stars, although they exhibit a larger vertical scatter and, in some cases, large uncertainties. {Objects also appear systematically older than on the HR diagram in Fig.\,\ref{fig:TWA_HR_models}, with the bulk of stars at $T_{\rm eff}$$\sim$3000--4000~K being distributed between the 10~Myr and 20~Myr isochrones according to their $\log{g}$. This effect, already observed, for instance, in the X-shooter survey of Lupus (see Figs.\,4 and 6 of \citealp{frasca2017}), is possibly due to issues in the $\log{g}$ vs. age calibration of PMS evolutionary models (see also \citealp{herczeg2015}).}

\begin{table}
\caption{New estimates of the $\log{g}$ parameter for disk-free TWA objects investigated in \citet{stelzer2013}, except TWA\,26 and TWA\,29.}
\label{tab:new_logg_III}
\centering
\begin{tabular}{l c c}
\hline\hline
{Star} & {$\log{g}$} & {err} \\
\hline
{TWA\,2A} & {4.42} & {0.15} \\
{TWA\,6} & {4.8} & {0.3} \\
{TWA\,7} & {4.55} & {0.15} \\
{TWA\,9A} & {4.6} & {0.2} \\
{TWA\,9B} & {4.43} & {0.13} \\
{TWA\,13A} & {4.55} & {0.15} \\
{TWA\,13B} & {4.57} & {0.14} \\
{TWA\,14} & {4.78} & {0.18} \\
{TWA\,15A} & {4.53} & {0.16} \\
{TWA\,15B} & {4.74} & {0.19} \\
{TWA\,25} & {4.44} & {0.16} \\
\hline
\end{tabular}
\end{table}

\section{Classification of H$\alpha$ line profiles across the TWA sample} \label{sec:Halpha_analysis}

The classification of H$\alpha$ line profiles developed by \citet{reipurth1996} comprises four morphologically distinct main types: (i) Type~I, symmetric; (ii) Type~II, double-peaked, with the secondary peak being at least half as strong as the primary peak; (iii) Type~III, double-peaked, with the secondary peak being less than half the strength of the primary peak; (iv) Type~IV, which exhibits an absorption feature at the base of the emission line, with no secondary peak. Each of the types II, III, and IV are further distinguished into two subclasses, depending on the location of the secondary peak or absorption feature (bluewards or redwards of the primary peak). This variety of line profiles reflects contributions from matter inflows and outflows from the inner disk regions, filtered through the geometric configuration of the star-disk system with respect to the line of sight to the observer \citep[e.g.,][]{lima2010, kurosawa2011, tambovtseva2014}.

\begin{figure*}
\centering
\includegraphics[width=\textwidth]{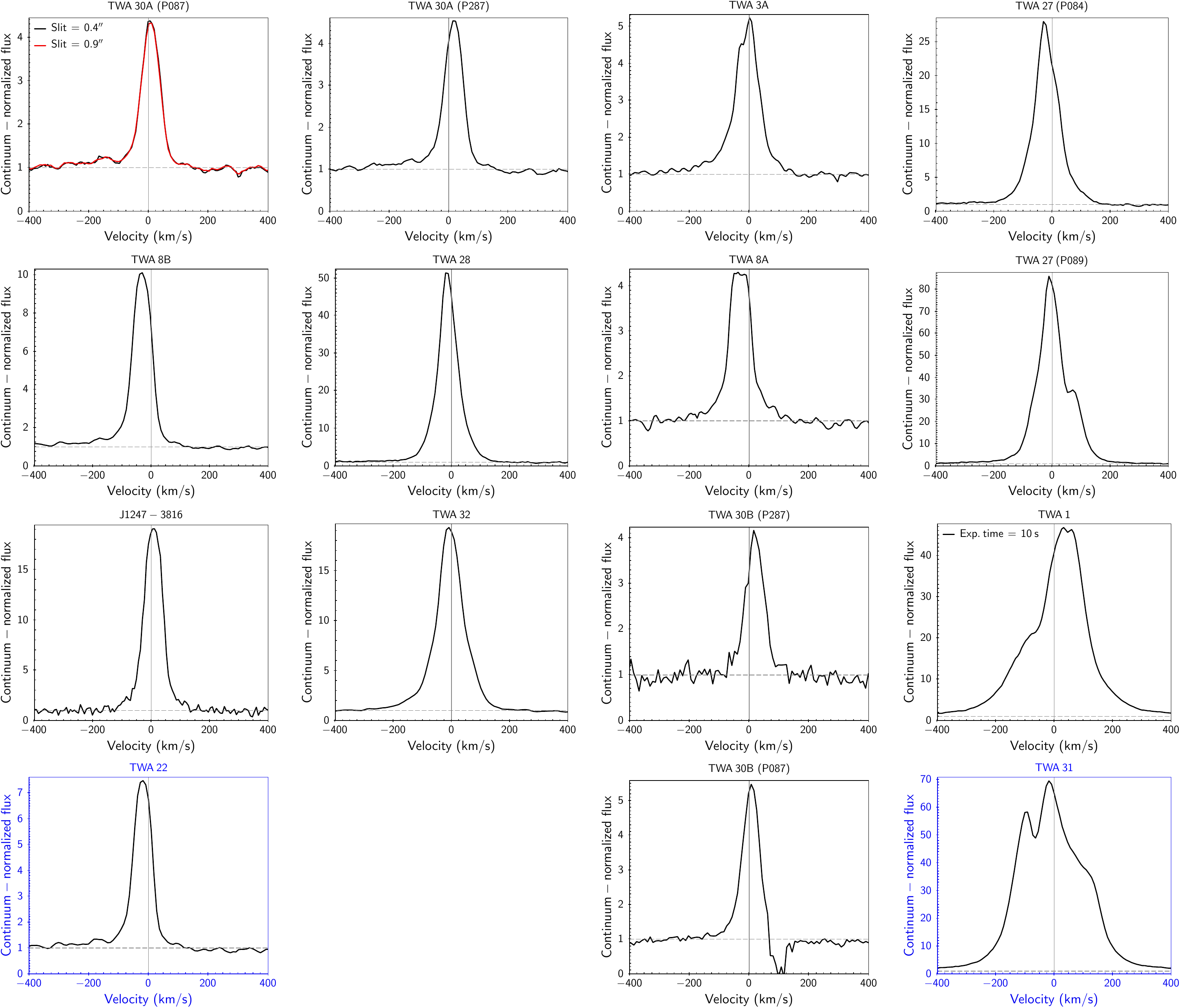}
\caption{H$\alpha$ line profiles for all objects in Table~\ref{tab:TWA_targets}, except TWA\,3B (which does not exhibit signatures of a disk up to 10~$\mu$m) and TWA\,40 (for which no line emission was detected). The profiles are sorted according to their qualitative classification following \citeauthor{reipurth1996}'s (\citeyear{reipurth1996}) scheme: Type~I (first and second columns from the left), Type~III (first three panels from the top in the third column from the left), Type~II/III (last column from the left), Type~IV (last panel from the top in the third column). Blue frames are adopted for the H$\alpha$ profiles of disputed members TWA\,22 and TWA\,31. Only the H$\alpha$ profile of the spectrum obtained with lower integration time is shown for TWA\,1, because the other spectrum is likely saturated in H$\alpha$. When multiple spectra were acquired for the same star at different observing epochs (TWA\,27, TWA\,30A, TWA\,30B), the corresponding H$\alpha$ profiles are illustrated in separate panels.}
\label{fig:TWA_Ha_atlas}
\end{figure*}

In Fig.\,\ref{fig:TWA_Ha_atlas} we illustrate all H$\alpha$ line profiles observed for stars in our sample. To test the overall degree of symmetry of the line profiles, we computed how well each of them can be described as a Gaussian\footnote{The computation of the best-fit parameters was performed using the function {\it nls()} in the R environment.} of equation
\begin{equation} \label{eqn:gaussian}
y = a + k \cdot \exp{\left[-\frac{(x - \mu)^2}{2 \sigma^2}\right]}\,,
\end{equation}
where initial guesses for the free parameters $a$, $k$, $\mu$, and $\sigma$ were defined as the normalized continuum floor ($\equiv$1), the normalized peak height, the velocity of the peak, and half the measured full width at half maximum (FWHM), respectively. Strictly speaking, none of the line profiles shown in Fig.\,\ref{fig:TWA_Ha_atlas} (with the exception of the noisy profile associated with TWA\,30B\,[P287]) can be described satisfactorily\footnote{$p$-value $\leq$ 0.05.} in terms of a Gaussian curve, according to a $\chi^2$ test. However, the best Gaussian fit is sufficient to reproduce the overall H$\alpha$ profile for objects TWA\,3A, TWA\,8A, TWA\,8B, TWA\,22, TWA\,30A, and J1247-3816, with point-by-point differences between the observed profile and the fitting curve that are consistent with the $\pm$3\,$\sigma$ noise around the local continuum in the spectrum. These objects exhibit relatively narrow H$\alpha$ line profiles (FWHM\,$\sim$\,70--85~km/s). TWA\,8B, TWA\,22, TWA\,30A, and J1247-3816 exhibit a single-peak profile with no absorption features, hence falling into the Type~I definition of \citet{reipurth1996}. TWA\,28 and TWA\,32 also exhibit a Type~I profile, although diverging from a Gaussian description in the broad wings at velocities $\sim 100$~km/s. TWA\,3A, TWA\,8A, and TWA\,30B~[P287], instead, exhibit a structured peak, with a secondary blueshifted peak appearing above the mean intensity level in the first and the fourth cases; these objects therefore classify as Types~II in \citeauthor{reipurth1996}'s scheme. Among the remaining objects, TWA\,27 exhibits a secondary redshifted peak or hump, oscillating between Type~II and Type~III; TWA\,1 and TWA\,31 exhibit very wide and multi-structured profiles, also oscillating between Type~II and Type~III; TWA\,30B\,[P087] exhibits an inverse P~Cygni profile, therefore qualifying as Type~IV. 

\section{[O\,I\,]\,6300\AA\mbox{} emission measured in the TWA}

Table~\ref{tab:forbidden_line_emission} reports the detected [O\,I]\,6300\AA\mbox{} line luminosities for TWA\,1, TWA\,30A, TWA\,30B, and TWA\,31, decomposed into LVC and HVC. The associated uncertainties account for the continuum noise around the line and for the uncertainties on the line profile fit parameters.

\begin{table}
\caption{[O\,I]\,6300\AA\mbox{} line luminosities measured across our sample, decomposed into LVC and HVC, when detected.}
\label{tab:forbidden_line_emission}
\centering
\begin{tabular}{l c c}
\hline\hline
Star & $L_{LVC}\, \pm$ err & $L_{HVC}\, \pm$ err \\
 & $[\times 10^{-7} L_\odot]$ & $[\times 10^{-7} L_\odot]$ \\
\hline
{\small TWA\,1\,{\it [10~s]}} & ${107\pm7}$ & -- \\
{\small TWA\,1\,{\it [60~s]}} & ${99\pm7}$ & -- \\
{\small TWA\,30A\,{\it [P087,\,ns]}} & ${11.9\pm0.9}$ & ${2.4\pm0.8}$ \\
{\small TWA\,30A\,{\it [P087,\,ls]}} & ${10.0\pm0.5}$ & ${2.8\pm0.5}$ \\
{\small TWA\,30A\,{\it [P287]}} & ${1.18\pm0.07}$ & -- \\
{\small TWA\,30B\,{\it [P087]}} & ${0.515\pm0.015}$ & ${0.0629\pm0.0005}$ \\
{\small TWA\,30B\,{\it [P287]}} & ${1.57\pm0.11}$ & ${0.187\pm0.019}$ \\
{\small TWA\,31} & ${0.21\pm0.04}$ & -- \\
\hline
\end{tabular}
\end{table}

\begin{figure*}
\centering
\includegraphics[width=\textwidth]{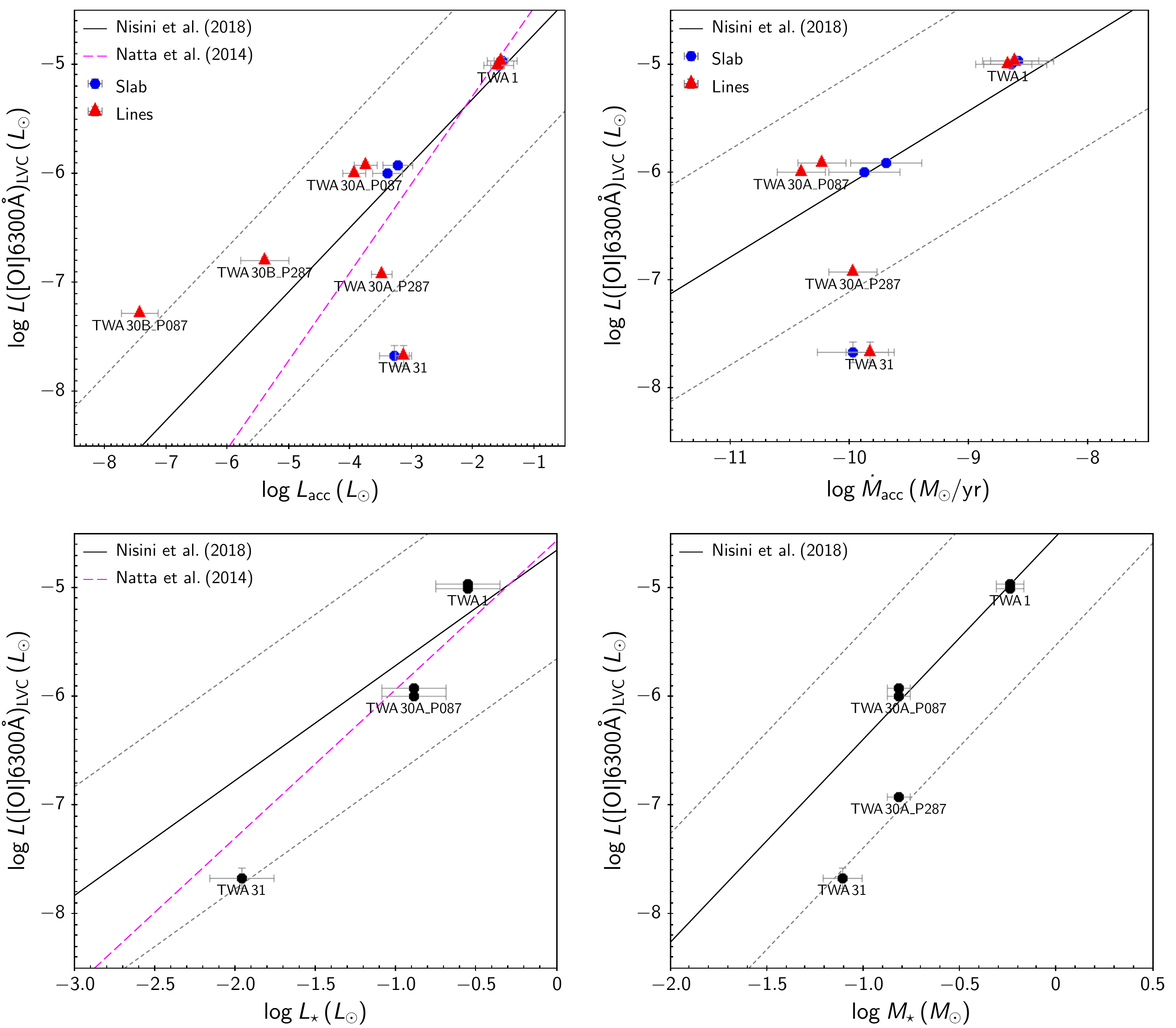}
\caption{Logarithmic luminosity of the [O\,I]\,6300\,\AA\mbox{} LVC as a function of $\log{L_{\rm acc}}$ (upper left panel), $\log{\dot{M}_{\rm acc}}$ (upper right panel), $\log{L_\star}$ (lower left panel), and $\log{M_\star}$ (lower right panel) for stars with detected forbidden line emission in our sample (labeled on each plot). In the upper panels, accretion parameters derived from the Balmer continuum diagnostics (see Sect.\,\ref{sec:multicomponent_fit}) are marked as blue dots, while those derived from the luminosity of emission lines (see Sect.\,\ref{sec:Lacc_Lline}) are marked as red triangles. On each panel, the correlation relationship between the relevant quantities derived by \citet{nisini2018} is traced as a solid black line, while dotted gray lines mark a range of $\pm$1~dex around the calibration (which corresponds to the vertical spread of points observed on the corresponding diagrams in \citealp{nisini2018}). The correlation trends obtained by \citet{natta2014} are also shown as dashed magenta lines for comparison (left panels).}
\label{fig:TWA_OI6300_param}
\end{figure*}

In Fig.\,\ref{fig:TWA_OI6300_param} we illustrate the luminosity measured for the [O\,I]\,6300\AA\mbox{} LVC in our sample (TWA\,1, TWA\,30A, TWA\,30B, and TWA\,31) as a function of their stellar ($L_\star$, $M_\star$) and accretion ($L_{\rm acc}$, $\dot{M}_{\rm acc}$) parameters. Although the paucity of our sample hampers any detailed considerations with respect to the correlation trends extracted from younger YSO populations, we note that the distribution of points derived here on each panel in Fig.\,\ref{fig:TWA_OI6300_param} is globally consistent with earlier results \citep{nisini2018, natta2014}, within the associated dispersions. TWA\,30B appears to be an outlier, {at least in one epoch}, with more intense [O\,I]\,6300\AA\mbox{} LVC than expected, in analogy with \citeauthor{nisini2018}'s (\citeyear{nisini2018}) results, based on the estimated $L_{\rm acc}$. This is likely due to a severe underestimate of the $L_{\rm acc}$ of the object, caused by its edge-on configuration and consequent partial concealing of its accretion features. Indeed, as noted earlier, the observed $L_\star$ would place this source well below the main sequence turnoff on the HR diagram in Fig.\,\ref{fig:TWA_HR_models}. \citet{kastner2016} reported a $\log{L_\star}\,[L_\odot] = -1.28$ for TWA\,30B, which indicates that our measurement in Table~\ref{tab:TWA_parameters} may be underestimated by three orders of magnitude. The strongest accretors in our sample exhibit $L_{\rm acc} \sim 0.1 \times L_\star$, while the bulk of the objects are distributed between $L_{\rm acc} \sim 0.01 \times L_\star$ and $L_{\rm acc} \sim 0.001 \times L_\star$. Therefore, the actual $\log{L_{\rm acc}}\,[L_\odot]$ on TWA\,30B may be of the order of $-3.5$, which would project the star within the stripe around the $\log{L(\mbox{[O\,I]\,6300\AA})_{LVC}}-\log{L_{\rm acc}}$ prescriptions in Fig.\,\ref{fig:TWA_OI6300_param}. Conversely, the $L(\mbox{[O\,I]\,6300\AA})_{LVC}$ measured for TWA\,31 is lower than expected based on the measured $L_{\rm acc}$. We note that the detection of [O\,I]\,6300\AA\mbox{} emission for this latter object is somewhat more uncertain than for the other objects, due to a low peak-to-continuum contrast and a noisy line profile, as shown in Fig.\,\ref{fig:TWA_OI630}.

\section{Tables of emission line luminosities} \label{sec:Lline_data}

\begin{figure*}
\centering
\includegraphics[width=0.5\textwidth]{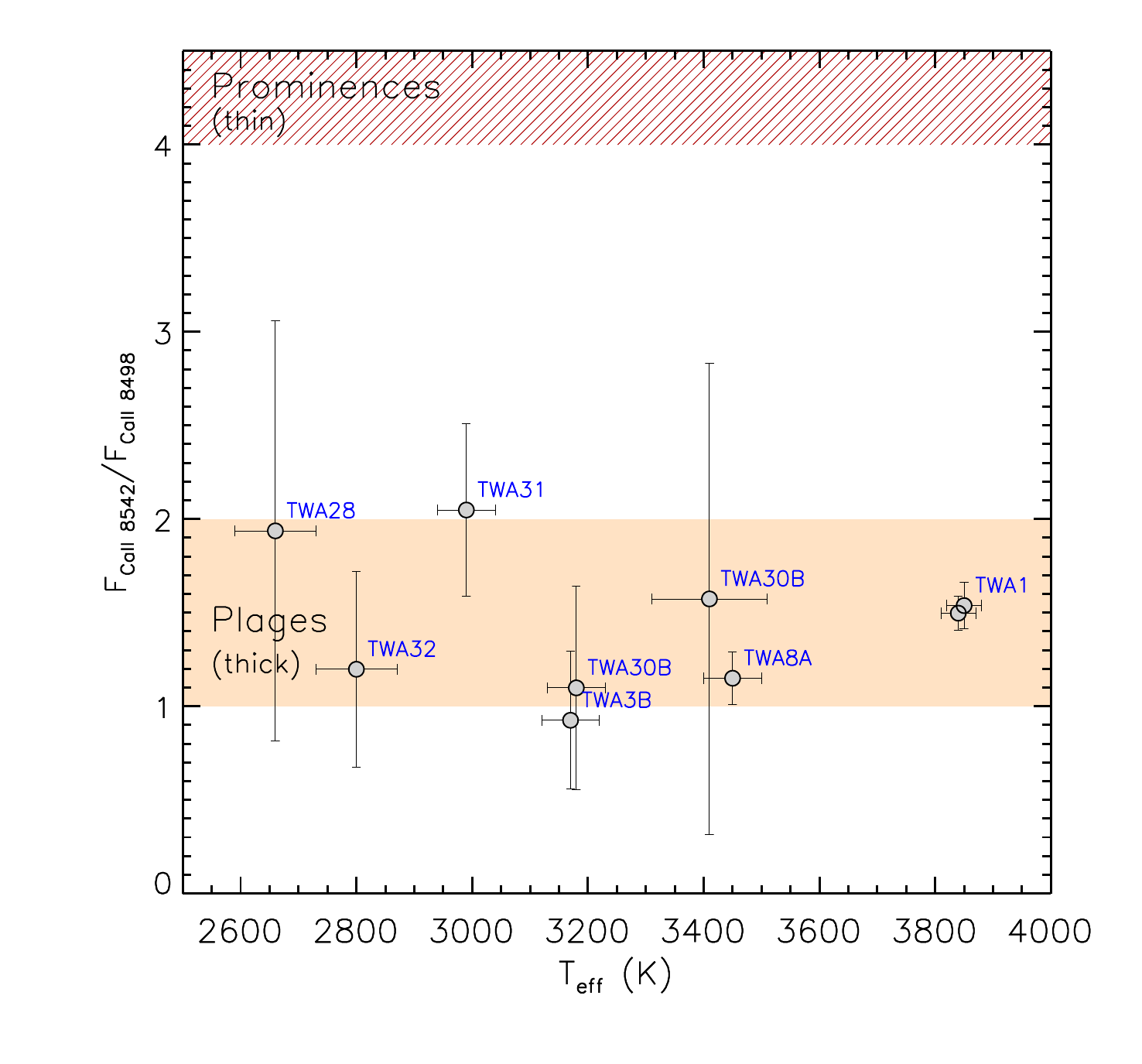}
\caption{$F_{\rm Ca\,II\,8542}/F_{\rm Ca\,II\,8498}$ ratio measured for TWA targets with detected Ca\,II IRT emission, as a function of stellar $T_{\rm eff}$. The range of values typical for solar plages and prominences are marked as a filled strip and a cross-hatched strip, respectively, for comparison purposes.}
\label{fig:fratio_CaIRT_xsho_teff}
\end{figure*}

Table~\ref{tab:Balmer_series} lists the measured luminosities of the Balmer series lines from H$\alpha$ to H15, with relative uncertainties, which were estimated from the rms scatter of the continuum around the position of the line in the spectrum. Table~\ref{tab:other_em_series} lists the luminosities of additional lines that were used to estimate $\widetilde{L}_{\rm acc}^{lines}$, when detected. Fig.\,\ref{fig:fratio_CaIRT_xsho_teff} illustrates the $F_{\rm Ca\,II\,8542}/F_{\rm Ca\,II\,8498}$ line ratio as a function of $T_{\rm eff}$ for our targets.

\begin{sidewaystable*}
\caption{Luminosities of the Balmer series lines between H$\alpha$ and H15, when detected.}
\label{tab:Balmer_series}
\centering
\resizebox{\textwidth}{!} {
\begin{tabular}{l c c c c c c c c c c c c c}
\hline\hline
Star & $L_{H\alpha}\, \pm$ err & $L_{H\beta}\, \pm$ err & $L_{H\gamma}\, \pm$ err & $L_{H\delta}\, \pm$ err & $L_{H\epsilon}\, \pm$ err & $L_{H8}\, \pm$ err & $L_{H9}\, \pm$ err & $L_{H10}\, \pm$ err & $L_{H11}\, \pm$ err & $L_{H12}\, \pm$ err & $L_{H13}\, \pm$ err & $L_{H14}\, \pm$ err & $L_{H15}\, \pm$ err \\
 & $[\times 10^{-7} L_\odot]$ & $[\times 10^{-7} L_\odot]$ & $[\times 10^{-7} L_\odot]$ & $[\times 10^{-7} L_\odot]$ & $[\times 10^{-7} L_\odot]$ & $[\times 10^{-7} L_\odot]$ & $[\times 10^{-7} L_\odot]$ & $[\times 10^{-7} L_\odot]$ & $[\times 10^{-7} L_\odot]$ & $[\times 10^{-7} L_\odot]$ & $[\times 10^{-7} L_\odot]$ & $[\times 10^{-7} L_\odot]$ & $[\times 10^{-7} L_\odot]$ \\
\hline
{\small TWA\,1\,{\it [10~s]}} & $42600\pm1900$ & $4300\pm200$ & $2000\pm300$ & $1500\pm200$ & $740\pm170$ & $1010\pm160$ & $750\pm90$ & $560\pm70$ & $430\pm130$ & $440\pm130$ & $340\pm110$ & $260\pm30$ & $190\pm30$ \\
{\small TWA\,1\,{\it [60~s]}} & $30400\pm1200$ & $4020\pm190$ & $2000\pm300$ & $1370\pm150$ & $710\pm190$ & $920\pm50$ & $690\pm110$ & $530\pm60$ & $550\pm190$ & $370\pm140$ & $290\pm130$ & $210\pm70$ & $180\pm50$ \\
{\small TWA\,3A} & $250\pm50$ & $56\pm8$ & $36\pm12$ & $28\pm9$ & $16.4\pm1.8$ & $19\pm4$ & $15\pm4$ & $10\pm3$ & $9\pm3$ & $6\pm4$ & $3\pm2$ & $2.6\pm1.6$ & $3\pm3$ \\
\small{TWA\,8A} & $450\pm120$ & $180\pm18$ & $101\pm19$ & $72\pm11$ & $50\pm6$ & $48\pm9$ & $31\pm7$ & $25\pm5$ & $22\pm6$ & $14\pm3$ & $11\pm3$ & $9\pm3$ & $14\pm7$ \\
{\small TWA\,8B} & $75\pm12$ & $15.2\pm0.7$ & $8.0\pm1.1$ & $5.2\pm1.1$ & $3.8\pm1.6$ & $3.3\pm1.2$ & $3\pm2$ & $1.7\pm1.4$ & $1.8\pm1.3$ & $1.0\pm0.8$ & $0.8\pm0.7$ & -- & -- \\
{\small TWA\,22} & $35\pm11$ & $9.0\pm1.1$ & $5.6\pm0.6$ & $4.0\pm0.6$ & $2.7\pm0.2$ & $2.8\pm0.4$ & $2.1\pm0.6$ & $1.3\pm0.3$ & $1.2\pm0.3$ & $0.85\pm0.17$ & $0.43\pm0.07$ & $0.36\pm0.13$ & $0.6\pm0.2$ \\
{\small TWA\,27\,{\it [P084]}} & $6.8\pm0.5$ & $2.8\pm0.2$ & $3.0\pm0.3$ & $2.6\pm0.2$ & $1.7\pm0.4$ & $1.6\pm0.3$ & $1.2\pm0.7$ & $0.8\pm0.5$ & $0.7\pm0.5$ & $0.5\pm0.3$ & $0.3\pm0.3$ & -- & -- \\
{\small TWA\,27\,{\it [P089]}} & $14.9\pm0.3$ & $2.46\pm0.09$ & $1.56\pm0.04$ & $1.20\pm0.07$ & $0.82\pm0.03$ & $0.83\pm0.06$ & $0.60\pm0.08$ & $0.44\pm0.06$ & $0.31\pm0.06$ & $0.29\pm0.08$ & $0.24\pm0.07$ & $0.17\pm0.03$ & $0.11\pm0.04$  \\
{\small TWA\,28} & $4.77\pm0.18$ & $1.92\pm0.09$ & $0.91\pm0.08$ & $0.61\pm0.09$ & $0.44\pm0.09$ & $0.47\pm0.14$ & $0.27\pm0.08$ & $0.19\pm0.14$ & $0.17\pm0.10$ & $0.17\pm0.11$ & -- & -- & -- \\
{\small TWA\,30A\,{\it [P087,\,ns]}} & $170\pm40$ & $36\pm4$ & $19\pm5$ & $12\pm6$ & $12\pm10$ & -- & -- & -- & -- & -- & -- & -- & -- \\
{\small TWA\,30A\,{\it [P087,\,ls]}} & $140\pm30$ & $24\pm3$ & $13\pm2$ & $7.1\pm1.6$ & $7\pm2$ & $7\pm2$ & -- & -- & -- & -- & -- & -- & -- \\
{\small TWA\,30A\,{\it [P287]}} & $18\pm4$ & $59\pm8$ & $33\pm6$ & $23\pm4$ & $20\pm5$ & $18\pm7$ & -- & -- & -- & -- & -- & -- & -- \\
{\small TWA\,30B\,{\it [P087]}}  & $0.043\pm0.006$ & $0.014\pm0.004$ & $0.008\pm0.003$ & $0.005\pm0.003$ & $0.0030\pm0.0013$ & --  & -- & -- & -- & -- & -- & -- & -- \\
{\small TWA\,30B\,{\it [P287]}} & $0.37\pm0.09$ & -- & -- & -- & -- & -- & -- & -- & -- & -- & -- & -- & -- \\
{\small TWA\,31}  & $488\pm9$ & $129\pm3$ & $73\pm4$ & $55\pm2$ & $26.8\pm1.2$ & $34\pm2$ & $30\pm3$ & $24\pm4$ & $20\pm2$ & $16.3\pm1.4$ & $14.9\pm1.3$ & $11.2\pm1.1$ & $7.7\pm0.6$  \\
{\small TWA\,32} & $198\pm14$ & $51\pm3$ & $37.8\pm1.5$ & $30.2\pm1.0$ & $22.4\pm0.8$ & $19.8\pm0.9$ & $14.8\pm0.7$ & $11.4\pm0.7$ & $8.9\pm0.3$ & $7.1\pm1.1$ & $5.7\pm0.3$ & $3.7\pm0.4$ & $2.5\pm0.3$ \\
{\small TWA\,40} & -- & -- & -- &  -- & -- & -- & -- & -- & -- & -- & -- & -- & -- \\
{\small J1247-3816} & $1.16\pm0.12$ & $0.36\pm0.09$ & $0.27\pm0.03$ & $0.35\pm0.09$ & -- & -- & -- & -- & -- & -- & -- & -- & -- \\
\hline
\end{tabular}
}
\end{sidewaystable*} 

\begin{table*}
\caption{Luminosities of other emission lines used as tracers of accretion, when detected.}
\label{tab:other_em_series}
\centering
\begin{tabular}{l c c c c}
\hline\hline
Star & $L_{Ca\,II\,\lambda3934}\, \pm$ err & $L_{He\,I\,\lambda4026}\, \pm$ err & $L_{He\,I\,\lambda5876}\, \pm$ err & $L_{Pa\beta}\, \pm$ err \\
 & $[\times 10^{-7} L_\odot]$ & $[\times 10^{-7} L_\odot]$ & $[\times 10^{-7} L_\odot]$ & $[\times 10^{-7} L_\odot]$  \\
\hline
{\small TWA\,1\,{\it [10~s]}} & $460\pm40$ & $110\pm50$ & $360\pm70$ & $2320\pm70$ \\
{\small TWA\,1\,{\it [60~s]}} & $450\pm30$ & $90\pm40$ & $330\pm130$ & $2250\pm100$ \\
{\small TWA\,3A} & $23\pm3$ & -- & $6\pm3$ & -- \\
\small{TWA\,8A} & $135\pm2$ & $8\pm5$ & $21\pm11$ & -- \\
{\small TWA\,8B} & $9.6\pm0.7$ & -- & $2.0\pm0.9$ & -- \\
{\small TWA\,22} & $5.4\pm0.4$ & -- & $1.2\pm0.3$ & -- \\
{\small TWA\,27\,{\it [P084]}} & $0.6\pm0.2$ & $0.20\pm0.17$ & -- & -- \\
{\small TWA\,27\,{\it [P089]}} & $0.21\pm0.03$ & $0.06\pm0.02$ & -- & -- \\
{\small TWA\,28} & $0.51\pm0.07$ & -- & -- & -- \\
{\small TWA\,30A\,{\it [P087,\,ns]}} & $33\pm12$ & -- & -- & -- \\
{\small TWA\,30A\,{\it [P087,\,ls]}} & $20\pm2$ & -- & -- & -- \\
{\small TWA\,30A\,{\it [P287]}} & $18\pm6$ & -- & $0.8\pm0.2$ & -- \\
{\small TWA\,30B\,{\it [P087]}}  & $0.034\pm0.004$ & -- & -- & --  \\
{\small TWA\,30B\,{\it [P287]}} & $0.35\pm0.09$ & -- & -- & -- \\
{\small TWA\,31}  & $15.1\pm0.8$ & $3.0\pm0.7$ & $11.1\pm0.6$ & $17.0\pm1.9$  \\
{\small TWA\,32} & $12.5\pm0.3$ & $1.1\pm0.2$ & $6.2\pm0.5$ & -- \\
{\small TWA\,40} &-- & -- & -- & -- \\
{\small J1247-3816} & -- & -- & -- & -- \\
\hline
\end{tabular}
\end{table*} 

\section{Additional figure: Results from the multicomponent fit}

\begin{figure*}
\centering
\includegraphics[width=\textwidth]{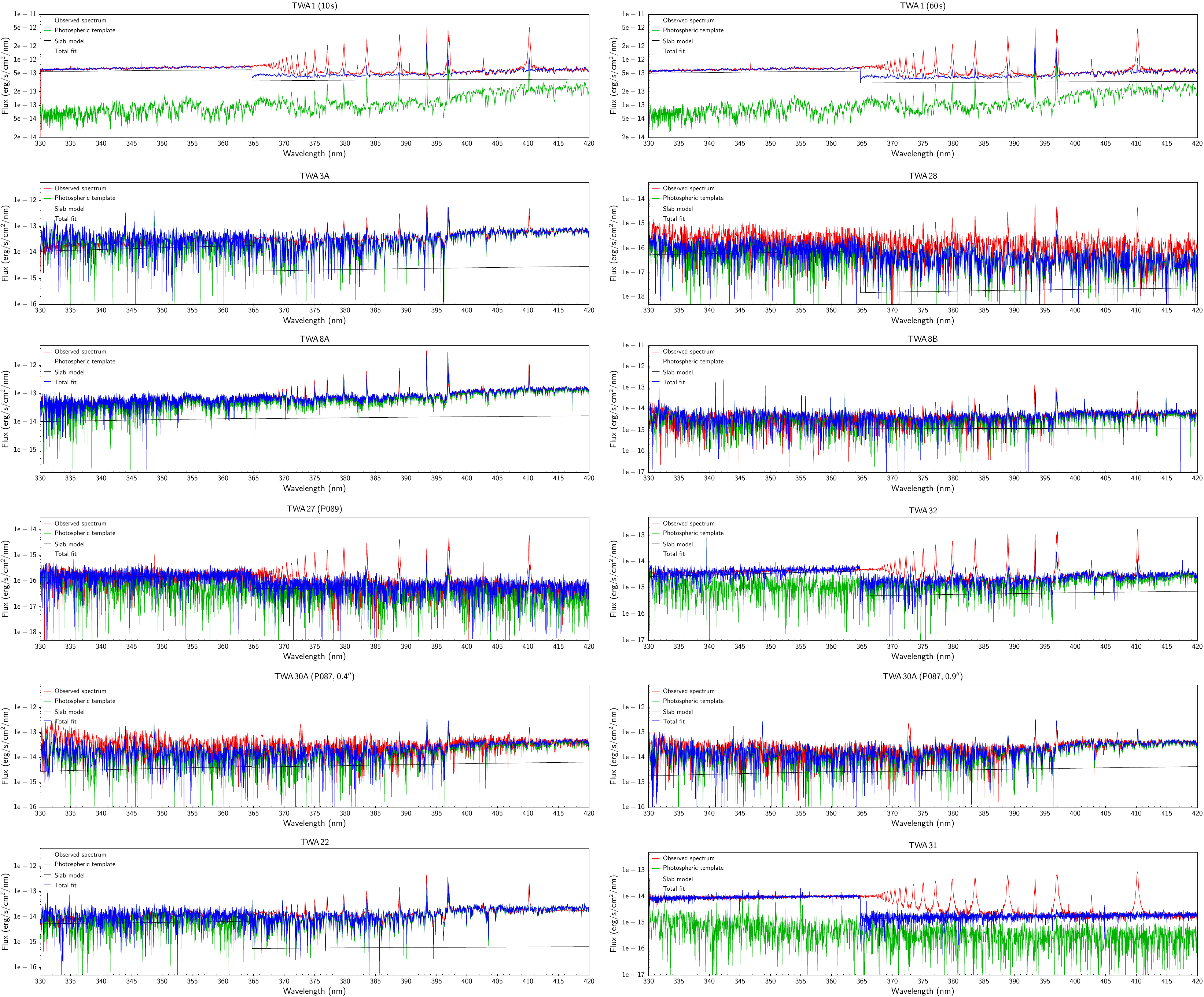}
\caption{Results of the multicomponent fit to the observed spectra of our targets (red) around the Balmer jump region. The best-fit photospheric template is illustrated in green, the continuum emission from the hydrogen slab is shown in black, and the total fit is traced in blue.}
\label{fig:TWA_fit_atlas}
\end{figure*}

\end{appendix}

\end{document}